\documentclass{pasa}%

\usepackage{graphicx}

\title[The Radioactive Universe]{The Radioactive Nuclei $^{\textbf{26\!}}$Al and $^{\textbf{60}}$Fe in the Cosmos and in the Solar System}

\author[Diehl et al.]{
R.~Diehl$^1$, 
M.~Lugaro$^{2,3,4}$, 
A.~Heger$^{4,5,6,7}$, 
A.~Sieverding$^{8,9}$, 
X.~Tang$^{10}$, 
K.~A.~Li$^{10}$, 
E.~T.~Li$^{11}$, 
C.~L.~Doherty$^{2,4}$,  
M.~G.~H.~Krause$^{12}$, 
A.~Wallner$^{13,14}$, 
N.~Prantzos$^{15}$,
H.~E.~Brinkman$^{2,16}$, 
J.~W.~den~Hartogh$^2$,  
B.~Wehmeyer$^{2,12}$, 
A.~Yag\"ue~L\'opez$^2$,
M.~M.~M.~Pleintinger$^1$,  
P.~Banerjee$^{17}$, 
W.~Wang$^{18,19}$
\affil{$^1$Max Planck Institut f\"ur extraterrestrische Physik, D-85748 Garching, Germany}%
\affil{$^2$Konkoly Observatory, 
E\"{o}tv\"{o}s Lor\'and Research Network (ELKH), H-1121 Budapest, Konkoly Thege M. \'ut 15-17, Hungary}
\affil{$^3$ELTE E\"{o}tv\"{o}s Lor\'and University, Institute of Physics, Budapest 1117, P\'azm\'any P\'eter s\'et\'any 1/A, Hungary}
\affil{$^4$School of Physics and Astronomy, Monash University, VIC 3800, Australia}
\affil{$^{5}$Australian Research Council Centre of Excellence for Gravitational Wave Discovery (OzGrav), Clayton, Vic 3800, Australia}%
\affil{$^{6}$Center of Excellence for Astrophysics in Three Dimensions (ASTRO-3D), Australia}%
\affil{$^{7}$Joint Institute for Nuclear Astrophysics, 1 Cyclotron Laboratory, NSCL, 
Michigan State University, East Lansing, MI 48824
, USA}%
\affil{$^8$School of Physics and Astronomy, University of Minnesota, Minneapolis, MN 55455 , USA}
\affil{$^9$Physics Division, Oak Ridge National Laboratory, Oak Ridge, TN 37831, USA}
\affil{$^{10}$Institute of Modern Physics, Chinese Academy of Sciences, Lanzhou, P.R. China}
\affil{$^{11}$College of Physics and Optoelectronic Engineering, ShenZhen University, P.R. China}
\affil{$^{12}$Centre for Astrophysics Research, 
University of Hertfordshire, College Lane, Hatfield, Hertfordshire, AL10 9AB, UK}
\affil{$^{13}$Helmholtz-Zentrum Dresden-Rossendorf, Institute of Ion Beam Physics and Materials Research, 01328 Dresden, Germany}
\affil{$^{14}$Research School of Physics, Australian National University, Canberra, ACT 2601, Australia}%
\affil{$^{15}$Institut d'Astrophysique, Paris, France}%
\affil{$^{16}$Graduate School of Physics, University of Szeged, Dom ter 9, Szeged, 6720 Hungary}%
\affil{$^{17}$Discipline of Physics, Indian Institute of Technology Palakkad, Kerala, India 678557}
\affil{$^{18}$School for Physics and Technology, Wuhan University, Wuhan 430072, P.R. China}
\affil{$^{19}$WHU-NAOC Joint Center for Astronomy, Wuhan University, Wuhan 430072, P.R. China}
}%

\jid{PASA}
\doi{10.1017/pas.\the\year.xxx}
\jyear{\the\year}

\usepackage{aas_macros}
\usepackage{hyperref}
\hypersetup{colorlinks,citecolor=blue,linkcolor=blue,urlcolor=blue}

\newcommand{\about}{\ensuremath{\sim}}
\newcommand{\iso}[2]{\hbox{${}^{#1}{\rm #2}$}}
\newcommand{\Msun}{\ensuremath{{M}_{\odot}}}
\newcommand{\Msol}{\ensuremath{{M}_{\odot}}}
\newcommand{\ms}{M$_\odot$}
\newcommand{\Al}{$^{26\!}$Al}
\newcommand{\Fe}{$^{60}$Fe}
\newcommand{\al}{{$^{26\!}$Al~}}
\newcommand{\alu}{{$^{26\!}$Al}}
\newcommand{\fe}{{$^{60}$Fe~}}
\newcommand{\feu}{{$^{60}$Fe}}
\newcommand{\HI}{ H{\sc{i}}~}
\newcommand{\NaI}{ Na{\sc{i}}~}
\newcommand{\msun}{{M$_\odot$}}

\newcommand{\kms}{{km~s$^{-1}$}}

\newcommand{\nar}{New Astronomy Reviews}

\begin{document}

\begin{frontmatter}
\maketitle

\begin{abstract}
The cosmic evolution of the chemical elements from the Big Bang to the present time is driven by nuclear fusion reactions inside stars and stellar explosions.  A cycle of matter recurrently re-processes metal-enriched stellar ejecta into the next generation of stars.
The study of cosmic nucleosynthesis and of this matter cycle requires the understanding of the physics of nuclear reactions, of the conditions at which the nuclear reactions are activated inside the stars and stellar explosions, of the stellar ejection mechanisms through winds and explosions, and of the transport of the ejecta towards the next cycle, from hot plasma to cold, star-forming gas.
Due to the long timescales of stellar evolution, and because of the infrequent occurrence of stellar explosions, observational studies are challenging, as they have biases in time and space as well as different sensitivities related to the various astronomical methods.
Here, we describe in detail the astrophysical and nuclear-physical processes involved in creating two radioactive isotopes useful in such studies, $^{26\!}$Al and $^{60}$Fe.
Due to their radioactive lifetime of the order of a million years these isotopes are suitable to characterise simultaneously the processes of nuclear fusion reactions and of interstellar transport.
We describe and discuss the nuclear reactions involved in the production and destruction of $^{26\!}$Al and $^{60}$Fe, the key characteristics of the stellar sites of their nucleosynthesis and their interstellar journey after ejection from the nucleosynthesis sites.
This allows us to connect the theoretical astrophysical aspects to the variety of astronomical messengers presented here, from stardust and cosmic-ray composition measurements, through observation of $\gamma$ rays produced by radioactivity, to material deposited in deep-sea ocean crusts and to the inferred composition of the first solids that have formed in the Solar System.
We show that considering measurements of the isotopic ratio of $^{26\!}$Al to $^{60}$Fe eliminate some of the unknowns when interpreting astronomical results, and discuss the lessons learned from these two isotopes on cosmic chemical evolution.
This review paper has emerged from an ISSI-BJ Team project in 2017--2019, bringing together nuclear physicists, astronomers, and astrophysicists in this inter-disciplinary discussion.

\end{abstract}

\end{frontmatter}

\section{Introduction}
\label{sec:intro}

Understanding the cosmic evolution of the composition of matter from the Big Bang until the present time requires tracing the ensemble of atomic nuclei through their nuclear transformations on their journey across space and time.
These transformations are called \emph{nucleosynthesis}:  nuclear reactions that rearrange how protons and neutrons are grouped into the different isotopes of the chemical elements.
In nature, nuclear reactions may occur through collisions or disintegration of nuclei in hot and energetic environments, such as the Big Bang, stellar explosions, the hot interiors of stars, and the interstellar space where they involve accelerated cosmic-ray particles.
Rearrangements of nucleons through nuclear reactions therefore drive the change of elemental and isotopic composition in the Universe from the almost pure H and He made in the Big Bang to the current rich variety of elements, including C to U, that also enables biological life.
This process is called \emph{chemical evolution}\footnote{Although there is no chemistry involved in the evolution of elemental and isotopic abundances.}.  In this review, we will disentangle the processes involved by picking specific nuclei as examples, and tracing their origins and cosmic journey to us.



\begin{figure}
\centering
\includegraphics[width=\columnwidth]{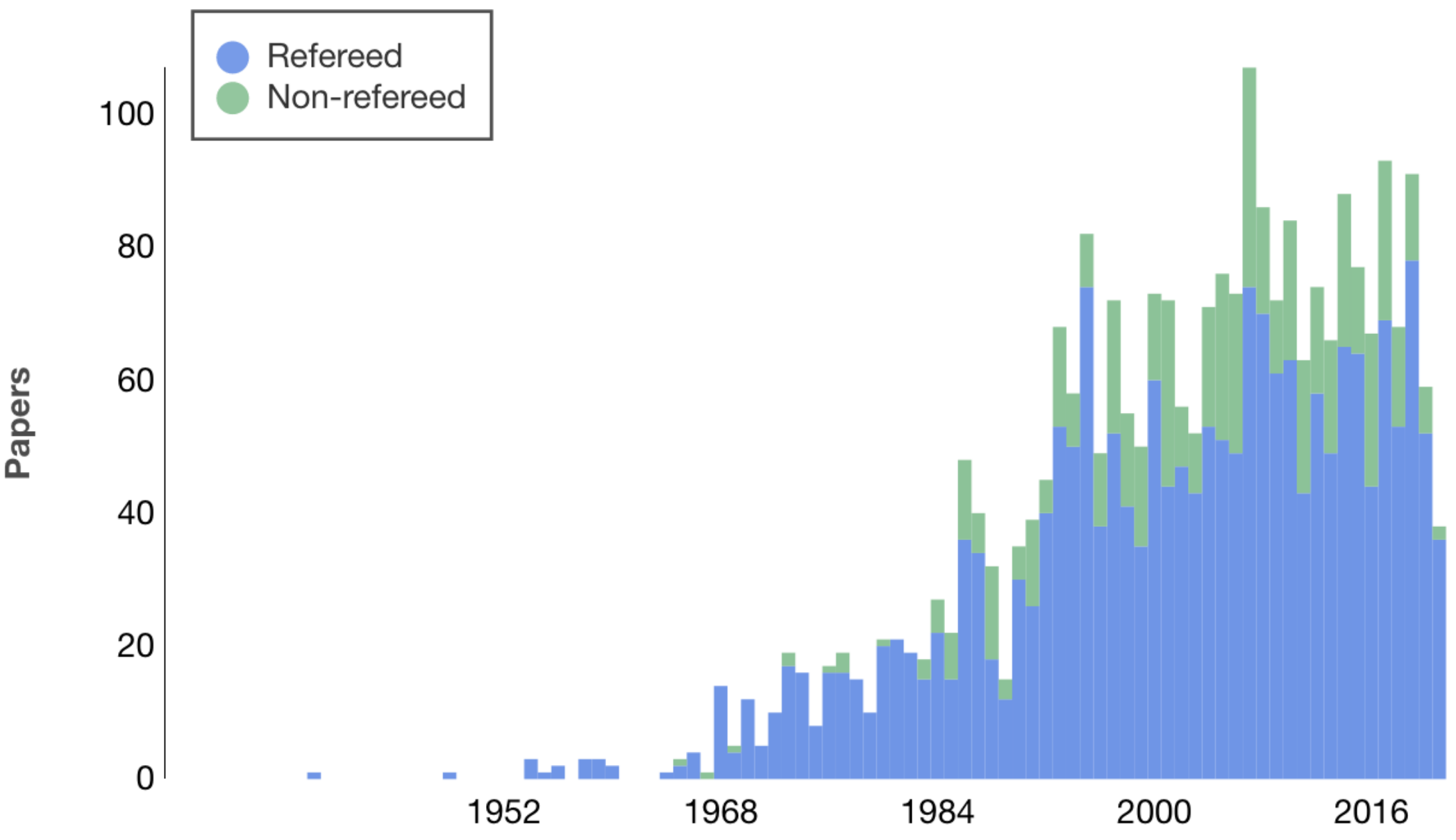}
\includegraphics[width=\columnwidth]{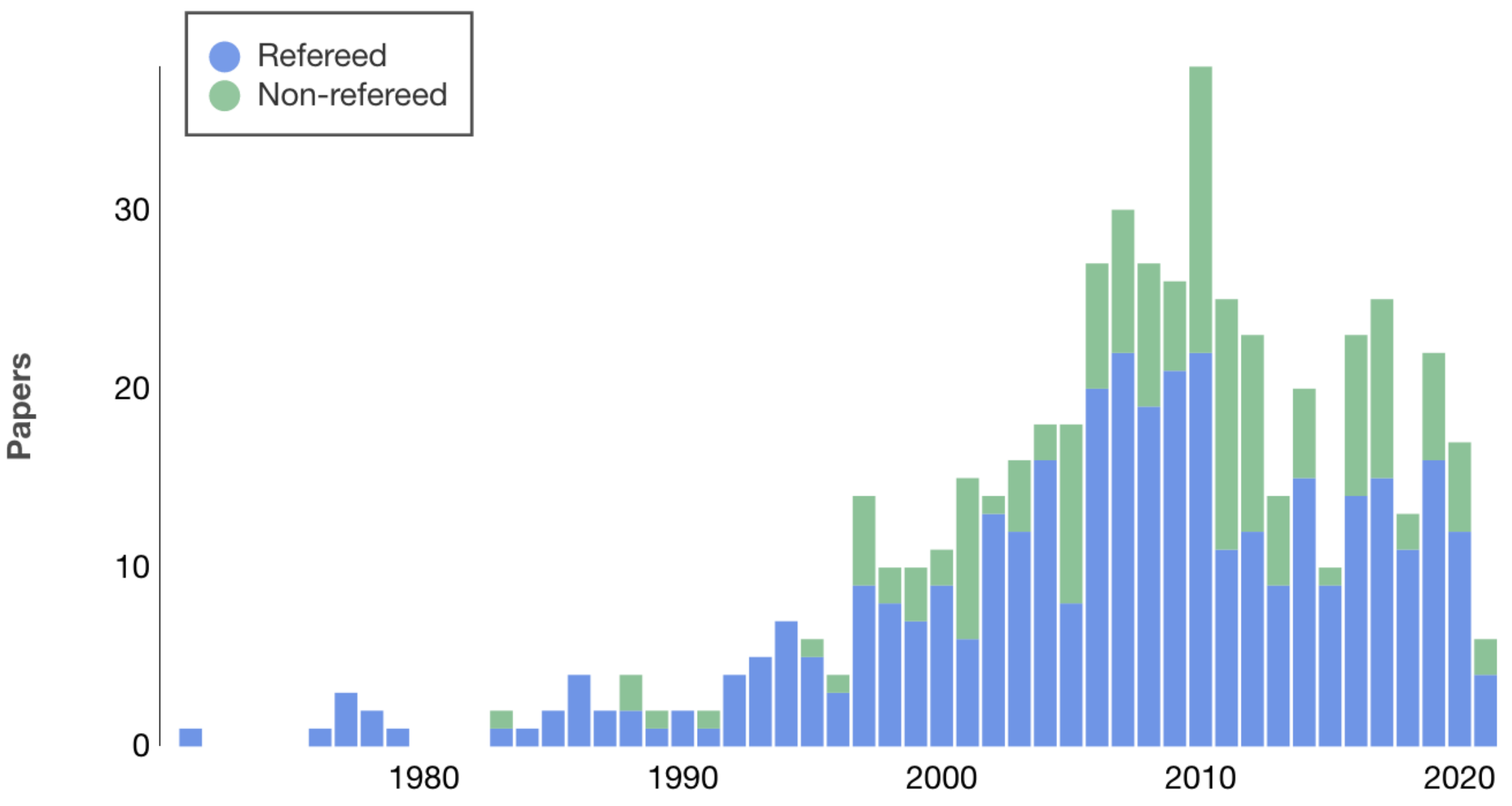}
\caption{Scientific publications per year, addressing $^{26}$Al (above) and $^{60}$Fe (below). A total of $>$2,000 refereed papers with $>$25,000 citations and $>$300,000 reads (for $^{26}$Al) represent the size of the community involved in these topics. (Data and plots from NASA ADS). }
\label{fig_publicationRecord}
\end{figure}

The relative abundances of different isotopes in a given material are the result of the nucleosynthetic episodes that such an ensemble of nucleons and isotopes has experienced along its cosmic trajectory.
First, we have to understand the nucleosynthesis processes themselves, within stars and stellar explosions, that modify the nuclear composition; the nuclear reactions here mostly occur in low-probability tails at energies of tens of keV, which in many cases is far from what we can study by experiments in terrestrial laboratories, so that often sophisticated extrapolations are required.
Beyond these nuclear reactions and their sites, we have to understand how nuclei are transported in and out of stellar nucleosynthesis sites and towards the next generation of stellar nucleosynthesis sites throughout the Galaxy.
A key ingredient is the path through the interstellar matter towards newly-forming stars, after nuclei have been ejected from the interior of a star by a stellar wind or a stellar explosion.

It is possible to measure interstellar isotopes and their relative abundances directly, by suitably capturing cosmic matter and then determining its isotopic composition, e.g., using mass spectrometry.  In fact, cosmic matter rains down onto Earth continuously in modest but significant quantity -- the discovery of live radioactive $^{60}$Fe isotopes in Pacific ocean crusts \citep{Knie:2004} and in galactic cosmic rays \citep{Binns2016} have demonstrated this. It is a major challenge for astronomical instrumentation, however, to determine abundances of cosmic nuclei for regions that are not accessible through material transport or spacecraft probes, i.e., in different parts of our current and past Universe.  For example, in starlight spectra only some isotopic signatures may be recognised, and only when measuring at extremely high spectral resolution.

Astronomical abundance measurements are subject to biases,  in particular, because atomic nuclei appear in different phases, such as plasma, neutral or partially-ionized atoms, or molecules. Therefore, observational signals differ from each other. For example, an elemental species may be accelerated as cosmic rays or condensed into dust, depending on how a meteoric inclusion, such as a pre-solar dust grain, had been formed, or how an ion mixture may generate an observable spectral line in the atmosphere of a star, characteristically absorbing the starlight originating from the interiors of stars.
Observations of cosmic isotopes are rather direct if radioactive isotopes can be seen via their radioactive-decay signatures outside stars, i.e., without biases and distortions from absorption. This is possible when characteristic $\gamma$-ray lines are measured from such radioactive decay.  The detection of characteristic $^{26\!}$Al decay $\gamma$ rays \citep{Mahoney:1982} was the first direct proof that nucleosynthesis must be ongoing within the current Galaxy, because this isotope has a characteristic decay half-life of $0.72\,$Myr, much shorter than the age of the Galaxy, more than 10 Gyr.
$^{26\!}$Al, and, similarly, $^{60}$Fe (with a half-life of 2.62~Myr), both probe recent nucleosynthesis and ejecta transport.   They have been measured in $\gamma$ rays from interstellar space, have been found in terrestrial deposits, and have even been inferred to exist in specific abundance in the first solids that formed in the Solar System 4.6~Gyr ago.
These two isotopes exemplify a new approach to cosmic chemical evolution studies, which involves a wide community, from nuclear physicists through Solar System scientists, astrophysical theorists, and astronomers working on a broad range of topics.  As a result, there is a significant diversity of scientific publications addressing these two isotopes, with discussions increasing in intensity over the past two decades (Figure~\ref{fig_publicationRecord}).
This review focuses on discussion of these two specific isotopes, in relation to the nuclear and astrophysical processes involved in the cycle of matter that drives cosmic chemical evolution.

In this paper, we assemble and combine the different views on this theme from a working group on ``Radioactive Nuclei in the Cosmos
and in the Solar System''
that met at ISSI-Beijing\footnote{The International Space Science Institute ISSI has its main home in Bern, Switzerland, and a satellite institute in Beijing. Scientific workshops and working groups are one main asset of the ISSI in support of the scientific community.} in 2018 and 2019.  The team included astronomers, theorists in various aspects of astrophysics and nuclear physics, as well as nuclear physics experimentalists.  The members of the working group covered a variety of different expertises and interests
and we chose to exploit this diversity to describe the journey of cosmic isotopes from a nuclear astrophysics perspective using the two isotopes $^{26\!}$Al and $^{60}$Fe as examples.  We describe the properties of these nuclei and their reactions with other nuclei, the astrophysical processes involved in their production, and how observations of their abundance ratio can be exploited to learn about which nuclear transformations happen inside stars and their explosions.

The main goal of this paper is to pose the scientific questions in all their detail, not to provide ultimate consensus nor answers.
We aim to illuminate the approximations and biases in our way of arguing and learning, as this is important for all theory, observations, and experiment.  Ideally, we wish to identify critical observations, experiments, and simulations that can help to validate or falsify these approximations, towards a better understanding of the physical processes involved in transforming the initial H and He during cosmic evolution into the material mix that characterises our current, life-hosting Universe.

In Section~\ref{sec:Al26}, we focus on the case of $^{26\!}$Al and carry this discussion from nuclear properties and reaction physics through cosmic nucleosynthesis sites to interstellar transport and creation of astronomical messengers. Section~\ref{sec:Fe60} discusses the case of $^{60}$Fe and what is different from the case of $^{26\!}$Al in relation to each of those processes for $^{60}$Fe. Section~\ref{sec:60Fe26Alratio} shows how investigation of the abundance ratio of these two isotopes allows to eliminate some of the unknowns in astrophysical modelling and interpretation. Our conclusions (Section~\ref{sec:conclusions}) summarise the nuclear physics, astrophysics, astronomical, and methodological issues, and the lessons learned as well as the open questions from the study of $^{26\!}$Al and $^{60}$Fe in the context of cosmic chemical evolution.

\section{The cosmic trajectory of $^{26}$Al}
\label{sec:Al26}
\subsection{Nuclear properties, creation and destruction reactions}
\label{sec:al26nuclear}



\begin{figure}
\centering
	\includegraphics[width=\linewidth]{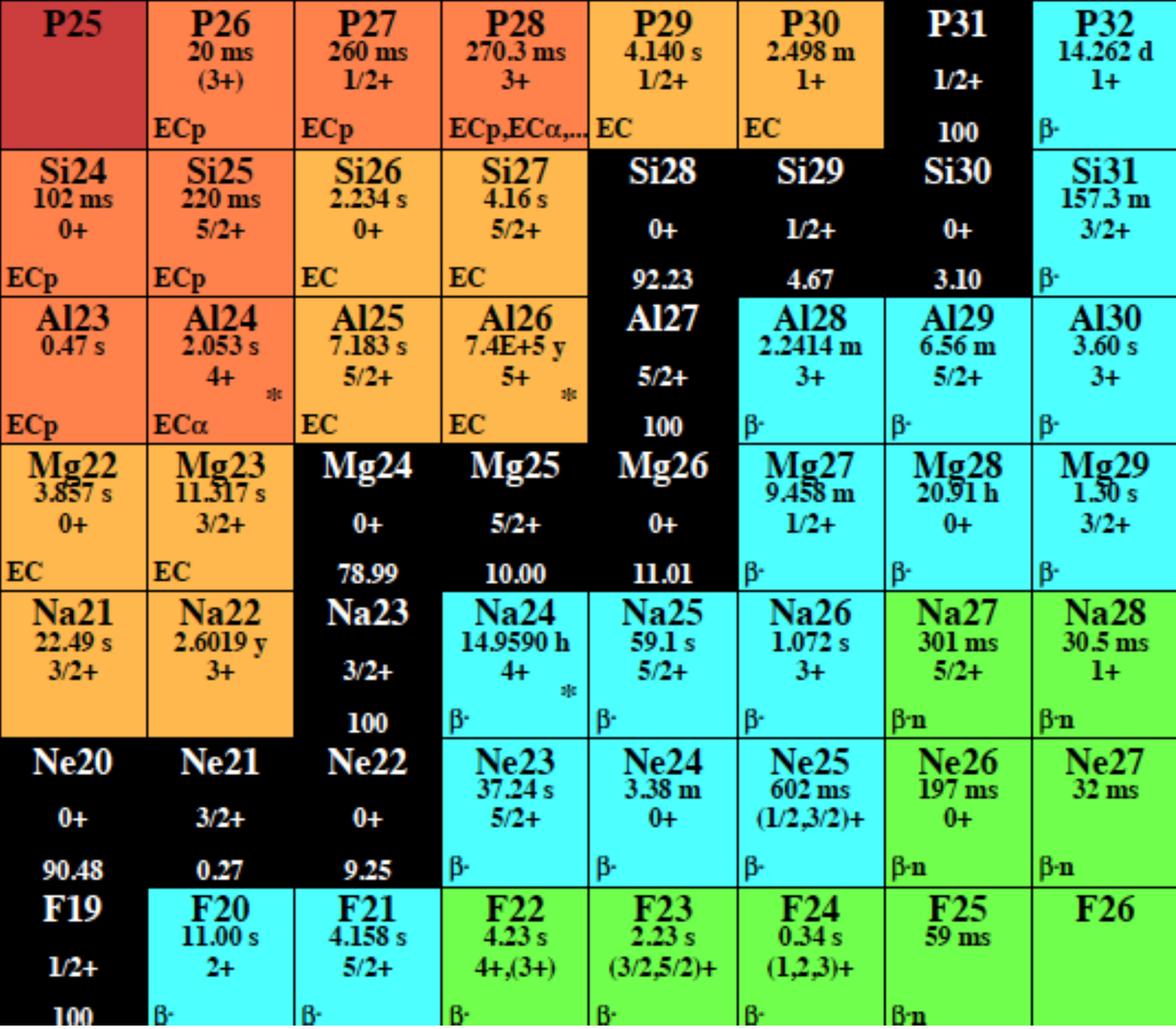}
	\caption{The table of isotopes in the neighbourhood of ${}^{26\!}$Al.
	Each isotope is identified by its usual letters and the total number of nucleons, with stable  isotopes and black and unstable isotopes in colored boxes.  The second line for unstable isoptopes indicates the lifetime. The third line lists spin and parity of the nucleus ground state.
The primary decay channel is indicated in the bottom left. The stable elements have their abundance fractions on Earth in the last row. (extracted from Karlsruher Nuklidkarte, original by the JRC of the EU)
	}
	\label{fig:isotopes_Al-region}
\end{figure}

\subsubsection{Nuclear properties of $^{26}$Al}

\begin{figure*}
\centering
\includegraphics[width=1.5\columnwidth]{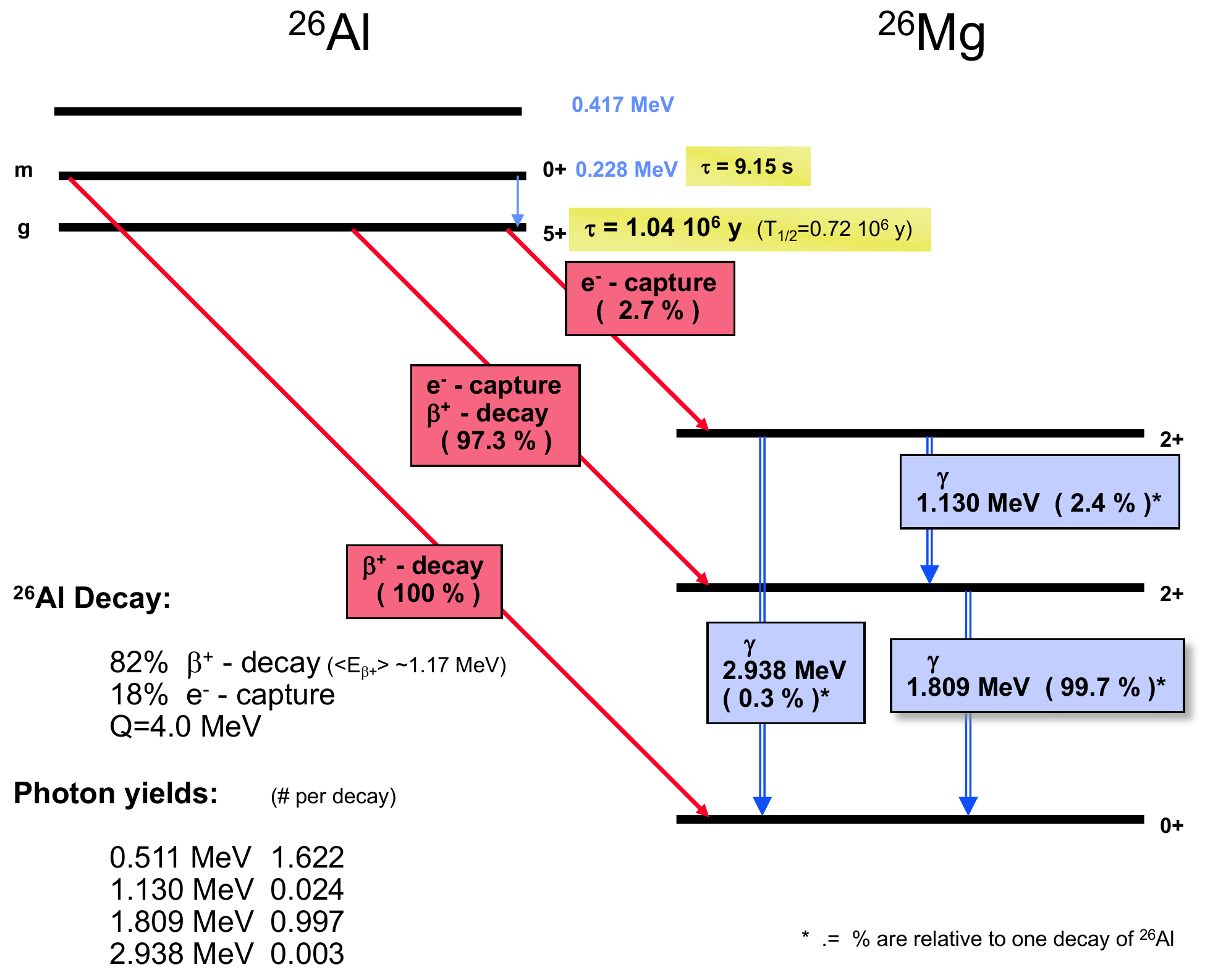}
\caption{The nuclear level and decay scheme of $^{26}$Al (simplified). $\gamma$~rays are listed as they arise from decay of $^{26}$Al, including annihilation of the positrons from $\beta^+$-decay.}
\label{fig_26AlDecay}
\end{figure*}

Figure~\ref{fig:isotopes_Al-region} shows the \al isotope within its neighbouring nuclides, with $^{27}$Al as the only stable isotope of Al.
The ground state of $^{26}$Al ($^{26}$Al$^g$) (see Figure~\ref{fig_26AlDecay}) has a spin and parity of  $5^+$ and a $\beta^+$-decay half-life of 0.717 Myr.
It decays into the  first excited state of $^{26}$Mg ($1809$~keV; $2^+$), which then undergoes $\gamma$-decay to the ground state of $^{26}$Mg producing the characteristic $\gamma$ ray at 1808.63~keV. The first excited state of $^{26}$Al at 228~keV ($^{26}$Al$^m$) is an isomeric state with a spin and parity of $0^+$.
It is directly connected to the $^{26}$Al$^g$ state via the highly-suppressed $M5$ $\gamma$-decay with a half-life of 80,500~yr according to shell model calculations \citep{coc+2000,ref3_PhysRevC.97.065807}. $^{26}$Al$^m$ decays with a half-life of just 6.346~s almost exclusively to the ground state of $^{26}$Mg via super-allowed $\beta^+$-decay\citep{ref1_Audi_2017}, with a branching ratio of 100.0000$^{+0}_{-0.0015}$ \citep{ref4_PhysRevC.85.055501}.

In cosmic nucleosynthesis, the correct treatment of $^{26}$Al$^m$ and $^{26}$Al$^g$ in reaction network calculations is crucial \citep{runkle+2001,ref12_PhysRevC.64.025805}. When ${}^{26}$Al is produced by a nuclear reaction, it is produced in an excited state, which rapidly decays to the isomeric and/or ground states by a series of $\gamma$-ray cascades. At low temperatures ($T\lesssim$ 0.15~GK), communication between $^{26}$Al$^m$ and $^{26}$Al$^g$ can be ignored due to the negligibly-low internal transition rates. Therefore, $^{26}$Al$^m$ and $^{26}$Al$^g$ can be treated as two distinct species  with their separate production and destruction reaction rates \citep{ref10_Iliadis_2011}.

At higher temperatures ($T\gtrsim$ 0.4~GK), instead, higher excited states of $^{26}$Al can be populated on very short timescales by photo-excitation of $^{26}$Al$^g$ and $^{26}$Al$^m$ resulting in thermal equilibrium where the abundance ratio of the states are simply given by the Boltzmann distribution.
In this case, it is sufficient to have just one species of $^{26}$Al in reaction network calculations defined by its thermal equilibrium ($^{26}$Al$^t$), with suitable reaction rates that take into account the contributions from all the excited states that are populated according to the Boltzmann distribution  \citep{ref10_Iliadis_2011}.

The situation becomes complicated at intermediate temperatures (0.15~GK $\lesssim$ T $\lesssim$ 0.40~GK). Although, $^{26}$Al$^g$ and $^{26}$Al$^m$ can still communicate with each other via the higher excited states, the timescale required to achieve thermal equilibrium becomes comparable or even longer than the timescale for $\beta^+$-decay for $^{26}$Al$^m$ (as well as $\beta^+$-decay of higher excited states).
Thus, neither the assumption of thermal equilibrium nor treating $^{26}$Al$^g$ and $^{26}$Al$^m$ as two separate species are viable options \citep{ref3_PhysRevC.97.065807,misch+2021}. In this case, it becomes necessary to treat at least the lowest four excited states as separate species in the reaction network, along with their mutual internal transition rates, in order to calculate the abundance of $^{26}$Al accurately \citep{ref10_Iliadis_2011}. However, as will be discussed below, it turns out that the production of $^{26}$Al in stars happen mostly either in the low or the high temperature regime, and the problematic intermediate temperature regime is rarely encountered.

\subsubsection{Production and destruction of $^{26}$Al}
\label{sec:rates}


\begin{figure}
	\includegraphics[width=\columnwidth]{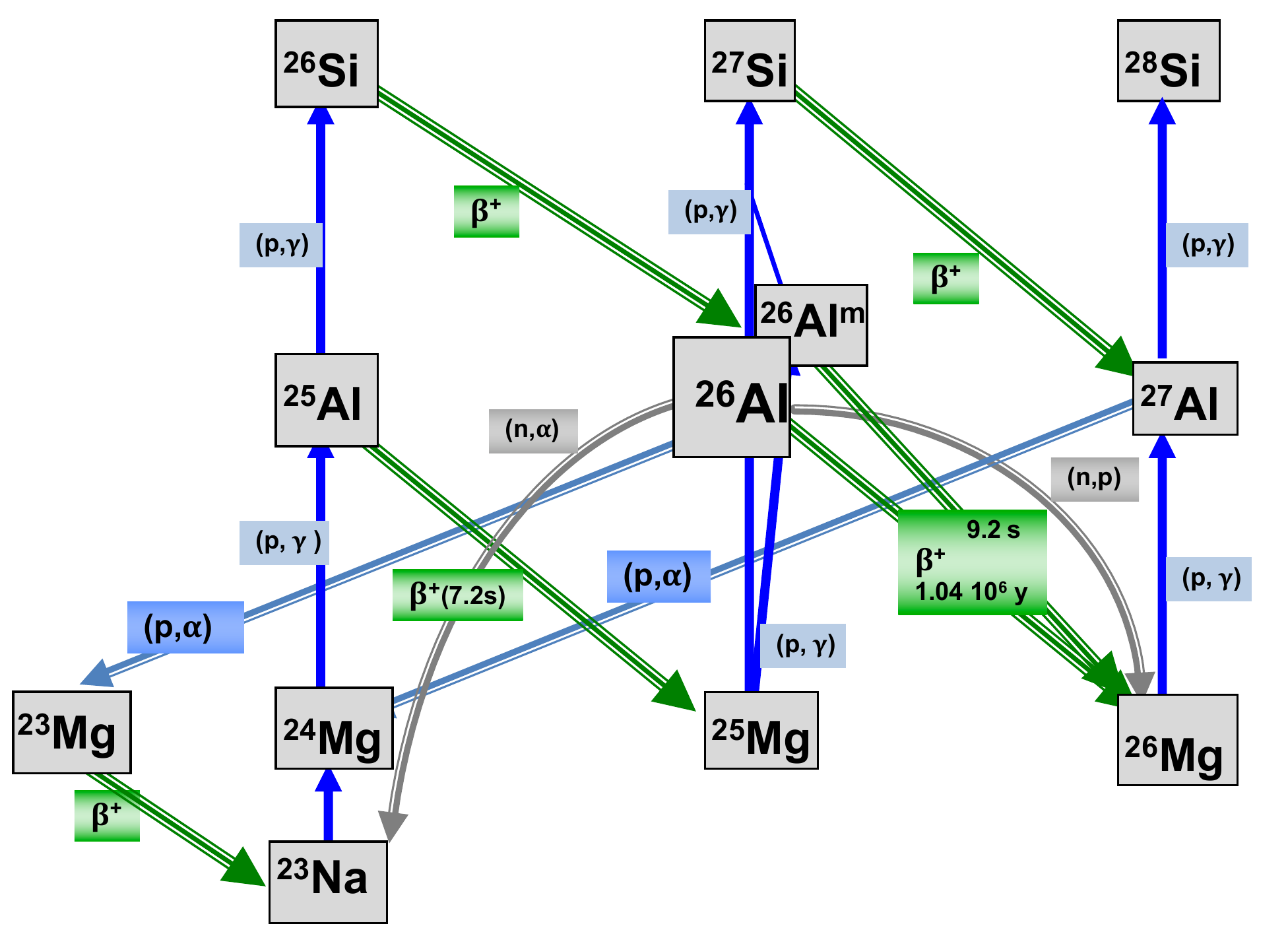}
	\caption{The Na-Mg-Al cycle encompasses production and destruction reactions, and describes $^{26}$Al in stellar environments.}
	\label{fig:Na-Al_cycle}
\end{figure}

$^{26}$Al is expected to be primarily produced in the hydrostatic burning stages of stars through p-capture reactions on $^{25}$Mg. These occur in massive stars during core hydrogen burning, hydrostatic/explosive carbon/neon shell burning, and in the hydrogen-burning shell, in some cases located at the base of convective envelope, of asymptotic giant branch (AGB) stars. Explosive oxygen/neon shell burning probably also contributes to the production of this isotope. All these sites are be described in more detail in Section~\ref{sec:stars}. The typical temperatures of the H-burning core in massive stars and the H-burning shell of AGB stars are T = 0.04 - 0.09 GK.
 In these environments, ${}^{26}$Al is produced by $^{25}$Mg($p$,$\gamma$)$^{26}$Al$^{g,m}$ acting on the initial abundance of \iso{25}Mg within the MgAl cycle shown in Figure~\ref{fig:Na-Al_cycle}. \iso{25}Mg can also be produced by the   $^{24}$Mg($p$,$\gamma$)$^{25}$Al($\beta^+$)$^{25}$Mg reaction chain at the temperature above 0.08GK.
At such low temperatures, there is no communication between $^{26}$Al$^{g}$ and $^{26}$Al$^{m}$. $^{26}$Al$^{g}$ may be destroyed by  $^{26}$Al$^{g}$($p$,$\gamma$)$^{27}$Si and by the $\beta^+$-decay.

\begin{figure}
	\includegraphics[width=\columnwidth]{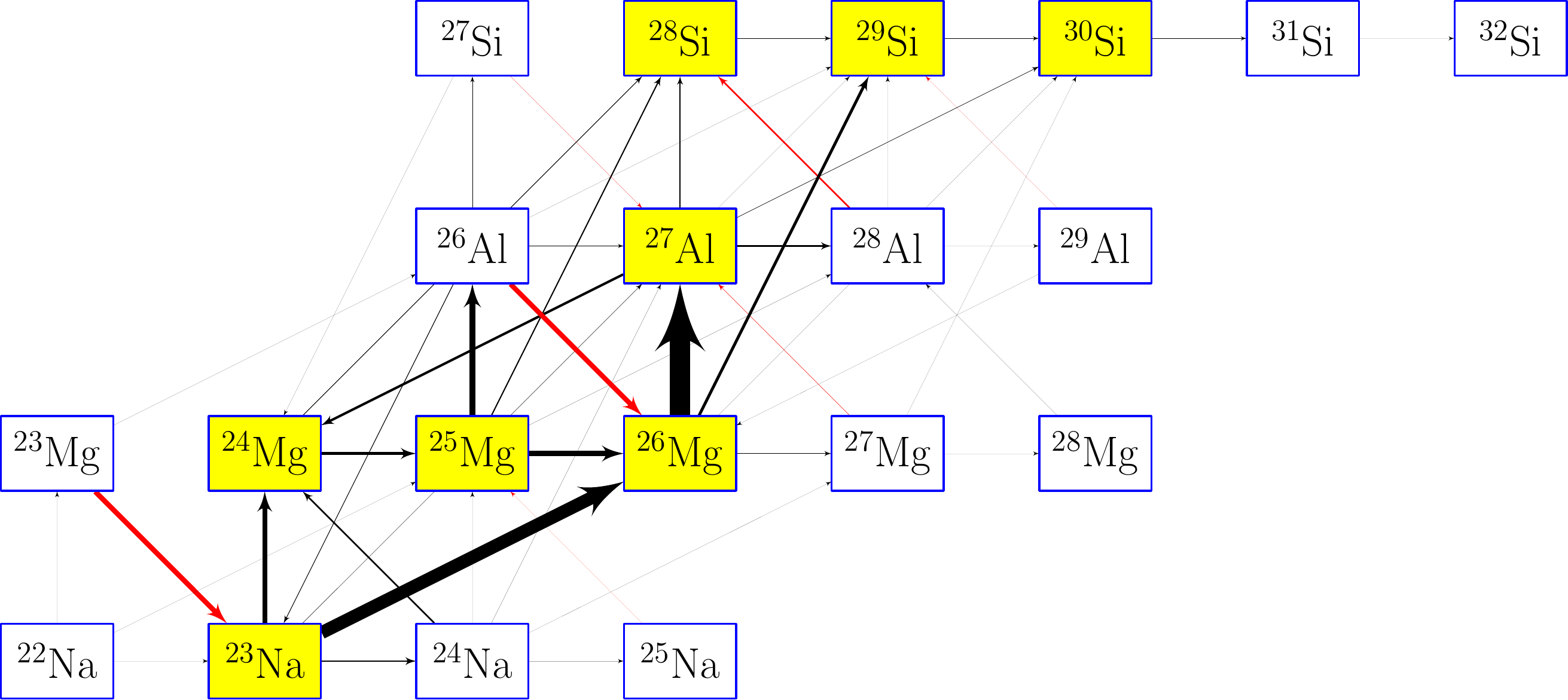}
	\caption{Integrated reaction flow for the hydrostatic C/Ne shell burning calculated with the NUCNET nuclear network code. The thickness of the arrows correspond to the intensities of the flows; red and black arrows show $\beta$ interactions and nuclear reactions, respectively. Here $^{26}$Al is at its thermal equilibrium. Only a fraction of the flows of Na, Mg, Al and Si are displayed. The neutron source reactions, such as $^{12}$C+$^{12}$C and $^{22}$Ne($a$,$n$), are not shown.}
	\label{fig:flow_shell_burning}
\end{figure}

\begin{figure}
	\includegraphics[width=\columnwidth]{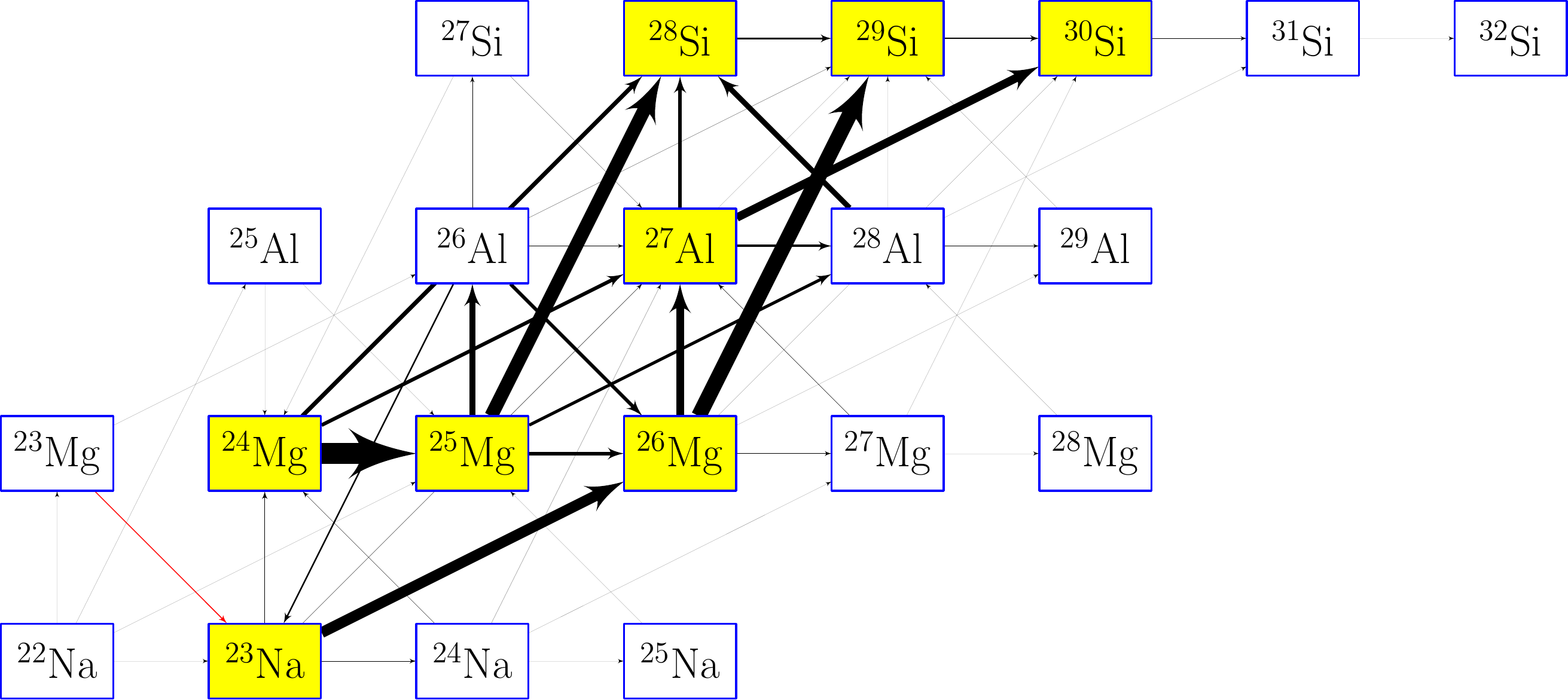}
	\caption{Same as Figure~\ref{fig:flow_shell_burning} but for C/Ne explosive burning.}
	\label{fig:flow_expl_burning}
\end{figure}

Hydrostatic C/Ne shell burning occurs at a temperature around 1.2 GK.
Here,  $^{26}$Al is produced by the $^{24}$Mg($n$,$\gamma$)$^{25}$Mg($p$,$\gamma$)$^{26}$Al$^{t}$ reaction chain. The detailed flow chart is shown in Figure~\ref{fig:flow_shell_burning}.
At the temperature of C/Ne shell burning, $^{26}$Al reaches thermal equilibrium and can be treated at a single species, $^{26}$Al$^{t}$ (see above).
Destruction of $^{26}$Al mostly occurs through neutron capture reactions.
The main neutron sources are the $^{22}$Ne($\alpha$,n)$^{25}$Mg and ${}^{12}$C(${}^{12}$C,n)${}^{23}$Mg reactions. ${}^{26}$Al$^{t}$ is also destroyed by the $\beta^+$-decay process in C/Ne shell burning.
The explosive C/Ne shell burning may raise the temperature up to 2.3~GK and then quickly cool down to 0.1~GK within a time scale of 10~s.
The detailed flow chart in these conditions is shown in Figure~\ref{fig:flow_expl_burning}. ${}^{26}$Al is produced by the same process as during hydrostatic C/Ne shell burning, except that the ${}^{23}$Na($\alpha$,p)${}^{26}$Mg reaction competes with ${}^{23}$Na($p$,$\gamma$)${}^{24}$Mg and the $^{25}$Mg($\alpha$,$n$)$^{28}$Si reaction competes with ${}^{25}$Mg($p$,$\gamma$)$^{26}$Al$^t$.
These two $\alpha$-induced reactions bypass the the production of ${}^{26}$Al$^t$. ${}^{26}$Al$^t$ is primarily destroyed by ${}^{26}$Al$^t$($n$,$p$)$^{26}$Mg instead of $\beta^+$-decay.

In an explosive proton-rich environment such as within a nova, the peak temperature may reach about 0.3~GK. Here, $^{26}$Al is produced by two sequences of reactions: $^{24}$Mg($p$,$\gamma$)$^{25}$Al($\beta^+$)$^{25}$Mg($p$,$\gamma$)$^{26}$Al$^{g,m}$, and  $^{24}$Mg($p$,$\gamma$)$^{25}$Al($p$,$\gamma$)$^{26}$Si($\beta^+$)$^{26}$Al$^{g,m}$, which favours the production of $^{26}$Al$^{m}$, therefore bypassing the observable $^{26}$Al$^{g}$.


$^{26}$Al can also be directly
produced in the core-collapse supernova $\nu$~process via $^{26}$Mg$(\nu_e,e^-)$
\citep{woosley:1990}, when the high-energy ($\sim 10$ MeV) neutrinos emitted
during the collapse and cooling of a massive star interact
with nuclei in the mantle that is processed by the explosion shock at the same time.
Neutrino-nucleus reactions that lead to proton emission also increase the production of $^{26}$Al via the reactions discussed above.
The contribution of the $\nu$~process to the total
supernova yield is expected to be at the $10\%$ level \citep{Sieverding:2017,Timmes:1995}; we caution that this value is subject to uncertainties in the neutrino physics and the details of the supernova explosion mechanism.


\subsubsection{Uncertainties in the relevant reaction rates }
\label{sec:al26_nuclear_uncertainties}

The uncertainties of the rates of main production reactions $^{25}$Mg(p,$\gamma$)$^{26}$Al$^{g}$ and $^{25}$Mg(p,$\gamma$)$^{26}$Al$^{m}$ are around 10\% at T$_9$>0.15; at lower temperatures, the uncertainties are even larger than 30\% \citep{iliadis_2010NuPhA_rate_compilation}.
Since there is little communication between $^{26}$Al$^g$ and $^{26}$Al$^m$ at T$_9$<0.15, these two reactions  need to be determined individually. Two critical resonance strengths, at center-of-mass energies $E_{c.m.}=$92 and 198 keV, have been measured using the LUNA underground facility with its accelerator \citep{mg25pg_luna_Strieder_2012}. However, due to the lack of statistical precision of the measurement and of decay transition information, the branching ratio of the ground state transition still holds a rather large uncertainty, in spite of
some progress from a recent measurement with Gammasphere at ANL \citep{mg25pg_anl_Kankainen_2021}.  Prospects to re-study this resonance with better statics are offered by the new JUNA facility in China, also extending the measurements down to the resonance at 58 keV with a more intense beam \citep{juna_Liu2016}.
Note that at such low energies, screening needs to be taken into account, in order to obtain the actual reaction rate in stellar environments \citep{mg25pg_luna_Strieder_2012}.

The effect of variations of other thermonuclear reaction rates on the ${}^{26}$Al
production in massive stars was investigated in detail by \citet{ref10_Iliadis_2011}, who performed
nucleosynthesis post-processing calculations for each
site by adopting temperature and density time profiles from astrophysical
models, and then applying reaction rates from the STARLIB compilation \citep{starlib_Sallaska_2013}.
The effect of $^{26}$Al$^m$ has also explicitly been taken into account. These authors identified the following
four reactions: $^{26}$Al$^{t}$($n$,$p$)$^{26}$Mg, $^{25}$Mg($a$,$n$)$^{28}$Si,
$^{24}$Mg($n$,$\gamma$)$^{25}$Mg and $^{23}$Na($a$,$p$)$^{26}$Mg to significantly affect
the $^{26}$Al production yield in massive stars. For a status review of these reaction rates
see \citet{ref10_Iliadis_2011}, who estimate a typical reaction-rate uncertainty of a factor two.

Recently, four direct measurements of $^{23}$Na($a$,$p$)$^{26}$Mg
have been performed \citep{na23ap_almaraz,na23ap_howard,na23ap_tomlinson,na23ap_avila}.
The reaction rate
in the key temperature region, around 1.4 GK, was found to be consistent within 30\%
with that predicted by the statistical model \citep[][NON-SMOKER]{RAUSCHER20001}. This level of precision in
the $^{23}$Na($a$,$p$)$^{26}$Mg reaction rate should allow useful comparisons between observed and simulated
astrophysical $^{26}$Al production.

The determination of the $^{26}$Al$^{t}$($n$,$p$)$^{26}$Mg reaction rate
actually requires the independent measurements of two reactions: $^{26}$Al$^{g}$($n$,$p$)$^{26}$Mg
and $^{26}$Al$^{m}$($n$,$p$)$^{26}$Mg. Two direct measurements of
$^{26}$Al$^{g}$($n$,$p$)$^{26}$Mg have been published up to now, using $^{26}$Al$^g$ targets~\citep{al26np_trautvetter1986,al26np_koehler1997}.
Their results differ by a factor of 2, calling for more experimental work.
The preliminary result of a new measurement of $^{26}$Al($n$,$p$)$^{26}$Mg performed by the n$\_$TOF collaboration is a promising advance \citep{nTOF:2019}.
Production of a $^{26}$Al$^m$ target is not feasible due to the short lifetime of $^{26}$Al$^m$.
So, indirect measurement methods appear promising, such as the Trojan Horse Method \citep{Tribble_2014}.

On top of the main reactions discussed above, the ${}^{12}$C+${}^{12}$C fusion reaction drives C/Ne burning and therefore the production of \iso{26}Al there. Herein,  $^{12}$C($^{12}$C,$\alpha$)$^{20}$Ne
and $^{12}$C($^{12}$C,$p$)$^{23}$Ne are two major reaction channels.
Measurements of these have been performed at energies above E$_{c.m.}$ = 2.1~MeV, and
three different extrapolation methods have been used to estimate the reaction cross section at lower energies \citep{beck_2020EPJA}.
Comparing to the standard rate CF88 from \citet{cf88}, the indirect measurement using the Trojan Horse Method
suggests an enhancement of the reaction rate due to a number of potential resonances
in the unmeasured energy range \citep{Tumino_2018Nature}, while the phenomenological \emph{hindrance model} suggests a
greatly suppressed and lower rate \citep{Jiang2018_hindrance}.
Normalizing the rates to the CF88 standard rate, the Trojan Horse Method rate
and the hindrance rate are 1.8 and 0.3 at T$_9$=1.2 GK, respectively.
However, differences are reduced to less than 20\% at T=2.0~GK.
A systematic study of the carbon isotope system suggests that the reaction rate is at most
a factor of 2 different from the standard rate, and that the hindrance model is not a valid model for the carbon isotope system \citep{Zhang_2020PhLB,li2020_ccfusion}.
Direct measurements are planned in both underground and ground-level labs to reduce the uncertainty.

The $^{12}$C($^{12}$C,$n$)$^{23}$Mg reaction is an important neutron source for C burning, and has been measured first at energies above E$_{c.m.}$ = 3~MeV;
with a recent experiment, measured energies are now extending down to the Gamow window.
At typical carbon shell burning temperatures, T = 1.1-1.3~GK, the uncertainty is less than 40\%, and reduced to 20\% at T = 1.9-2.1~GK, which is relevant for
explosive C burning \citep{Bucher_2015PhRvL}.

The $^{22}$Ne($\alpha$,$n$)$^{25}$Mg reaction is another important neutron source. It has been measured directly down to E$_{cm}$ = 0.57~MeV with an
experimental sensitivity of 10$^{-11}$b \citep{ne22an_Jaeger2001}.
For a typical C shell burning temperature T = 1.2~GK,
the important energies span from 0.84 to 1.86~MeV, and these are
fully covered by the experimental measurements.
Therefore, the uncertainty is less than $\pm$6\%
for both hydrostatic and explosive C burning. At the temperatures of the He shell burning, the uncertainty would be as large as 70\%; while this is not relevant for the production of \iso{26}Al \citep{iliadis_2010NuPhA_rate_compilation}, it is very crucial for the production of \iso{60}Fe. During He burning, the $^{22}$Ne($\alpha$,$\gamma$) rate also affects the amount of $^{22}$Ne available for the production of neutrons.
A number of indirect measurements have obtained important information on the nuclear structure of $^{26}$Mg.
However, evaluations of the reaction rates following the collection of new nuclear data presently show differences of up to a factor of 500, resulting in considerable uncertainty in the resulting nucleosynthesis.
Detailed discussions can be found in the recent compilations  \citep{ne22an_ag_Longland2012,ne22an_ne22ag_Adsley2021}. Direct measurements of $^{22}$Ne+$\alpha$ in an underground laboratory are urgently needed to achieve accurate rates for astrophysical applications.

\begin{figure}
	\includegraphics[width=\linewidth]{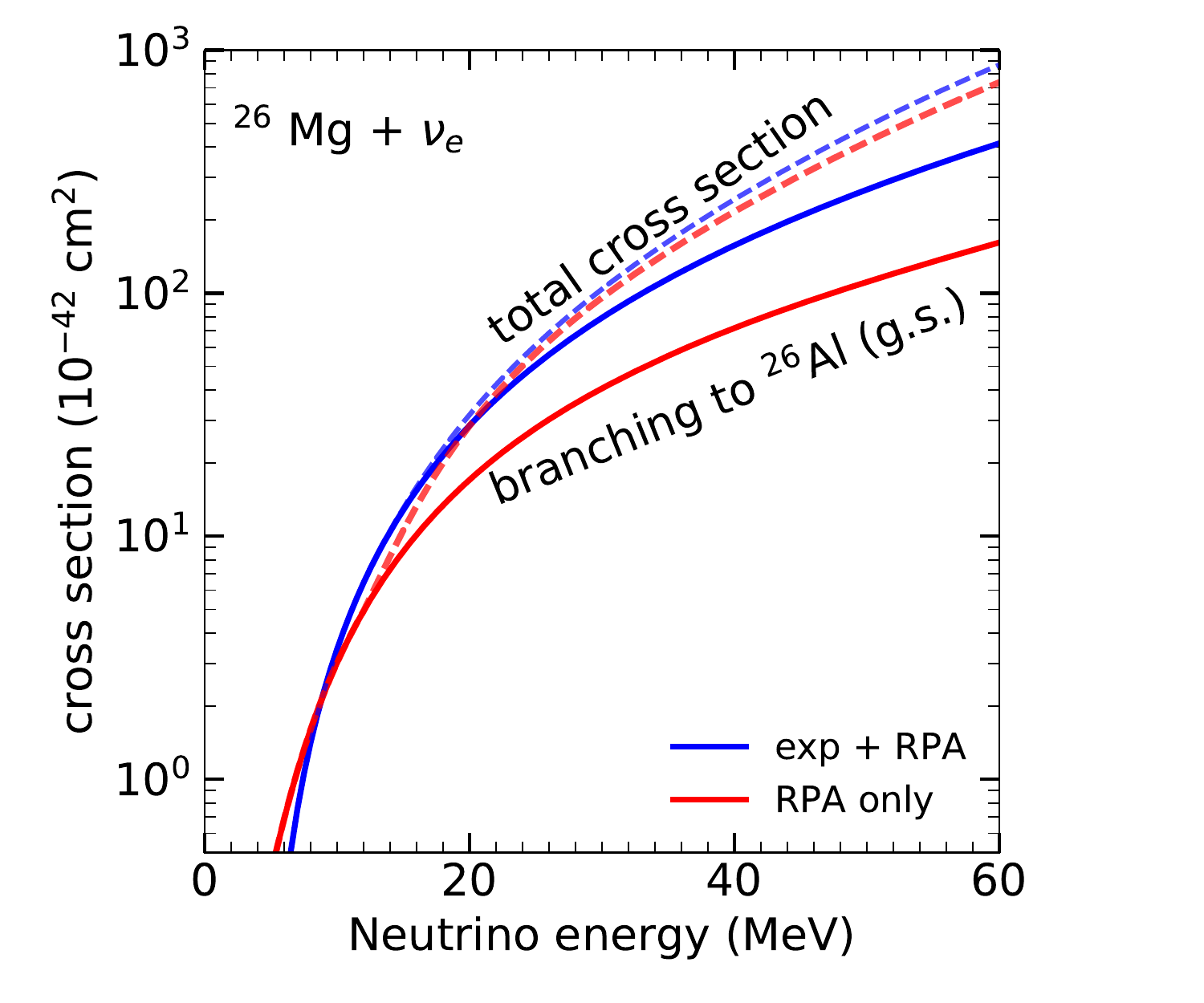}
	\caption{Cross section for the reaction $^{26}$Mg$(\nu_e,e^-)$
	comparing results based entirely on theoretical calculations (red lines) and
	results based on the experimentally measured Gamow-Teller strength distribution (blue lines).
	The experimentally determined distribution increases the strength at
	low energies and gives a larger cross section for the transitions to
	the $^{26}$Al ground state.  }
	\label{fig:nue_mg26_csect}
\end{figure}

Finally, the cross section for $^{26}$Mg$(\nu_e,e^-)^{26}$Al is dominated by the transition to the isobaric analog state
of the $^{26}$Mg ground state at $228.3\,\mathrm{keV}$ and further contributions
from a number of Gamow-Teller (GT) transitions at low energies. \citet{zegers:2006} have used charge
exchange reactions to determine the GT strength distribution of
$^{26}$Mg. \citet{sieverding:2018} have calculated the cross section based
on these experimental results with forbidden transitions at higher energies.
Figure \ref{fig:nue_mg26_csect} shows a comparison between the
theoretical cross section based on the \emph{Random Phase Approximation} and the values using the experimentally determined strength at low energies.
The particle emission branching has been calculated with a statistical model
code \citep{loens:2010,rauscher:2000}. While the theoretical model captures the total
cross section quite well, the values for transition to the \iso{26}Mg ground state
are substantially underestimated in the calculations.

\subsection{Cosmic Nucleosynthesis Environments} 
\label{sec:stars}
Here we address stellar nucleosynthesis, as we know it from models and theoretical considerations, in greater detail first for stars that are not massive enough to end in a core collapse, then for the different nucleosynthesis regions within massive stars and their core-collapse supernovae; and finally, we comment on other explosive sites such as novae and high-energy reactions in interstellar matter.


\subsubsection{Low- and Intermediate-Mass Stars}  
\label{agb1}
Low and intermediate mass stars (of initial masses $\approx$~0.8$-$8~\msun) become asymptotic giant branch (AGB) stars after undergoing core H and He burning. An AGB star consists of a CO core, H and He burning shells surrounded by a large and extended H-rich convective envelope. These two shells undergo alternate phases of stable H burning and repeated He flashes (thermal pulses) with associated convective regions. Mixing events (called \emph{third dredge ups}) can occur after thermal pulses, whereby the base of the convective envelope penetrates inwards, dredging up material processed by nuclear reactions from these deeper shell burning regions into the envelope. Mass is lost through a stellar wind and progressively strips the envelope releasing the nucleosynthetic products into the interstellar environment (see \cite{Karakas:2014} for a recent review of AGB stars.).

The production of $^{26}$Al\footnote{Note that in this and the following sections, for sake of simplicity, the notation $^{26}$Al represents $^{26}$Al$^g$, unless noted otherwise.} within AGB stars has been the focus of considerable study (e.g., \citealt{norgaard1980,forestini1991,mowlavi2000,karakas2003,siess2008,lugaro2008,ventura2011}). Here we do not attempt a review of the extensive literature, but briefly summarize the relevant nucleosythesis, model uncertainties, stellar yields, and the overall galactic contribution.

The main site of $^{26}$Al production in low-mass AGB stars is within the H-burning shell. Even in the lowest mass AGB stars, temperatures are such ($\geq$ 40~MK), that the MgAl chain can occur and the $^{26}$Al is produced via the $^{25}$Mg(p,$\gamma$)$^{26}$Al reaction.
The H burning ashes are subsequently engulfed in the thermal pulse convective zone, with some $^{26}$Al surviving and later enriching the surface via the third dredge up. 
In AGB stars of masses $\geq$ 2-3~\msun (depending on metallicity) the temperature within the thermal pulse is high enough ($>$ 300~MK) to activate the $^{22}$Ne($\alpha$,n)$^{25}$Mg reaction. The neutrons produced from this reaction efficiently destroy the $^{26}$Al (via the $^{26}$Al(n,p)$^{26}$Mg and $^{26}$Al(n,$\alpha$)$^{23}$Na channels), leaving small amounts to be later dredged to the surface.

In more massive AGB stars another process is able to produce $^{26}$Al: the hot bottom burning. This hot bottom burning takes place when the base of the convective envelope reaches high enough temperatures for nuclear burning ($\sim$ 50-140~MK). Due to the lower density at the base of the convective envelope than in the H burning shell, higher temperatures are required here to activate the Mg-Al chain of nuclear reactions. The occurrence of hot bottom burning is a function of initial stellar mass and metallicity, with higher mass and/or lower metallicity models reaching higher temperatures. The lower mass limits for hot bottom burning (as well as its peak temperatures) also depend on stellar models, in particular on the treatment of convection (e.g.\citealt{ventura2005}). Values from representative  models of the Monash group \citep{karakas10a} are $\sim$~5~\msun\ at metallicity Z=0.02, decreasing to $\sim$ 3.5~\msun~ at Z=0.0001.
Typically, there is larger production of $^{26}$Al by hot bottom burning when temperatures at the base of the envelope are higher and the AGB phase is longer. The duration of the AGB phase is set by the mass loss rate, which is 
a major uncertainty in the predicted $^{26}$Al yields \citep{mowlavi2000,siess2008,Hofner:2018}.

As the temperature at the base of the convective envelope increases two other reactions become important. First, at $\sim$ 80 MK, $^{24}$Mg is efficiently destroyed via $^{24}$Mg($p,\gamma$)$^{25}$Al($\beta^{+}$)$^{25}$Mg leading to more seed $^{25}$Mg for $^{26}$Al production, Second, at above 100~MK, the $^{26}$Al itself is destroyed via $^{26}$Al($p$,$\gamma$)$^{27}$Si($\beta^{+}$)$^{27}$Al. This last reaction has the largest nuclear reaction rates uncertainty within the Mg-Al chain, variations of this rate within current uncertainties greatly modify the AGB stellar $^{26}$Al yield \citep{izzard2007,vanraai08}.

\begin{figure}
	\includegraphics[width=\columnwidth]{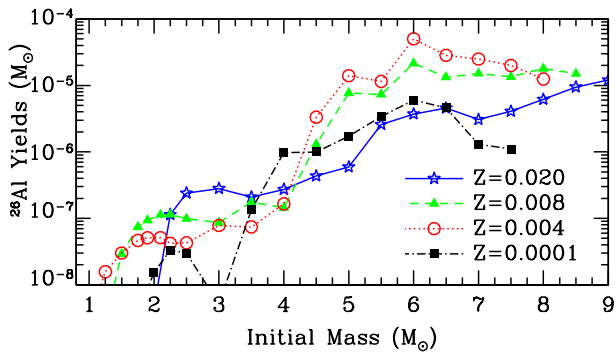}
 \caption{AGB star yields of $^{26}$Al for the range of metallicities (Z = 0.02 - 0.0001) as a function of initial mass. Results taken from \cite{karakas10a} and \cite{doherty2014a,doherty2014b}}
\label{fig:yieldagb}
\end{figure}

Figure~\ref{fig:yieldagb} shows the $^{26}$Al yields for a range of metallicites (Z = 0.02 - 0.0001) as a function of initial mass from the Monash set of models of \cite{karakas10a} and \cite{doherty2014a,doherty2014b}.
The relative efficiency of the two different modes of production are evident: in the lower mass models, where $^{26}$Al is enhanced only by the third dredge-ups of the H-shell ashes, show a low yield of $\approx$ $10^{-8}-3~\times~10^{-7}$~\msun. The more massive AGB stars that undergo hot bottom burning, instead, have substantially higher yield of $\approx$ $10^{-6}-10^{-4}$~\msun.

Metallicity also has an impact to the AGB $^{26}$Al yield, in particular for intermediate-mass AGB stars. The larger yields at Z=0.004 and 0.008, when compared to Z=0.02, are primarily due to their higher temperatures and longer AGB phases.
At the lowest metallicity (Z=0.0001) the seed $^{25}$Mg nuclei are not present in sufficient amounts to further increase the $^{26}$Al yield even with a higher temperature and similar duration of the AGB phase. This is the case even thought the majority of the initial envelope $^{24}$Mg has been transmuted to $^{25}$Mg, and the intershell $^{25}$Mg is efficiently dredged-up via the third dredge-ups.
The decreasing trend in $^{26}$Al yield for the most massive metal-poor models is due to their shorter AGB phase, less third dredge-up and higher hot-bottom burning temperatures, which activate the destruction channel $^{26}$Al($p$,$\gamma$)$^{27}$Si.

The contribution from AGB stars to the galactic inventory of $^{26}$Al has been estimated at between 0.1-0.4~\msun\ (e.g., \citealt{mowlavi2000}). More recently \cite{siess2008} also included super-AGB stars\footnote{Super-AGB stars are the most massive AGB stars ($\approx$~7$-$10~\msun) which have undergone central C burning prior to the super-AGB phase - for a recent review, see \citet{doherty2017}.} yields in this contribution, and also their impact seems to be rather modest. Even when factoring in the considerable uncertainties impacting the yields, AGB stars are expected to be of only minor importance to the Galactic $^{26}$Al budget at solar metallicity.
However, \cite{siess2008} noted that at lower metallicity, around that of the Magellanic clouds (Z=0.004-0.008), the contribution of AGB and super-AGB stars may have been far more significant.

\subsubsection{Massive Stars and their core-collapse supernovae}  
\label{sec:al26_from_CCSN}
Massive stars are defined as stars with main-sequence masses of more than $8-10\,M_\odot$.
They are characterized by relatively high ratios of temperature over density ($T/\rho$) throughout their
evolution. Due to this, such stars tend to be more luminous.
Unlike lower-mass stars, they avoid electron degeneracy in the core during most of their evolution.
Therefore, core contraction leads to a smooth increase of the temperature.
This causes the ignition of all stable nuclear burning phases, from H, He, C, Ne, and O burning up
to the burning of Si both in the core and in shells surrounding it.
The final Fe core is bound to collapse, while Si burning continues in a shell and keeps on increasing the mass of the core.
During this complex sequence of core and shell burning phases, many of the
elements in the Universe are made.
A substantial fraction of those newly-made nuclei are removed from the
star and injected into the interstellar medium  by the core-collapse supernova explosion, leaving behind a neutron star or a black hole.
The collapse of the core is accompanied by the emission of a large number of neutrinos.
The energy spectrum of these neutrinos reflects the high temperature environment from which they originate,
with mean energies of $10-20\,\mathrm{MeV}$. The fact that these neutrinos could be observed in
Supernova 1987A is a splendid confirmation of our understanding of the
the lives and deaths of massive stars \citep{Burrows:1987,Arnett:1987}. 

The mechanism that ultimately turns the collapse of a stellar core into a supernova explosion is an active field of research.
In our current understanding, a combination
of neutrino heating and turbulent fluid motion are crucial components for successful explosions \citep[see][for reviews of the status of core-collapse modeling]{janka12,Burrows:2021}.
Due to the multi-dimensional nature and multi-physics complexity of this
problem, simulations of such explosions from first principles are  still in their
infancy \citep{Muller:2016a}.
Parametric models, however, have proven to be able to explain many properties of supernovae,
although they need to be fine-tuned accordingly \citep{Burrows:2021}.

The supernova explosion expels most of the stellar material that had been enriched
in metals by the hydrostatic burning and the explosion shock itself.
Before the explosion, strong winds already take away some of the outer envelopes of these massive stars, especially in the luminous blue variable and Wolf-Rayet phases of evolution (as will be discussed in detail see below).
This ejected material also contains a range of radioactive isotopes, including some with
lifetimes long enough to be observable long after the explosion has faded, such
as $^{26}$Al and $^{60}$Fe. In this section we describe the various ways in which $^{26}$Al is made in massive
stars and the ensuing supernova explosion.

\begin{figure}
	\includegraphics[width=\linewidth]{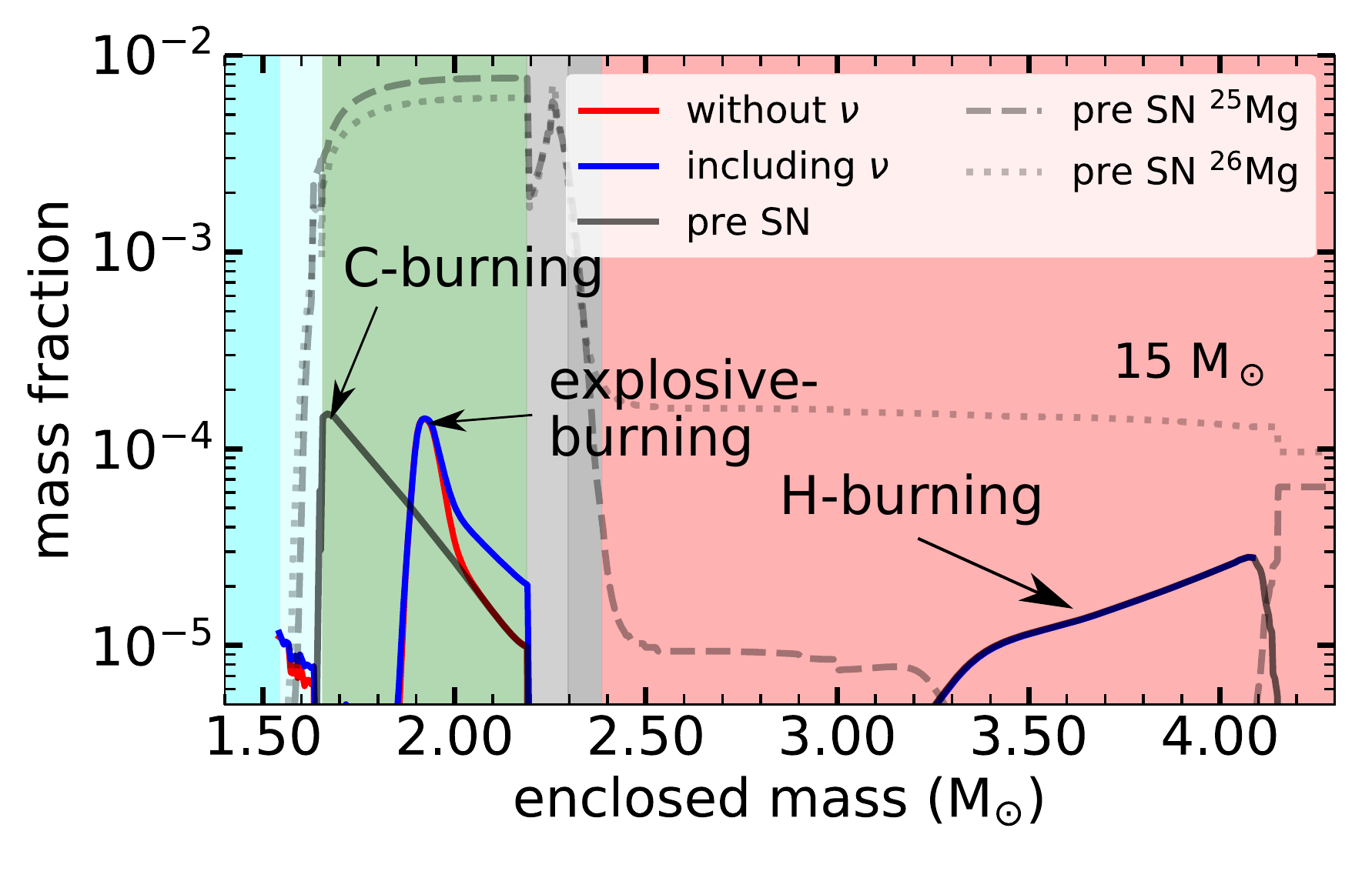}
	\caption{Mass fraction profiles of $^{26}$Al indicating regions of different production mechanisms. }
	\label{fig:as_al26_profile}
\end{figure}

The production of $^{26}$Al always operates through the $^{25}$Mg(p,$\gamma$)$^{26}$Al reaction, which is active during different epochs of the stellar evolution.
We can distinguish four main phases that contribute to the production of $^{26}$Al during massive star evolution and the supernova explosion \citep{Limongi:2006a}.

\begin{enumerate}
 \item In H core and shell burning $^{26}$Al is produced from the $^{25}$Mg that is
	 present due to the initial metallicity.
 \item During convective C/Ne shell burning $^{26}$Al is produced from the $^{25}$Mg
	 that results from the MgAl cycle with protons provided by the C
		fusion reactions.
 \item The supernova explosion shock initiates explosive C/Ne burning and
		$^{26}$Al is efficiently produced in the
	 region of suitable peak temperature around $2.3\,\mathrm{GK}$. As we will show, this is the dominant
		contribution for stars in the mass range $10-30\,M_\odot$.
\item Neutrino interactions during the explosion can also affect the abundance of \iso{26}Al.
\end{enumerate}

Figure~\ref{fig:as_al26_profile} shows the  profile of the \Al\ mass fraction for a
$15\,M_\odot$ stellar model, calculated with the KEPLER hydrodynamics code in
spherical symmetry. The pre-supernova as well as the post-explosion abundance profiles
are shown, and the production mechanisms  indicated. We now discuss in detail each of the four main mechanisms listed above.

\noindent {\bf H core and shell burning:}
The production in the convective core H burning during the main
sequence mostly depends on the size of the convective core and the initial
amount of $^{25}$Mg. $^{26}$Al in the region that undergoes core He burning
is destroyed due to neutron-capture reactions, but some of it may survive in
the layers outside of the burning region. The $^{26}$Al produced during H core burning is also threatened by the
lifetime of the star. Since the post-main-sequence, i.e., post-H-burning,
evolution of a star can take more than 0.1 Myr, due to the exponential radioactive decay most of this early made $^{26}$Al decays
before it can be ejected by a supernova explosion.
In H-shell burning,
$^{26}$Al is also produced and it is more likely to survive until it is
ejected.  In cases in which the H-burning contribution is important for the
final $^{26}$Al yield, this component is sensitive to H burning conditions and in
particular to the treatment of convection.

Another way for $^{26}$Al from H burning to contribute to the ejecta is mass loss. For single stars, mass is lost via stellar winds driven by radiation pressure \citep{Cassinelli:1979,Vink:2011a}. Thus, it is stronger for more luminous, more massive stars. Stellar mass loss has been a subject of study for a long time \citep{1999ApJLamers,Vink:2011a}, but the details of the implementation in models still gives rise to significant uncertainties \citep{Farrell:2020}.
The H-burning contribution to $^{26}$Al is most-important
for massive stars with initial mass $>30\,M_\odot$ for which stellar winds are strong enough to remove material from the H burning regions below the H envelope \citep{Limongi:2006a}.


Stellar rotation may significantly increase mass loss and the mixing efficiency \citep{groh:2019,Ekstrom:2012}, which has significant impact on the \iso{26}Al yields. Stellar-evolution models that include a description of rotation have been developed for decades \citep[see][for extensive reviews]{2000ARA&AMaeder,2000ApJHeger}. However, the effects are still not well-understood. A major challenge is to model the transport of angular momentum within stars \citep{2019ARA&AAerts}. This determines how fast the internal regions of the star rotate at different radii and different latitudes. Friction from laminar and turbulent flows between layers of different velocity transports angular momentum, and Coriolis forces add complexity.
It is, therefore, far from straightforward to determine how much rotation-induced mixing happens in different regions of a star. This affects transport of heat and of material, and thus where and how nuclear burning may occur.

A wealth of information has become available on internal rotation rates of low-mass stars \citep{2019ARA&AAerts}, thanks to asteroseismology studies, e.g., with data from the Kepler and TESS spacecrafts \citep{2010SciBorucki,2015TESS}. These internal rotation rates can help us to investigate the stellar interiors directly. This led, for example, to the insight that rotation has a negligible effect on the $slow$ neutron-capture process nucleosynthesis in low-mass AGB stars \citep{denHartogh:2019}.
Information on the internal rotation rates of massive stars is more sparse, while there is information available on the rotation rates of black holes and neutron stars, which are the final phases of massive star evolution.Recently, \citet{2020A&ABelczynski} investigated how to match the LIGO/Virgo-derived compact-star merger rates, and their black hole masses and spins.
They concluded that massive stars transport angular momentum more efficiently than predicted by current stellar evolution models, and thus slow down their rotation rate.
This was attributed by these authors to the effect of magnetic fields via the Tayler-Spruit dynamo \citep{2002A&ASpruit} or similar processes \citep[e.g.][]{2019fuller}. None of the current published massive star yields include this effect so far, which means that the currently available yields from rotating massive stars may likely overestimate the effects of rotation.

\begin{figure}
	\includegraphics[width=\linewidth]{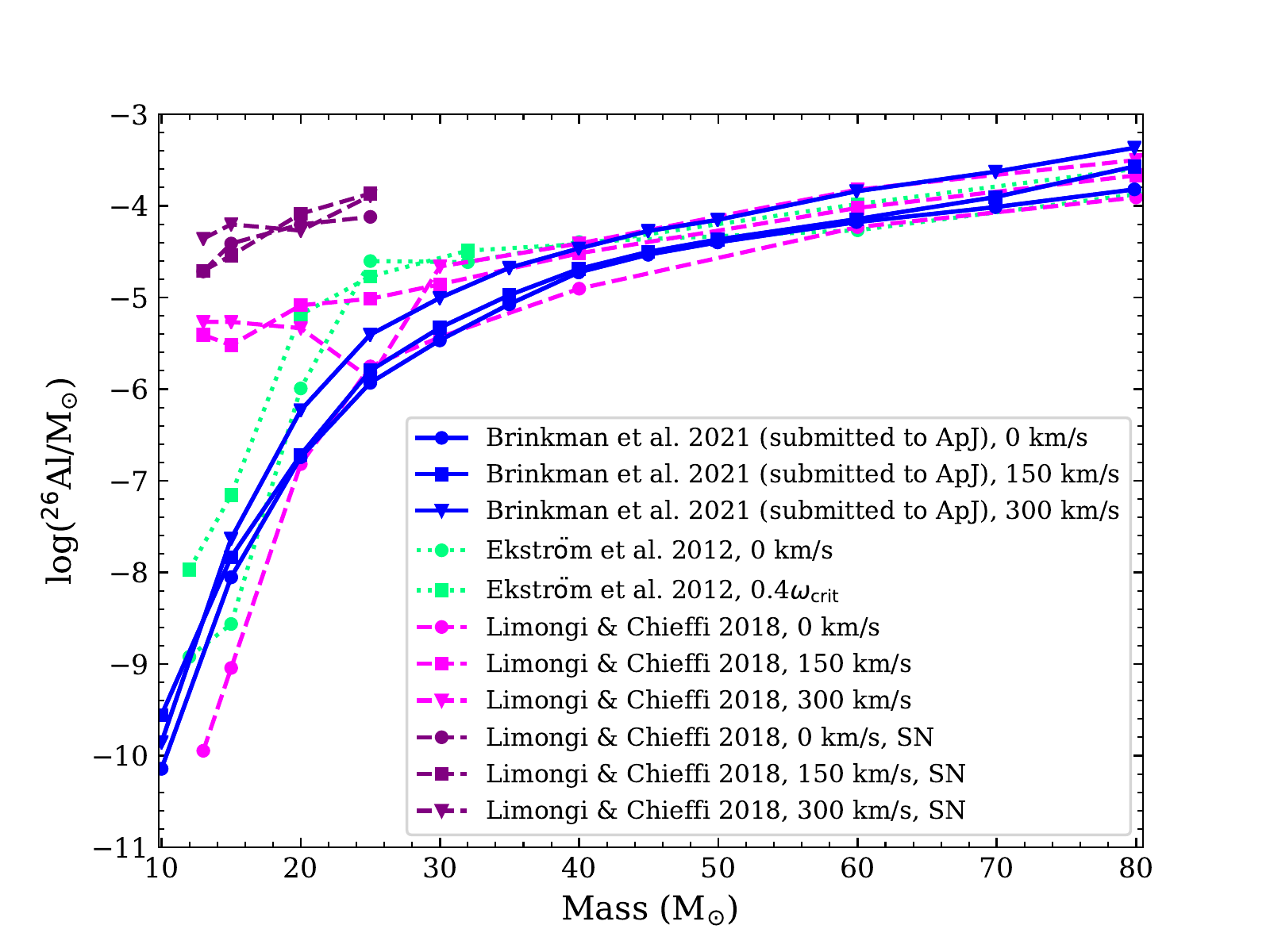}
	\caption{\iso{26}Al yields from three stellar evolution codes with different implementations of stellar rotation. Shown are contributions from winds of  solar metallicity stars \citep{Ekstrom:2012,Limongi2018,Brinkman:2021}, and supernova yields \citep{Limongi2018}. Initial rotation rates of 0 (non-rotating), 150, and 300~km~s$^{-1}$ are considered, as indicated in the legend.  Yields are in units of \msun. Based on Figure 4b of \citet{Brinkman:2021}.}
	\label{fig:LC2018_al26_yields}
\end{figure}

Recent nucleosynthesis models including stellar rotation allow us to get estimates of the impact of rotational mixing on the stellar yields.
Figure~\ref{fig:LC2018_al26_yields} illustrates several characteristic cases. Rotation generally is found to increase $^{26}\mathrm{Al}$ yields, due to the fact that the H-burning convective core is more extended and therefore more \iso{25}Mg is burnt into \iso{26}Al, which is also mixed up more efficiently due to rotation, and due to the fact that these stars experience more mass loss than their non-rotating counterparts.
For the lowest-mass models, 13 and 15 \msun, \citet{Limongi2018} find a large increase in the yields of rotating models, which is due to a significant increase in the mass-loss. This large increase is however not found in the other two studies. For the higher mass-end, 30~\Msol\  and up, the mass-loss is less affected by stellar rotation, and the yields only increase slightly, compared to the non-rotating models. This is the same for all three studies.
The supernova yields for the lowest-mass stars, shown in Figure \ref{fig:LC2018_al26_yields}, are another factor of 10-100 higher than the rotating single-star yields from the same set.
Supernova yields for higher masses are zero in the scenario discussed by \cite{Limongi2018}, because stars with an initial mass higher than 25M$_{\odot}$ are assumed to collapse completely into black holes,  and therefore do not eject $^{26}$Al in their supernova.
In Section~\ref{sec:popsyn}, we will also consider this comparisons in the light of population synthesis for both $^{26}$Al and $^{60}$Fe (Figures~\ref{fig:mp_vrot} and \ref{fig:mp_exp2}).


\begin{figure}
    \centering
    \includegraphics[width=\linewidth]{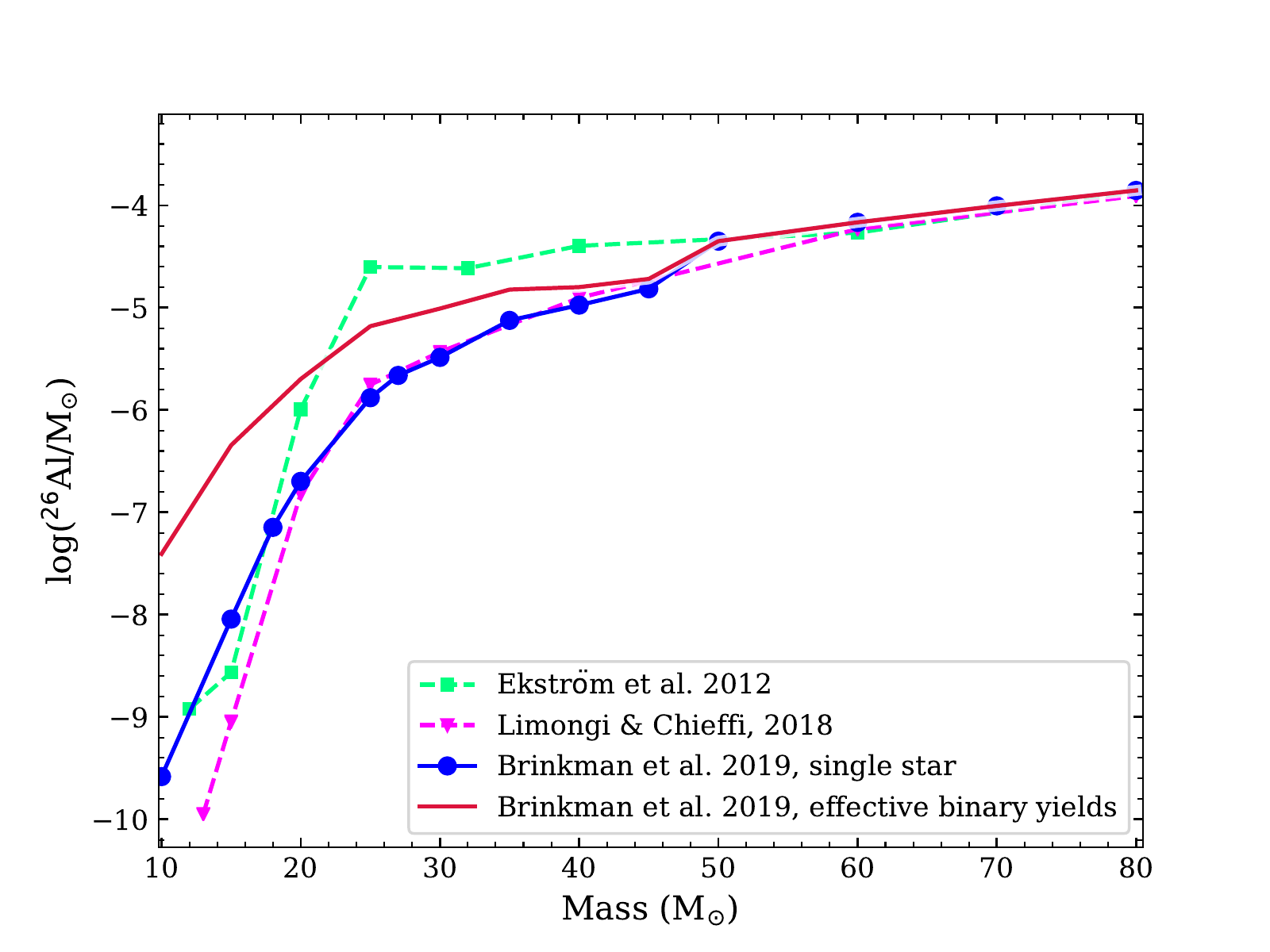}
    \caption{Yields of \iso{26}Al for various single star studies, as well as the effective binary yields defined as the average increase of the yield from a single star to the primary star of a binary system, when considering a range of periods \citep[see][for details]{Brinkman2019}.}
    \label{fig:effectiveyields}
\end{figure}

The presence of a binary companion may also have a significant impact on the \iso{26}Al yields from massive stars, because binary interactions can affect the mass loss. As shown by \citet{Sana:2012}, massive stars are rarely single stars: most, if not all, are found in binary or even multiple systems.
If close enough, the stars within such a system can interact and the gravitational pull between the stars affects the mass loss, known as Roche lobe overflow. In turn the mass loss affects the internal structure and thus further evolution of the star.
Figure~\ref{fig:effectiveyields} shows how binarity may affect $^{26}$Al yields across the stellar-mass range \citep{Brinkman2019}.
When the binary interaction takes place during the main sequence or shortly after, but before helium is ignited in the core, the impact on the amount of the ejected \iso{26}Al can be significant, and mostly prominent at the lower mass-end of massive stars (10-35\,M$_{\odot}$).
Single stars in this mass range lose only a small fraction of their whole H envelope, leaving a significant amount of \iso{26}Al locked inside. However, when part of a binary system, much more of the envelope can be stripped off because of mass transfer, which exposes the deeper layers of the star, those that were once part of the H-burning core, and now are the regions where most of the \iso{26}Al is located \citep{Brinkman2019}.
For more massive stars (M$_{*}\geq$35\,M$_{\odot}$), instead, mass loss through the stellar winds is strong even for single stars: these are the Wolf-Rayet stars observed to expose their He, C, N, or O-rich regions to the surface. These stars lose their H envelope and reveal deeper layers during their main-sequence evolution or shortly after even without having a companion, and the impact of binarity is found to be insignificant, especially for initial masses of 50M$_{\odot}$ and higher.

However, there are still many uncertainties concerning mass loss in general, and even more so in the combination of binary evolution and mass transfer, as orbital separations change in response to stellar evolution, and different phases of mass transfer may occur \citep{Podsiadlowski:2004,Sana:2012}. For wide binaries, little may change with respect to single-star evolution; but for close binaries, the impact on stellar evolution may be large \citep{Sana:2012}. Moreover, the coupling of binary evolution and mass transfer with rotation and its effect on the core-collapse explosive yields have not been explored yet.

\noindent {\bf Convective C/Ne shell burning:}
The production of $^{26}$Al in this region in the pre-supernova stage can be substantial, as shown in
Figure~\ref{fig:as_al26_profile} where the mass fraction
can reach values up to above
$10^{-4}$. The production of $^{26}$Al in C/Ne burning requires the existence of a convective shell that burns
C at a sufficiently high temperature. Convection is necessary for the supply of
fresh $^{25}$Mg, to which the C-burning reactions provide the protons. At the same
time, convection moves $^{26}$Al out of the hottest burning regions,
where it is destroyed quickly.
However, the $^{26}$Al produced in this process does not contribute to the final yield because it is later destroyed by
the high temperatures induced by the explosion shock (Figure~\ref{fig:as_al26_profile}).
Different models for the same mass range, however, may obtain almost no $^{26}$Al produced during
convective C-burning, while the production later during explosive burning still may lead
to very similar overall $^{26}$Al yields.
In principle, this depends on the explosion dynamics and
in particular on the explosion energy. The peak temperature, however, only
scales very weakly with the explosion energy, and therefore, even very weak
explosions with energies of the order of $10^{50}$ ergs
produce enough $^{26}$Al to dominate over
the contribution from C/Ne shell burning to the total yield.

\begin{figure}
	\includegraphics[width=\linewidth]{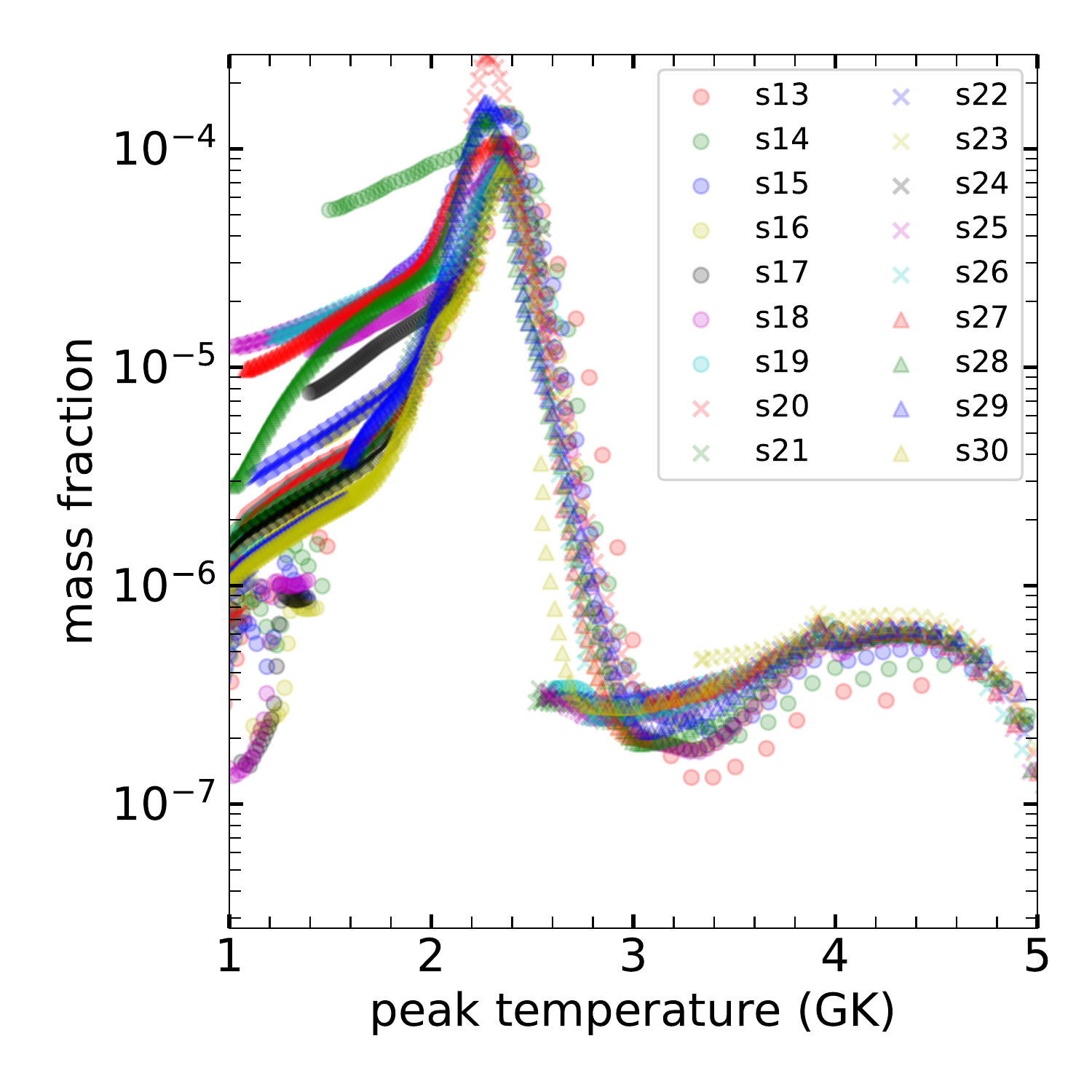}
	\caption{Mass fraction of $^{26}$Al for mass shells from a range of core-collapse supernova models. Results are shown for
a whole series of models with initial masses between $13\,M_\odot$ and $30\,M_\odot$ (as indicated in the legend). The
        temperature that leads to the largest $^{26}$Al mass fractions is very similar in all the models.}
	\label{fig:as_al26_tpeak}
\end{figure}

\noindent {\bf  Explosive contributions:}
For the explosive contribution of core-collapse supernovae two main quantities
affect the $^{26}$Al production.  First, as a
pre-requisite, the amount of produced $^{26}$Al  scales with the
$^{24}$Mg mass fraction in the C/Ne layer, because the production proceeds
through $^{24}$Mg$(n,\gamma)^{25}$Mg$(p,\gamma)^{26}$Al. This depends on the
conditions of C core and shell burning during the hydrostatic evolution of the
star and the C-burning reaction rates, as discussed above (Section~2.1).
Second, the optimal peak temperature
for $^{26}$Al production is in a narrow range between $2.1\,\mathrm{GK}$ and
$2.5\,\mathrm{GK}$.  This depends on the reaction rates of
$^{25}$Mg(p,$\gamma$)$^{26}$Al, on the neutron capture reaction on $^{26}$Al, and
on the neutron sources.
If the temperature is above the optimal value, i.e., at smaller radii, charged
particle reactions efficiently destroy the produced $^{26}$Al.  If
the temperature is too low, overcoming the Coulomb barrier in
$^{25}$Mg$(p,\gamma)^{26}$Al is harder, reducing the production.
Figure~\ref{fig:as_al26_tpeak} shows the $^{26}$Al mass fraction profile
as a function of the peak temperature reached at a given radius.
Across the mass range of progenitor models, between
$13\,M_\odot$ and $30\,M_\odot$, the highest $^{26}$Al mass fraction is reached
for the same peak temperature. The maximum mass fraction, i.e., the
height of the peak in Figure~\ref{fig:as_al26_tpeak}, depends mostly on the
local mass fraction of $^{24}$Mg (which is required to produce $^{25}$Mg by neutron captures during the explosion).
Since the peak temperature at a given radius
depends on the explosion energy, different explosion energies move the peak to
different densities. This also changes the value of the maximum $^{26}$Al mass fraction.
Since charged-particle induced nuclear reactions are highly temperature-dependent,
the peak temperature is the most important quantity.
Density and seed abundance enter linearly, the range of peak $^{26}$Al mass fraction is relatively narrow, within a factor two.
The site of the  explosive production  of $^{26}$Al is relatively far away from
the stellar core, and therefore not very sensitive to the dynamics of the explosion
mechanism itself. The explosion energy, however, depends on the supernova engine, and thus affects the
position of the peak mass fraction. The amount of matter  exposed to the critical
optimal conditions can also change by asymmetries of the explosion.
While Figure \ref{fig:as_al26_tpeak} shows that the mechanism is
always qualitatively similar, the actual yield  significantly depends on the mass of material that
is contained in the region that reaches this temperature. This varies much more between  models,
and gives rise to the non-monotonic dependence of the  $^{26}$Al yields on the progenitor mass
shown in Figure~\ref{fig:as_al26_tpeak}.
The optimal temperature is largely determined by the nuclear
reaction rates and thus independent of the progenitor model (Figure~\ref{fig:as_al26_tpeak}).


\begin{figure}
    \centering
    \includegraphics[width=\linewidth]{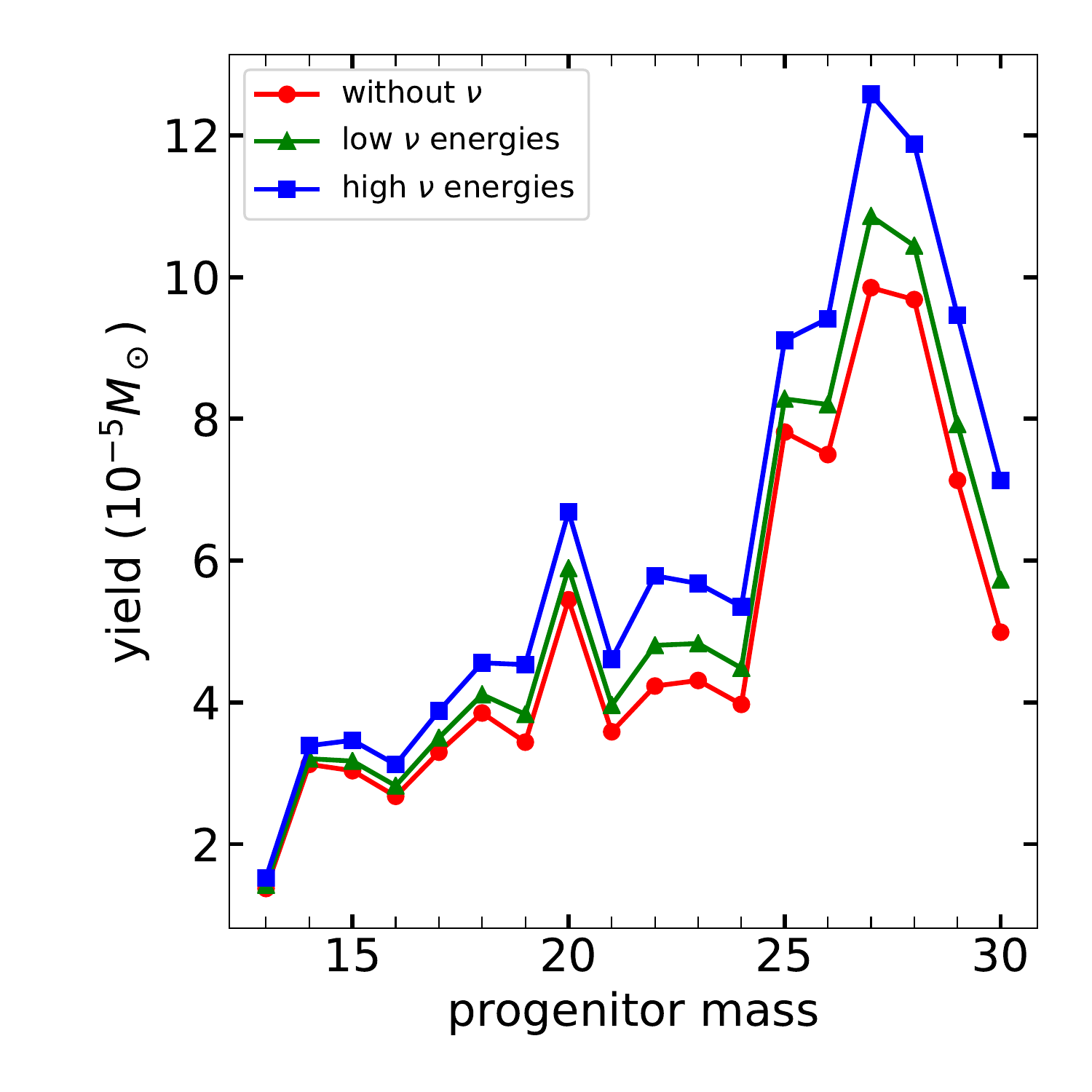}
    \caption{$^{26}$Al supernova yields from massive star progenitors in the range of 13-30 $M_\odot$ showing the also the contribution of the $\nu$ process and its dependence on the neutrino spectra \citep{sieverding:2018}. The progenitor models and explosion trajectories have been calculated with the KEPLER hydrodynamics code.}
    \label{fig:al26_yields_andre}
\end{figure}

\noindent {\bf  Neutrino interactions:}
$^{26}$Al yields are also coupled directly to the neutrino emission from
core collapse by the $\nu$ process, as mentioned in Section~\ref{sec:rates}.  The neutrinos that are copiously emitted
from the cooling proto-neutron star during a supernova explosions are sufficiently energetic
and numerous to induce nuclear reactions  in
the outer layers of the star.  Such reactions on the most abundant species have been found to be
responsible for, or at least contribute, to the solar abundances of a handful of
rare isotopes, including, $^{7}$Li, $^{11}$B, $^{19}$F, $^{138}$La, and
$^{180}$Ta.  The $\nu$ process also leaves traces in the supernova yields of
long-lived radioactive isotopes, such as
$^{10}$Be, $^{92}$Nb, and $^{98}$Tc and contributes to the explosive yield of $^{26}$Al.
This occurs in a direct and an indirect way.
A direct production
channel exists through $^{26}$Mg$(\nu_e,e^-)$.  Indirectly, neutral-current
inelastic neutrino scattering can lead to nuclear excitations that decay by
proton emission, i.e., reactions such as $^{20}$Ne$(\nu_x,\nu_x' p)$, where
$\nu_x$ includes all neutrino flavors. This provides another source of protons for
$^{25}$Mg$(p,\gamma)^{26}$Al to occur, and enhances production of $^{26}$Al outside of
the optimal temperature region.
With re-evaluated neutrino-nucleus cross
sections, a study of 1D explosions for progenitors in the mass range 13-30
M$_\odot$ has confirmed an early finding  \citep{Timmes:1995} that the $\nu$ process
increases the $^{26}$Al yields by up to $40\,\%$.
Previous studies, however, had assumed relatively large energies for the supernova neutrinos
that are not supported by current simulation results.  The contribution of the
$\nu$ process to $^{26}$Al is significantly reduced when such lower neutrino
energies are adopted.
For neutrino
spectra with temperature $T_{\nu_e}=2.8\,\mathrm{MeV}$ for $\nu_e$ and
$T_{\nu_x}=4\,\mathrm{MeV}$ for all other flavors, the increase of the
$^{26}$Al yield due to neutrinos is reduced to at most $10\,\%$ (Figure~\ref{fig:al26_yields_andre}).

There are still large uncertainties in the prediction of the
neutrino emission from a supernova explosion.  During the early phases of
vigorous accretion, the neutrino spectra can be much more energetic, increasing
the contribution of the $\nu$ process \citep{Sieverding:2018a}.
Neutrino flavor oscillations add additional uncertainty.
The production of $^{26}$Al occurs mostly at densities below the
critical density for neutrino flavor transformations due to the
MSW-resonance\footnote{Neutrino mass eigenstates propagate differently in matter with a density gradient, so that the neutrino flavor, the sum of the mass eigenstates, may change upon propagation. This \emph{MSW effect} was discovered in 1985 by Mikheev, Smirnov, and Wolfstein.}.
Due to collective non-linear effects, however, neutrino flavor
transformations may occur below the region relevant for the production of
$^{26}$Al.  This could lead to a significant increase of the $\nu_e$ spectral
temperature. Tentative calculations indicate that the $^{26}$Al yield may be increased
by up to a factor $2$, if a complete spectral swap between $\nu_e$ and the heavy
flavor neutrinos takes place below the O/Ne layer \citep{Sieverding:2020}.


%


\subsubsection{Other explosive events: Thermonuclear supernovae, Novae, X-ray bursts, and kilonovae}
\label{sec:otheral26}
Other cosmic environments are also plausible candidate to host the nuclear reactions that produce $^{26}$Al. Extremely hot plasma temperatures are likely when matter falls onto compact objects: the gravitational energy released by a proton that falls onto a neutron star is 2 GeV.
Therefore, nuclear reactions are expected on the surfaces of neutron stars and in the accretion disks that accompany newly-forming black holes. However, significant cosmic contributions to $^{26}$Al from these objects are unlikely for two reasons: (i) these events are so energetic that nuclei are decomposed into nucleons and $\alpha$ particles, and the $^{26}$Al abundance in such conditions will be low; and (ii) there is hardly significant material ejected from such compact regions.

One example of an exception, i.e., where significant material is ejected, is the recent observational confirmation of a kilonova \citep{Abbott:2017}. Here, the formation of a compact object after collision of two neutron stars has evidently led to brightening of the object from freshly-produced radioactivity, and the spectra of the kilonova light can be interpreted as a hint at overabundance in nuclei heavier than Fe \citep{Smartt:2017}.
If consolidated (in view of the significant uncertainties due to atomic-line unknowns and explosion asymmetries), this represents a potential signature of $rapid$ neutron capture ($r$-process) nucleosynthesis.
An ejected amount of such material in the range 10$^{-4}$ up to 10$^{-2}$~\msun\~
has been inferred \citep{Abbott:2019a}. However, being predominantly a result of neutron reactions, this ejected material is not expected to hold any significant amounts of $^{26}$Al.
Similarly, nuclear reactions that occur on the surfaces of neutron stars in binary systems are unlikely to contribute any significant cosmic $^{26}$Al. Such reactions have been observed in the form of Type-I X-ray bursts \citep{Bildsten:2000,Galloway:2008}.
These are thermonuclear runaway explosions after accretion of critical amounts of H and He on neutron star surfaces. Even He ashes may ignite and create super-bursts. The rapid proton capture (rp) process during an explosive hydrogen burst will process surface material up the isotope sequence out to Sm, and hence also include $^{26}$Al production. Characteristic afterglows have been observed, that are powered from the various radioactive by-products, likely including $^{26}$Al \citep{Woosley:2004}.

Another hot and dense nuclear-reaction site is the thermonuclear runaway in a white dwarf star after ignition of carbon fusion. This is believed to produce a supernova of type Ia.
Herein, temperatures of several GK and high densities of order 10$^{8-10}$~g~cm$^{-3}$ allow for the full range of nuclear reactions reaching nuclear statistical equilibrium \citep{Seitenzahl:2017}.
From such an equilibrium, one expects that the main products will be iron-group isotopes and elements, i.e., products near the maximum of nuclear binding energy, which is reached for $^{56}$Ni.
$^{26}$Al will also be produced herein. But the reaction paths are driven to tighter-bound nuclei under these circumstances. The results from models show  relatively low \al yields of $\sim10^{-8}$~\msun\ \citep{Iwamoto:1999,Nomoto:2018}. Therefore, we consider supernovae of Type Ia to be rather unimportant contributors to cosmic \al.

Novae are also potential contributors of \iso{26}Al in the Galaxy, as mentioned in Section~\ref{sec:rates}. Herein, hot hydrogen burning reactions can lead to significant \al production \citep{Jose:1998}, with \al mass fractions around 10$^{-3}$ for the more-massive O-Ne white dwarfs.
A major uncertainty in nova modelling is how the observationally inferred large ejected masses would be generated; this appears to ask for some currently unknown source of energy to make nova explosions more violent.
A total contribution from novae to Galactic \al of 0.1--0.4\msun\ have been estimated from self-consistent models \citep{Jose:1998}.
Higher values may possibly occur under favourable circumstances, with up to $\sim10^{-6}$~\msun\ for an individual nova \citep{Starrfield:1993}.


\subsubsection{Interstellar spallation reactions}
\label{sec:spallation}

\begin{figure}
\includegraphics[width=\columnwidth]{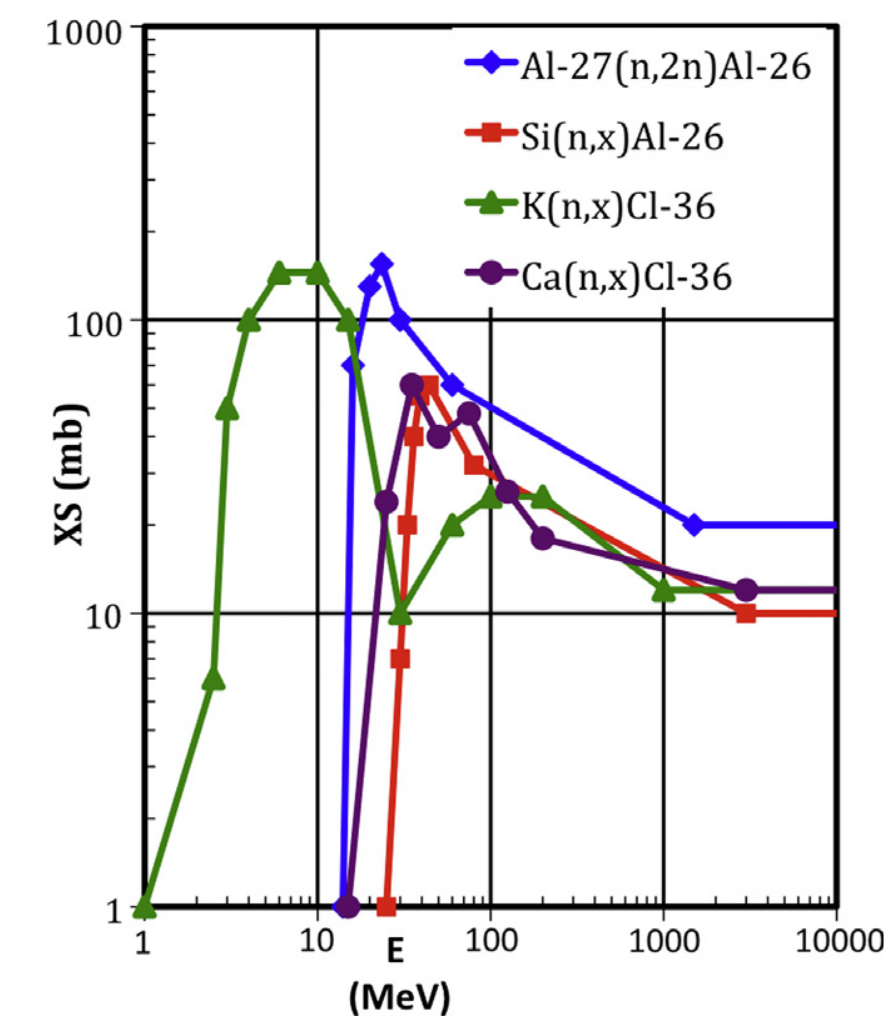}
\caption{Cross sections for spallation reactions of cosmic rays \citep[adapted from][]{Reedy:2013}. Reactions are indicated in the legend, and include $^{26}$Al production from neutron reactions. }
\label{fig:spallation}
\end{figure}

Cosmic-ray nuclei are characterised as relativistic particles by definition; therefore, when they collide with interstellar matter, the energies in the colliding-system coordinates exceed the threshold for nuclear reactions.
Since H is the most abundant nucleus in cosmic gas, nuclear fusion reactions between heavier nuclei are very rare. The most common reactions occurring with cosmic rays are fragmentations, or \emph{spallation} reactions, whereby the heavier nucleus is fragmented, or split up, into lighter daughter nuclei.
Because the abundances of nuclei heavier than iron in cosmic rays is negligible, such spallation is a significant production channel only for cosmic nuclei lighter than iron.
Therefore, spallation is among the candidate source processes for interstellar  $^{26}$Al \citep{Ramaty:1979}, but also in the early Solar System, via solar accelerated particles, as well as in the Earth's atmosphere at all times.
The cross section for production of $^{26}$Al from cosmic-ray neutrons, as an example, is about 150 mbarn, at its maximum of a collision energy of about 30 MeV, as shown in  Figure~\ref{fig:spallation} \citep{Reedy:2013}.
This figure also indicates which energy range of cosmic rays may contribute to interstellar spallation reactions contributing to  $^{26}$Al.
An estimate may be obtained from the known origins of Be and B from cosmic ray spallation of interstellar C, N, and O nuclei.
The solar abundance of CNO, a 10$^{-2}$ fraction by mass, and of Be+B, ~10$^{-9}$, imply an efficiency of 10$^{-7}$ for spallation of CNO mass  over the 10 Gy of the Galaxy's age.
Considering that present $^{26}$Al would have been produced only during its recent radioactive lifetime of 1~My and from even less-abundant heavier nuclei, its production from spallation would be $\sim$10$^{-3}$~\msun. This is well below of yields of stars and supernovae as discussed above; spallation is not a significant source of Galactic $^{26}$Al, even though its effect is observed in the abundance of $^{26}$Al in cosmic rays (Section~\ref{sec:cosmicrays}).

\subsection{From the Stellar Nucleosynthesis Sources into Galactic Medium and Stellar Systems} 



\subsubsection{Modelling the evolution of \iso{26}Al abundance in the galactic interstellar matter}
\label{sec:GCE}

Stars process the material out of which they were formed and return both the processed and unprocessed portions of their composition back to the interstellar medium when they die - except for the stellar mass that is buried in a compact remnant star or black hole.
 The processed and unprocessed material can then again be converted into stars. This procedure creates a cycle where matter is processed by multiple generations of stars. Overall, this gradually increases the fraction of processed gas in relation to unprocessed gas. Newly-formed nuclei are then enriched in the interstellar medium, as cosmic time proceeds. Since the processed matter mostly consists of metals, the metallicity of the simulation volume increases with time. Studies of galactic chemical evolution  investigate how the abundances of isotopes in stars and the interstellar medium evolve due to this cycle over the history of a galaxy.

Analytical descriptions of chemical evolution have been presented \citep{Tinsley:1974,Tinsley:1978,Clayton:1985,Pagel:1997,Matteucci:1989,Timmes:1995a,Chiappini:1997,Prantzos:1999,Chiappini:2001}, implementing assumptions about interstellar and source processes of various types in the descriptions of entire galaxies or different regions, such as the solar neighbourhood.
 Numerical chemcial-evolution models handle a given simulation volume with given resolution,
 and assume specific phases of interstellar matter and specific stellar objects (single and multiples) with their lifetimes for stellar and binary evolution.
 The consumption and re-ejection of matter from star formation and from nucleosynthesis ejecta, respectively, are evaluated as a function of time, to model the compositional changes in interstellar matter.
 For every time interval, assumptions are made on how interstellar matter is converted into stars, and how nucleosynthesis sources feed back the enriched ejecta.
 Current, sophisticated numerical  models implement  complex assumptions for the details of the modelling \citep[e.g.][]{cote19a}.
 An example is considering the evolution of the rate at which stars are formed, either intrinsically via star formation efficiencies which regulate the rate at which gas is converted into stars, or empirically, following known cosmic star formation rates.
 Other examples include the various galactic inflows and outflows, and delay times varied and adapted to a diversity of source types, from Type Ia supernovae to binary mergers.

 While a general growth with cosmic time holds for the overall metallicity, this does not necessarily apply to radioactive isotopes.
Consider a simple case: a one dimensional simulation volume converts matter and assumes stellar lifetimes in such a way that this results in constant time interval $\delta$ between stellar deaths of the same type, which eject a constant amount of $y_{\text{prod}}$ of a certain radioactive species. Then, the amount $y_{\text{obs}}$ of that species observed inside the volume ranges from $y_{\text{obs}}=y_{\text{mem}}$ to $y_{\text{obs}}=y_{\text{mem}} \mathrm{e}^{- \delta / \tau}$, with $\tau$ being the mean radioactive lifetime of the isotope, and $y_{\text{mem}}$ being a memory term with values that range from $y_{\text{mem}} = y_{\text{prod}}$ if $\tau \ll \delta$ to $y_{\text{mem}} = y_{\text{prod}} \times \tau/\delta$ if $\tau \gg \delta$.
If the memory term is large enough, a steady-state behaviour emerges, even in the case of a stochastic distribution for the $\delta$ values \citep{tsujimoto17, cote19b}.


Therefore, the abundance evolution of $^{26}$Al cannot be easily traced using 1D Galactic chemical-evolution models that impose instantaneous mixing
(therefore recycling) of the interstellar matter, since this
homogeneization approximation becomes less accurate the shorter the radioactive lifetime. In fact, if $\tau < \delta$, temporal heterogeneities become the dominant effect to determine the distribution of the abundance of \iso{26}Al in the interstellar medium, and spatial heterogeneities contribute also because a fraction of the radioactive nuclei decays as it travels interstellar distances.
The effect of these hetereogeneities has been explored both at the scale of the Galaxy \citep{Fujimoto:2018}) and of Giant Molecular Clouds  \citep{Vasileiadis:2013}).

Even when considering integrated, global effects, to compare to  \iso{26}Al data from steady diffuse $\gamma$-ray emission from the bulk of ejecta, the 1D homogeneization approximation is inherently insufficient.
For modelling any observed emission,  assumptions about the formation of its morphologies have to be made. This has been modelled using different approaches, for example by \citet{Pleintinger:2019}, \citet{Pleintinger:2020}, and \citet{Fujimoto:2020}. The specific case of star forming regions with their bubbles and superbubbles needs another more dedicated approach, as detailed below.



\subsubsection{\al and the role of superbubbles}
\label{superbubbles:text}


Short-lived massive stars shape their environment in star-forming regions via the mechanical and radiation energy they eject during their life, from main sequence and Wolf-Rayet winds to their core-collapse supernova.
This displaces the ambient interstellar medium and creates
a low-density region around the star, a bubble. If several bubbles overlap, which is rather the norm, as massive stars usually form in groups, a superbubble forms. At the time of the supernova explosion, the ejecta typically move into a superbubble with a diameter of the order of 100~pc
\citep[e.g.,][]{Krause:2013}
As outlined in Section~\ref{sec:stars}, 
\al is ejected from massive stars, and therefore, represents a tracer of their mass ejections.

Ejecta leave the sources in the form of hot, fast gas. Cooling quickly as they expand, dust may form and condense
some significant fraction of $^{26}$Al.
\citet{2018MNRAS.480.5580S} predict a dust mass of
$\approx 0.5$\msun for the core-collapse supernova SN1987A before the reverse
shock would have passed through the ejecta. This is compatible with dust mass
estimates from infrared-emission measurements with Herschel \citep{2011Sci...333.1258M} and Alma
\citep{2019ApJ...886...51C}. Dust components of various temperatures are
present, but the main component is at $\approx 20$~K
\citep{2019MNRAS.482.1715M}.
Much of this dust is, however, believed to be destroyed again later,
by sputtering, once the reverse
shock has fully traversed through and thermalised the ejecta. This happens on a timescale of $\approx 10^3$~yr,
and even earlier in higher density environments \citep{2008ApJ...682.1055N}.
In agreement with this expectation, observations of core-collapse supernova remnants show dust only in young supernova remnants and not in older ones,
and central dust only for remnants with ages $\lesssim 10^3$~yr \citep{2020MNRAS.493.2706C}, although some dust must survive, since we find stardust from core-collapse supernovae inside meteorites (Section~\ref{sec:observations26Al:stardust}).
Massive-star winds develop a similar reverse shock, which, however, now travels outwards in the
observer's frame \citep{Weavea77}. At late times, and in a realistic environments, shock heating of the ejecta
can take place at distances $\gtrsim 10$~pc from the ejecting star \citep{Krause:2013}.

\begin{figure*}[t] 
\centering
\includegraphics[width=1.6\columnwidth]{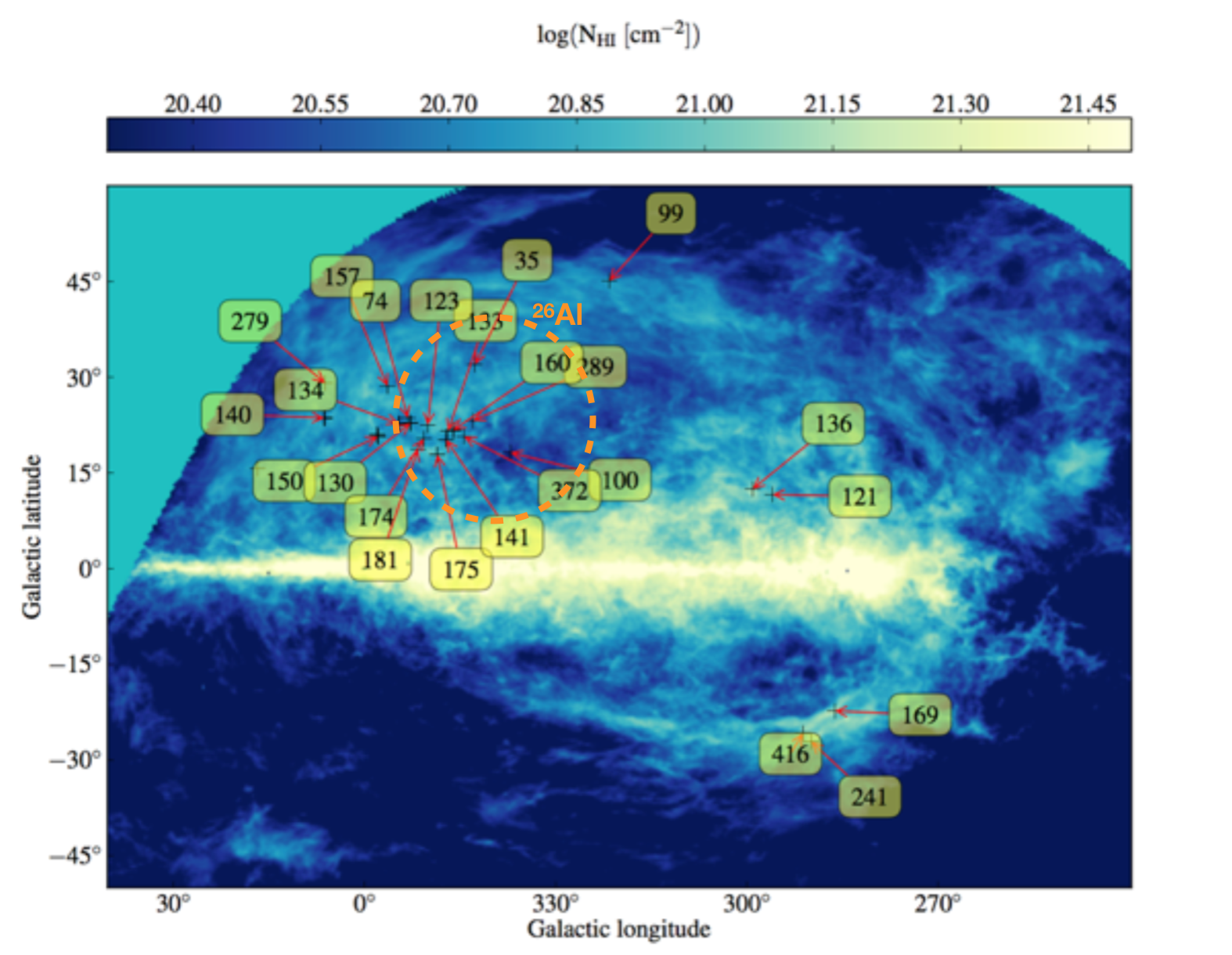}
\caption{The Sco-Cen superbubble from different astronomical constraints: This \HI image is taken in the 21~cm hyperfine-structure line from atomic hydrogen, integrated over the velocity range -20...0 km~s$^{-1}$ of HI generally approaching the observer. It shows the Scorpius-Centaurus supershell extending above and below the Galactic disk in the background. Yellow boxes show upper limits to the distances of the respective HI features, from \NaI absorption
against background stars. An orange circle represents the region of significant $^{26}$Al $\gamma$-ray emission as observed. 
Adapted from \citet{Krause:2018}.
\label{fig:scocen}}
\end{figure*} 

Massive stars are usually formed in OB associations or clusters together with other stars
\citep[e.g.,][]{ZY07}. For the path of \al through the interstellar medium, we need to consider
two distinct phases of star cluster evolution: the \emph{embedded} and the \emph{exposed} phases
\citep[see][for a recent review]{Krause:2020}.
In the embedded phase, the winds and eventual explosions occur in a dense, filamentary gas, so that
mixing is strong, and much of the \al may diffuse into cold, star-forming gas.
\al transport simulated by \citet{Vasileiadis:2013} finds for model times applicable to embedded star-forming regions
isotopic ratios of $10^{-6} <\, ^{26}$Al$/^{27}$Al$\, < 10^{-4}$.
This is compatible with meteoritic results for the early Solar System, where \iso{26}Al/\iso{27}Al $\simeq 5 \times 10^{-5}$ (see Section~\ref{sec:ESS}).

However we caution that the study of \citet{Vasileiadis:2013} used periodic boundary conditions, which prevents a
clearing of the dense gas on a larger scale, that would otherwise happen via formation of a superbubble.
Therefore, the results of \citet{Vasileiadis:2013} may not apply beyond the embedded phase. 
The embedded phase lasts for at most a few Myr, in clouds where massive stars are formed
\citep[e.g.,][]{Krause:2020}.
The feedback processes as a whole in processing gas of a giant molecular cloud typically encompass several tens of Myr.
For most of the time that a massive star group ejects \iso{26}Al, the group is exposed, rather than embedded. This means, a superbubble
has formed and the dense gas has mostly been dislocated into the supershell of compressed material forming the
outer edge of the superbubble.

\begin{figure} 
\includegraphics[width=\columnwidth]{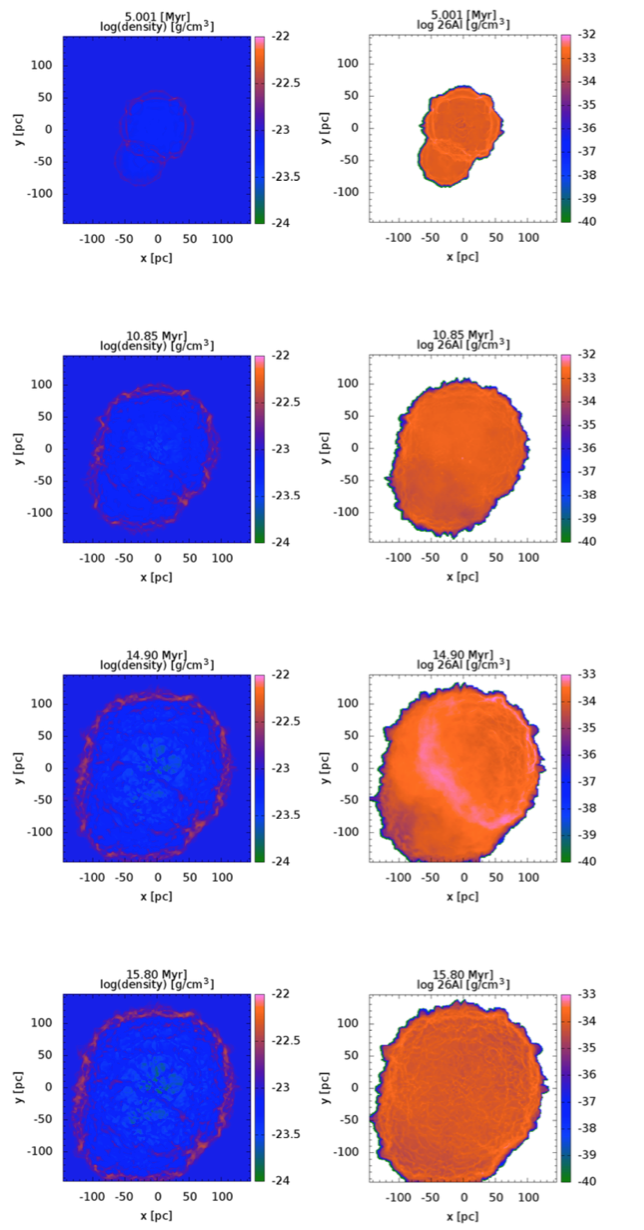}
\caption{3D hydrodynamics simulation of a superbubble. Time increases from top to bottom and
is indicated in the individual panels. Left: sightline-averaged gas density. Right: sightline-averaged
\al density. Adapted from \citet{Krause:2018}. \label{fig:sbevol}}
\end{figure}

In Figure ~\ref{fig:sbevol} we show
a 3D simulation
of the evolution of a superbubble \citep[see also][]{Krause:2018}, based on parameters for the nearby Sco-Cen massive-star region that are estimated from a variety of astronomical data.
After ejection, \al spreads quickly throughout the superbubble. Wind and shock velocities are of the order of 1000~km~s$^{-1}$.
The ejecta therefore traverse the entire superbubble, which can have a
diameter of typically $\approx 100 - 1000$~pc \citep{Krause:2015},
 in less than about 1~Myr which is the \al decay time. The simulation also shows local \alu-density variations within
the superbubble of the order of a factor ten, due to gas which may be sloshing back and forth  within the superbubble \citep{Krause:2014}, and due to supernova explosions
(snapshot at 14.9~Myr in Figure ~\ref{fig:sbevol}, pink, shell-like feature).

The diffusion of ejecta from bubbles and superbubbles into dense star-forming gas and the incorporation
into newly formed stellar systems has been modelled on the Galactic scale by \citet{Fujimoto:2018}.
They find
56\% of \al diffusing into molecular gas, and the isotopic ratio $^{26}$Al$/^{27}$Al
predicted for newly formed
stellar systems falls between $10^{-5}$ and $10^{-4}$. This would make the Sun a rather typical
star regarding its \al abundance, whereas the predicted \fe is orders of magnitude higher than observed in the early solar system \citep[this also is found in the study of][and other studies that are based on typical core-collapse supernova yields]{Vasileiadis:2013}.  It should be noted that in the study of \citet{Fujimoto:2018} they apply scaling factors\footnote{\citet{Fujimoto:2018} use the yields from \citet{Sukhbold:2016aa} and scale down $^{60}$Fe by 5 and $^{26}$Al up by 2, in order to adapt the $^{60}$Fe/$^{26}$Al yield ratio to the one observed in diffuse $\gamma$~rays \citep{Wang:2007}, which is adopted to be in steady state.} to massive-star yields, and they also adopt for simplicity a zero  delay time for the collapse of the presolar cloud, while delay may have a considerable effect due to interstellar mixing.

Gamma-ray observations of the \al decay in its 1809~keV line (see Section~\ref{sec:observations26Al-gammas}) support this picture of emission and transport of \al\ via bubbles and superbubbles in three ways:
first, \al has been directly detected (with a significance of $6\sigma$)
towards the Scorpius-Centaurus superbubble \citep{Krause:2018}.
Figure~\ref{fig:scocen} illustrates this: the HI image from the GASS HI survey
(velocity channels from -20~\kms to 0~\kms have been summed) shows
the nearby Scorpius-Centaurus supershell walls  above and below the bright HI ridge of the Galactic disk in the background; the nearby distance of the cavity is confirmed by upper limits that have been estimated by association of the \HI features in position and velocity with \NaI absorption
against background stars with distances known from HIPPARCOS or GAIA paralaxes, and these upper limits in pc are given in yellow boxes in the figure; the $\gamma$-ray emission from $^{26}$Al is overlaid as an orange circle.
Second, bulk Doppler shifts of the Galactic \al emission are observed to be different from cold-gas velocities by about 200~\kms
\citep{Kretschmer:2013}; this would be consistent with expectations for an outflow from the spiral arms into large inter-arm
superbubbles \citep{Krause:2015}. Third, the Galactic \al scale height is larger than for the dense gas components,
again suggesting an association with hot, tenuous superbubble gas \citep{Pleintinger:2019}.
The dynamics of hot gas sloshing-around within the bubble produces spatial variations in X-ray luminosity and spectra,
as observed, e.g., in the Orion-Eridanus superbubble \citep{Krause:2014a}.

\begin{figure}
\includegraphics[width=\columnwidth]{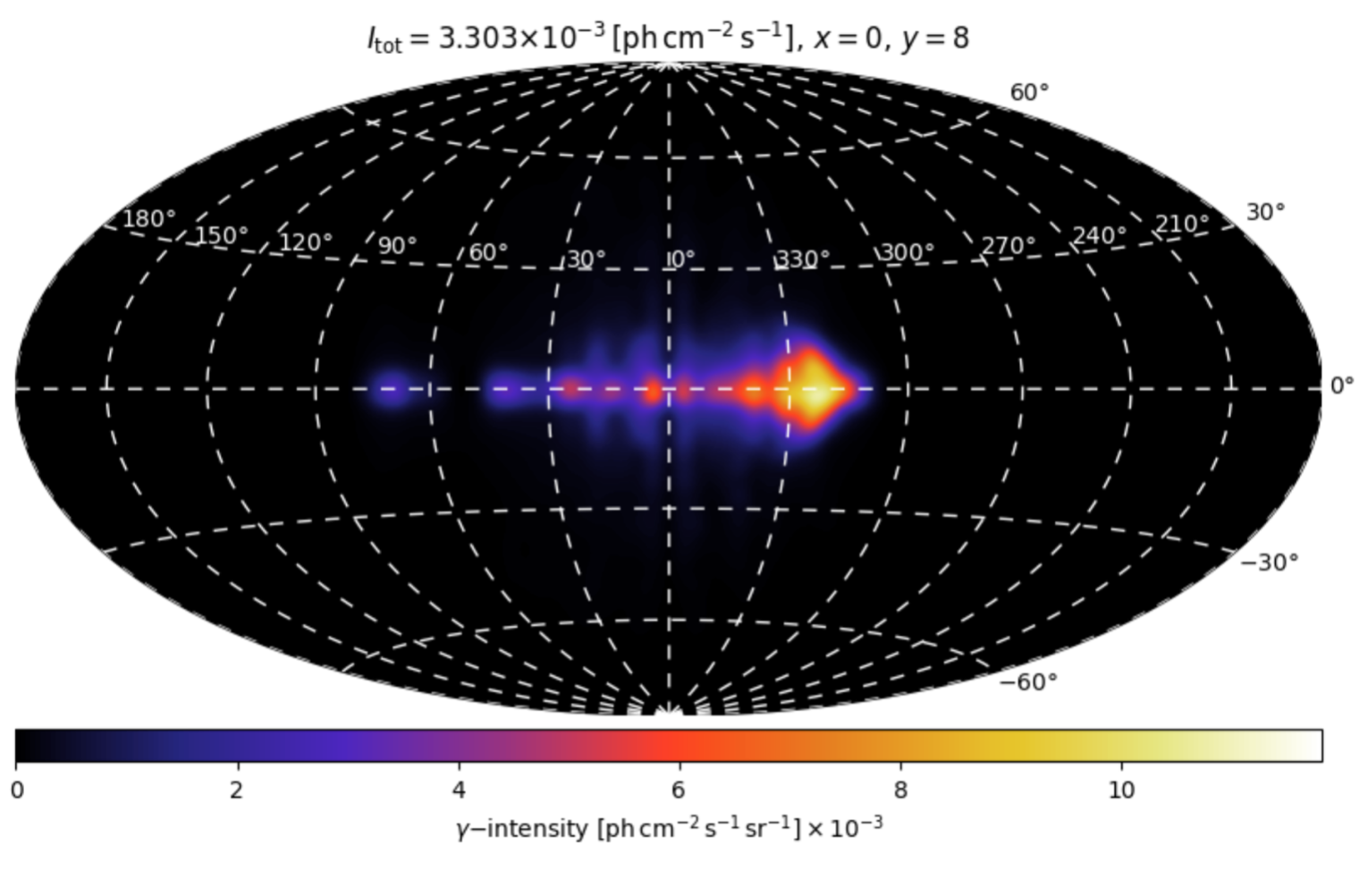}
\caption{Synthetic map for 1.8 MeV emission from radioactive decay of \alu. The map has been obtained from a whole-disc,
3D hydrodynamic simulation, with superbubbles concentrated towards spiral arms. The resolution is $4^\circ$,
as it is for the observed COMPTEL map.
Adapted from \citet{Rodgers-Lee:2019}
\label{fig:1P8synth}}
\end{figure}

A large part of \al may leave the Galactic disc and may trace metal loss into the halo.
The omnidirectional growth of superbubbles changes as bubbles reach the disk-halo interface above the Galactic disk, roughly at a scale height of gas of order 50-100~pc \citep[e.g.,][]{BB13}, which is a small extent for superbubbles. When superbubbles reach this diameter,
the shells are accelerated into the Galactic halo. Instabilities then destroy the shells, and the \alu-enriched
hot gas can stream at high velocity into the halo.
Due to large differences in gas densities between spiral arms and interarm regions, details of the diffusion of \al into the halo
depend on the properties of the spiral arms \citep{Fujimoto:2020}:
If the self-gravity of the disc stars is the dominant force of the system, one formes so called material arms, where the gas assembles at the centre of the spiral arm. Stars then form in the centres of the spiral arms and are surrounded by dense gas. This impedes the expansion of superbubbles and would hence
lead to little diffusion of \al out of the Galactic disc. For density-wave-type arms, which could arise from gravitational interaction of the disc with the bar, or interaction with infalling galaxies such as the Sagittarius dwarf galaxy, one expects the massive stars to be offset from dense gas in the spiral arms. Superbubbles can then expand comparatively easily beyond the scale height of the disc with a significant loss of \al to the Galactic halo.
3D hydrodynamics
simulations with externally-imposed spiral arms \citep{Rodgers-Lee:2019} model the latter situation and show that
\al extends up to several kpc into the halo. Synthetic maps from this work show diffuse, clumpy structure, similar to what is observed (Figure~\ref{fig:1P8synth}).
This component of \al may trace a disk outflow through chimneys, that may leak a significant fraction of the
Galactic \al production into the halo \citep{Krausea21a}.


\subsubsection{The present Solar System within the Local Bubble}
\label{sec:solarSysEnv}


The Solar System is currently embedded into an interstellar environment that had been shaped by supernovae and winds from massive stars creating a network of filaments, shells, and superbubbles
\citep[e.g.,][\citet{Krause:2021b} for a recent review]{Lallement:2018}.
Different interstellar environments may have existed during the past several million years, 
and, certainly, at the time of solar system formation 4.5 Gy ago.

The nearest star-forming regions relevant for shaping the local interstellar environment are the Scorpius-Centaurus groups (distance about 140~pc), the Perseus-Taurus complex with Perseus OB2 (300~pc), the Orion OB1  subgroups (about 400~pc), and the Cygnus complex (with several associations at 700-1500~pc and the massive OB2 cluster at about 1400~pc) \citep[see][for a literature summary]{Pleintinger:2020,Kounkel:2019}. It had been thought that the nearby star forming activity was related to the Gould Belt structure \citep{Poppel:2001} that had been attributed to a local distortion of the Galaxy by an infalling high-velocity cloud \citep{Olano:1982}. This interpretation of a causal connection and common origin of all these nearby star groups may have been an overinterpretation, and has been put in doubt recently \citep{Alves:2020}.

At least for the last 3 Myr and possibly for more than 10 Myr \citep{Fuchs:2009}, the Solar System has been located inside the Local Superbubble, a cavity with a density of $\sim$0.005 H atoms~cm$^{-3}$ (10$^{-26}$g$\cdot$cm$^{-3}$). This cavity extends for $\sim$ 60 -- 100 pc around the Sun within the galactic plane, and forms an open structure perpendicular to the plane, a galactic ``chimney'' \citep{Slavin:2017}.

The Solar System as a whole is moving relative to its also-moving dynamic surroundings: a velocity of sometimes exceeding 25 km$\cdot$s$^{-1}$ relative to the local structures within the Local Superbubble is observed.
In addition to Earth and the  Solar System experiencing such varying-density environments, the Solar System with Earth and Moon may have been exposed also to transient waves of supernova ejecta in its cosmic history.
A simple estimate using the average supernova rate of our Galaxy suggests that supernova explosions might occur on average at a rate of every few million years within a distance of $\leq$150 pc, which is considered the maximum distance from where supernova ejecta can directly penetrate into the Solar System. Such changing environments may also modulate the flux of interstellar dust and galactic cosmic rays at Earth \citep{Lallement:2014}.

Interstellar dust grains can penetrate deep into the Solar System: Space-born satellites such as STARDUST and ACE have been observing dust and particles near Earth with characteristics of origins outside the Solar System.
In contrast to interstellar-gas ions, only dust grains with larger diameters above a size of 0.2 $\mu$m will overcome the ram pressure of the solar wind and not be deflected away by the interplanetary magnetic field. Thus there will be a significant mass filter on these dust grains as they approach Earth orbit.

In this way, some fraction of the $^{26}$Al that condensed into dust particles may penetrate deep into the Solar System. By using these larger dust grains as vehicles, interstellar $^{26}$Al may accumulate on Earth in archives such as deep-sea sediments and ferromanganese (FeMn) crusts and nodules.
These archives grow slowly over time, and thus include  interstellar particles collected over time-periods up to millions of years, while also including time information imprinted on the sedimentation from magnetic-field changes, and radio-isotopes produced by cosmic-ray in the atmosphere.
Therefore, a search for traces from interstellar material in solar-system deposits was suggested \citep{Ellis:1996}. 
These nuclides could be present as geological radioisotope anomalies, \emph{live}, i.e., before they decay into stable nuclides \citep{Feige:2012,Feige:2013}.
Note that highly-energetic cosmic ray particles also reach the  Solar System and Earth, and are detected by satellite-born instruments. But deposits from these cosmic rays in deep-sea layers are many orders of magnitudes less abundant \citep{Korschinek:2015}.

In summary, the direct detection of live radionuclides such as $^{26}$Al on Earth is an important tool in the studies of nucleosynthesis in massive stars, and of interstellar medium dust formation and its transport through interstellar space and into the  Solar System \citep{Fields:2019}.
Such a detection also holds information on the local interstellar environment and its history, i.e. within a distance of order $\sim$150 pc, and for a time window of order of a few life-times of the radionuclide (e.g., a few Myr for $^{26}$Al); this must be taken into account when interpreting signals from such deposits.

\subsection{Measurements of cosmic \iso{26}Al} 
\label{sec:observations26Al}

Measuring cosmic radioactive isotopes is (and has always been) inherently  \emph{multi-messenger astronomy}. In fact, cosmic radioactive isotopes may be measured either in the laboratory in the form of solid materials, or astronomically via remote sensing. For each of these types of measurement the available approaches are more or less direct. First, we begin our description with the materials that can be analysed in laboratories.
These are either samples of terrestrial materials found in sediments, ocean crusts, and meteorites, or samples acquired through space technology from the Moon, comets, cosmic dust, as well as cosmic-ray particles.
Such materials experienced a journey from their cosmic production site to where we can recover it and biases due to its transport and deposition need to be considered.
Only indirect conclusions can be drawn as radioactive isotopes may have decayed already, and the mix of parent/daughter products may have a complex composition and re-processing history, such as for example in the case of early Solar System materials.
Second, we describe the astronomical methods. Direct isotope information is obtained from characteristic $\gamma$-ray lines that accompany radioactive decay. Less direct, still valuable astronomical measurements are possible from atomic and molecular lines, where the isotopic information must be decoded from auxilliary information.

\subsubsection{Interstellar dust collected on Earth}  
\label{sec:observations26Al-particles}

The radioactive lifetime of $\sim$ 1 Myr defines the detectability of the presence of $^{26}$Al in the interstellar medium before it decays. If condensed into interstellar dust, a fraction of it may enter the Solar System, and may eventually become incorporated into terrestrial archives. Ice cores, deep-sea sediments and deep-sea FeMn crust material grow over time-periods of hundreds of thousand years to tens of million years, and thus provide a proper data archive with time resolved material samples.
In particular, deep-sea sediments with growth rates of less than cm~kyr$^{-1}$ and FeMn crusts with very low growth rates of only mm~Myr$^{-1}$ provide time-resolved information over time scales of million years; such archives cover a full time period during which most of $^{26}$Al would have decayed.

An enhanced concentration of $^{26}$Al in such archives may thus be a signature of an enhanced interstellar influx of $^{26}$Al.
However, $^{26}$Al is also naturally produced near Earth in the terrestrial atmosphere through cosmic-ray induced nuclear reactions. 
Minor additional $^{26}$Al contributions in these archives are in-situ production within the archives. Influx of interplanetary dust grains, which also contain spurious amounts of cosmic-ray produced spallogenic $^{26}$Al, provide a minor background to a search of $^{26}$Al of stellar nucleosynthesis origin.
Any interstellar $^{26}$Al influx must therefore be detected on top of this naturally existing and approximately constant Solar System $^{26}$Al production.
Simple estimates suggest that a reasonable interstellar influx could be of order a few \% and up to ~10\% relative to the terrestrial production \citep{Feige:2013,Feige:2018}.
For example, using measured $^{60}$Fe data as a proxy for $^{26}$Al influx from the interstellar gas
(see Section~\ref{sec:observations60Fe}),
deposition rates into terrestrial archives of only between a few to some 50 $^{26}$Al atoms per cm$^2$ and per yr are estimated \citep{Feige:2018}, corresponding to $^{26}$Al concentrations of some 1,000 atoms per gram in deep-sea sediment.
Consequently, detection of $^{26}$Al from interstellar gas requires an extremely sensitive and efficient detection technique in order to identify a signal above the terrestrial background.

So far, only accelerator mass spectrometry  has the required sensitivity for such studies.
In this approach, one directly counts the radionuclide of interest one by one by means of a particle detector. The sample material needs to be pre-processed, i.e., dissolved so that the radionuclide of interest will be chemically separated from the bulk material.
An accelerated ion beam of the Al separated in this way from the sample includes $^{26}$Al together with stable terrestrial  $^{27}$Al (typically 12 to 16 orders of magnitude higher in abundance). In this ion beam,
the $^{26}$Al ions will be filtered out with electrostatic and magnetic deflectors removing unwanted background, and eventually the $^{26}$Al nuclei can be identified essentially background-free with an energy-sensitive particle detector.
Detection efficiencies for $^{26}$Al are typically of order one atom out of 10,000.

\begin{figure}  
\centering
\includegraphics[width=1.0\columnwidth]{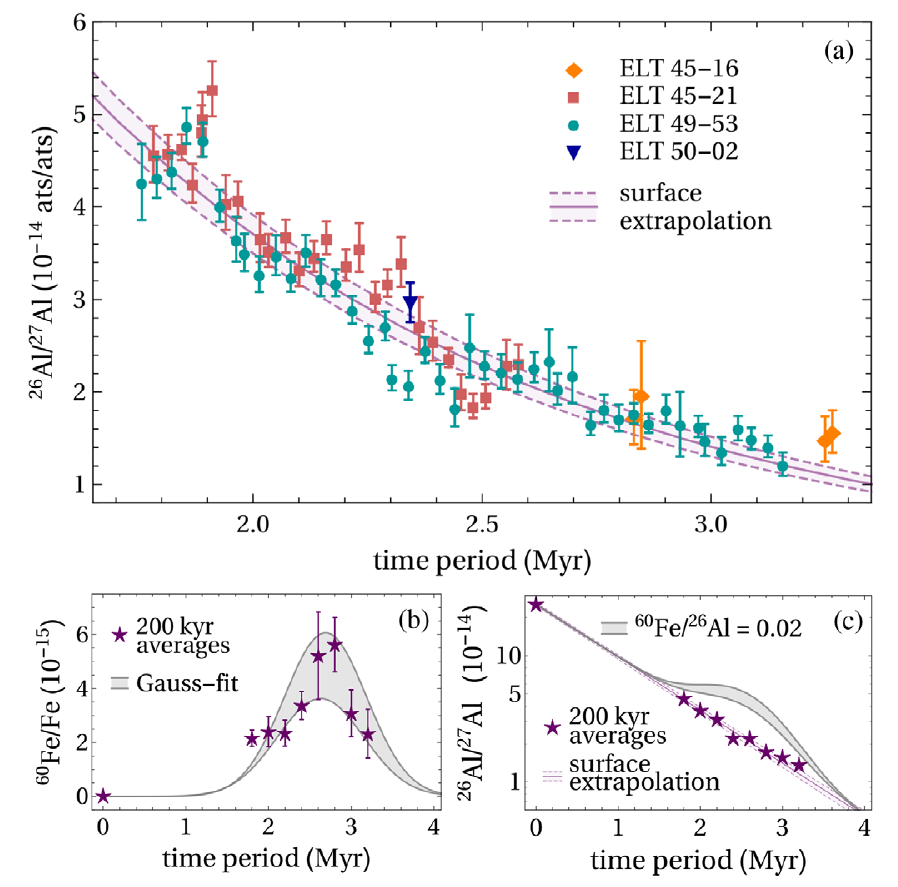}
\caption{The \Al\ search in terrestrial archives did not reveal a signal. The top graph shows \al counts versus age of the layer within the sample, with the grey band illustrating what would be expected from extrapolating the surface \al abundance due to radioactive decay. The  lower-left graph shows the \fe signal discussed in Section~\ref{sec:observations60Fe}, and the lower-right graph indicates how a ratio of 0.02 for \fe/\al would translate into an \al signal versus the actual data, as a decay-corrected version of the upper graph. \citep[From][]{Feige:2018}. }
\label{fig:Al_terrestrial}
\end{figure}   

Recently, \citet{Feige:2018} searched for presence of interstellar $^{26}$Al in an extensive set of deep-sea sediment samples that covered a time-period between 1.7 and 3.2 Myr (Fig~\ref{fig:Al_terrestrial}).
No significant $^{26}$Al above the terrestrial signal was found.
The data show an exponential decline of $^{26}$Al with the age of the samples that can be explained by radioactive decay of terrigenic $^{26}$Al.
Nevertheless, owing to the large number of samples analyzed, these data allowed to deduce an upper limit for the influx of interstellar  $^{26}$Al (see Section~\ref{sec:fe60obs}).
A much more favourable situation exists for measuring the radioisotope $^{60}$Fe from interstellar deposits in Earth, because terrestrial production is negligible for this isotope (see  Section~\ref{sec:observations60Fe}).

\subsubsection{Cosmic rays near Earth}
\label{sec:cosmicrays}

Along with dust grains, \emph{galactic cosmic rays}  provide a sample of matter from
outside the Solar system. Despite almost a century of active research,
the physics of cosmic rays (concerning their sources, acceleration and propagation
in the Galaxy) is not yet thoroughly understood. In particular, key questions
herein are related to the timescales of various processes, such as acceleration in one event or in a series of events, and their confinement in the Galaxy.
Radionuclides unstable to $\beta^{\pm}$ decay or $e$-capture, and with laboratory lifetimes
close to the $\sim$Myr timescales of interest for galactic cosmic rays, provide important probes of the
aforementioned processes \citep{Mewaldt:2001}.

\begin{figure}
\includegraphics[width=\columnwidth]{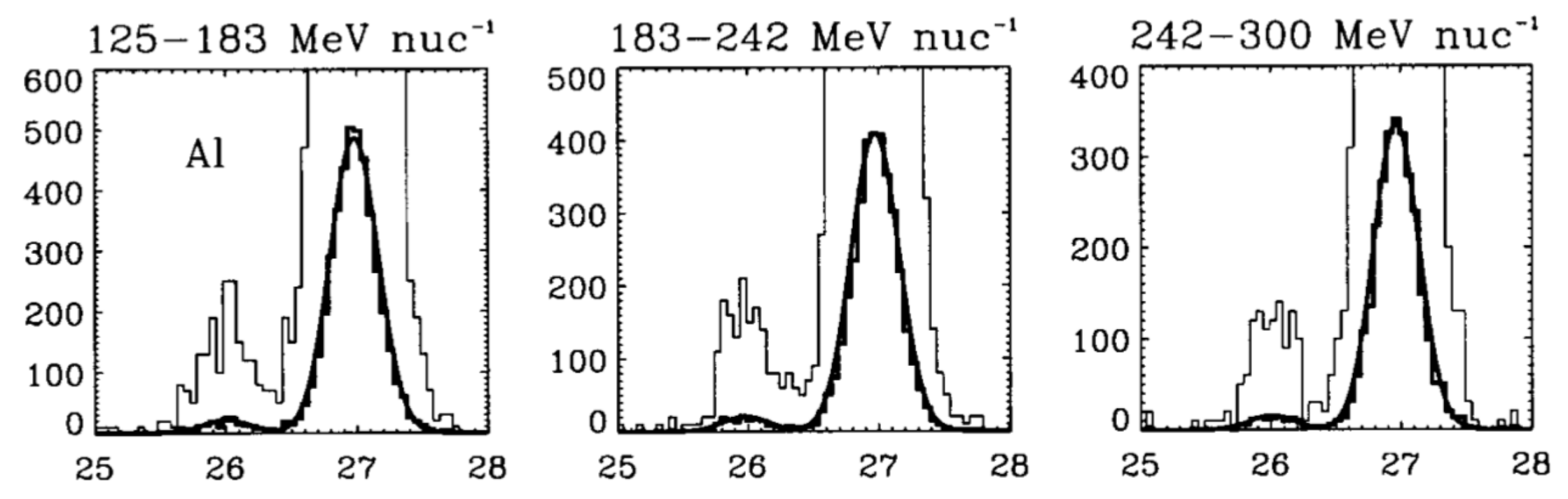}
\caption{Measurements of $^{26}$Al with the CRIS instrument on the ACE satellite, in different energy ranges of the observed cosmic ray particles \citep[adapted from][]{Yanasak2001}. The y axis shows CRIS counts, the x axis the calculated mass in AMU units. Note that this \al is most probably the result of interstellar spallations from heavier cosmic-ray nuclei, hence \emph{secondary}, and not of stellar nucleosynthesis, while contributing to observed abundances. 
}
\label{fig:26Al_CRIS}
\end{figure}

Despite more than a century of intense observational and theoretical investigation, the origin of cosmic rays is still unclear. Several steps are involved between the production of the Galactic cosmic-ray nuclides in stellar
interiors and their detection near Earth:
\begin{enumerate}
\item stellar nucleosynthesis,
\item ejection by stellar winds and explosions,
\item elemental fractionation,
\item acceleration of these {\it primary} nuclides, by shocks due to supernovae and  winds of massive stars,
\item propagation through the interstellar medium of the Galaxy, where those nuclides are fragmented (\emph{spallated} by collisions with the ambient gas), giving rise to {\it secondary} nuclei;
\item modulation at the heliospheric boundary, and
\item detection of cosmic rays near Earth.
\end{enumerate}
\noindent
In particular, cosmic-ray transport through interstellar space within the Galaxy has been studied with models of varying sophistication,
which account for a large number of astrophysical observables \citep[see the comprehensive review of][and references therein]{Strong2007}.
To describe the composition data, less sophisticated models
are sufficient, like e.g. the \emph{leaky-box} model, where
cosmic rays are assumed to fill homogeneously a cylindrical box - the Galactic disk - and  their
intensity in the interstellar medium is assumed to be in a steady state equilibrium between several
production and destruction processes. The former involve acceleration in cosmic-ray sources and production \emph{in-flight} trough fragmentation of heavier nuclides, while the latter include either physical losses from
the \emph{leaky box} (escape from the Galaxy) or losses in energy space (ionization) and in particle space (fragmentation, radioactive decay or pion production in the case of proton-proton collisions).
Most of the physical parameters describing these processes are well known,
although some spallation cross sections still suffer from considerable uncertainties.

The many intricacies of cosmic-ray transport
are encoded in a simple parameter, the {\it escape length}
$\Lambda_{\rm esc}$ (in g cm$^{-2}$): it represents the average column density traversed by
nuclei of Galactic cosmic rays  before escaping the Galactic leaky box.
The abundance ratio of a secondary to a primary nuclide depends essentially on the escape parameter $\Lambda_{esc}$.
Observations of LiBeB/CNO and ScTiV/Fe
in arriving Galactic cosmic rays, interpreted in this framework, suggest a mean escape length
$\Lambda_{\rm esc} \sim$7 g cm$^{-2}$. In fact, the observed secondary/primary ratios
display some energy dependence, which translates into an energy dependent $\Lambda_{\rm esc}(E)$,
going through a maximum at $E\sim$1 GeV~nucleon$^{-1}$ and decreasing both at higher and lower energies.
The observed energy dependence of $\Lambda_{\rm esc}(E)$
can be interpreted  in the framework of more sophisticated transport models and
provides valuable insight into the physics of transport (role of turbulent diffusion,
convection by a Galactic wind, re-acceleration, etc.); those same models can be used to
infer the injection spectra of cosmic rays at the source \citep[see][]{Strong2007}.
Once the key parameters of the leaky-box model are adjusted to reproduce the key
secondary/primary ratios, the same formalism may be used in order to evaluate the
secondary fractions (produced by fragmentation in-flight) of all Galactic cosmic-ray nuclides. Those fractions
depend critically on the relevant spallation cross sections, which are well known in most cases. Fractions close to 1 imply an almost purely secondary nature
while fractions close to 0 characterize primary nuclides (like, e.g.
$^{12}$C, $^{16}$O, $^{24}$Mg, $^{56}$Fe etc.). Contrary to the latter, the former are very sensitive to the
adopted $\Lambda_{\rm esc}$ \citep{Wiedenbeck2007}.

\Al\ has been measured with the CRIS instrument on the ACE satellite \citep{Yanasak2001} (Figure~\ref{fig:26Al_CRIS}).
This \al\ is plausibly \emph{secondary}, i.e., it has been produced from heavier cosmic-ray nuclei through spallation reactions. Therefore, it is not of stellar origin, and cosmic-ray \al measurements are not diagnostic towards \al sources beyond such interstellar spallation.
This is different for $^{60}$Fe, as discussed in Section~\ref{sec:observations60Fe}.

\subsubsection{Early Solar System materials}  
\label{sec:ESS}

\begin{figure}
\includegraphics[width=\columnwidth]{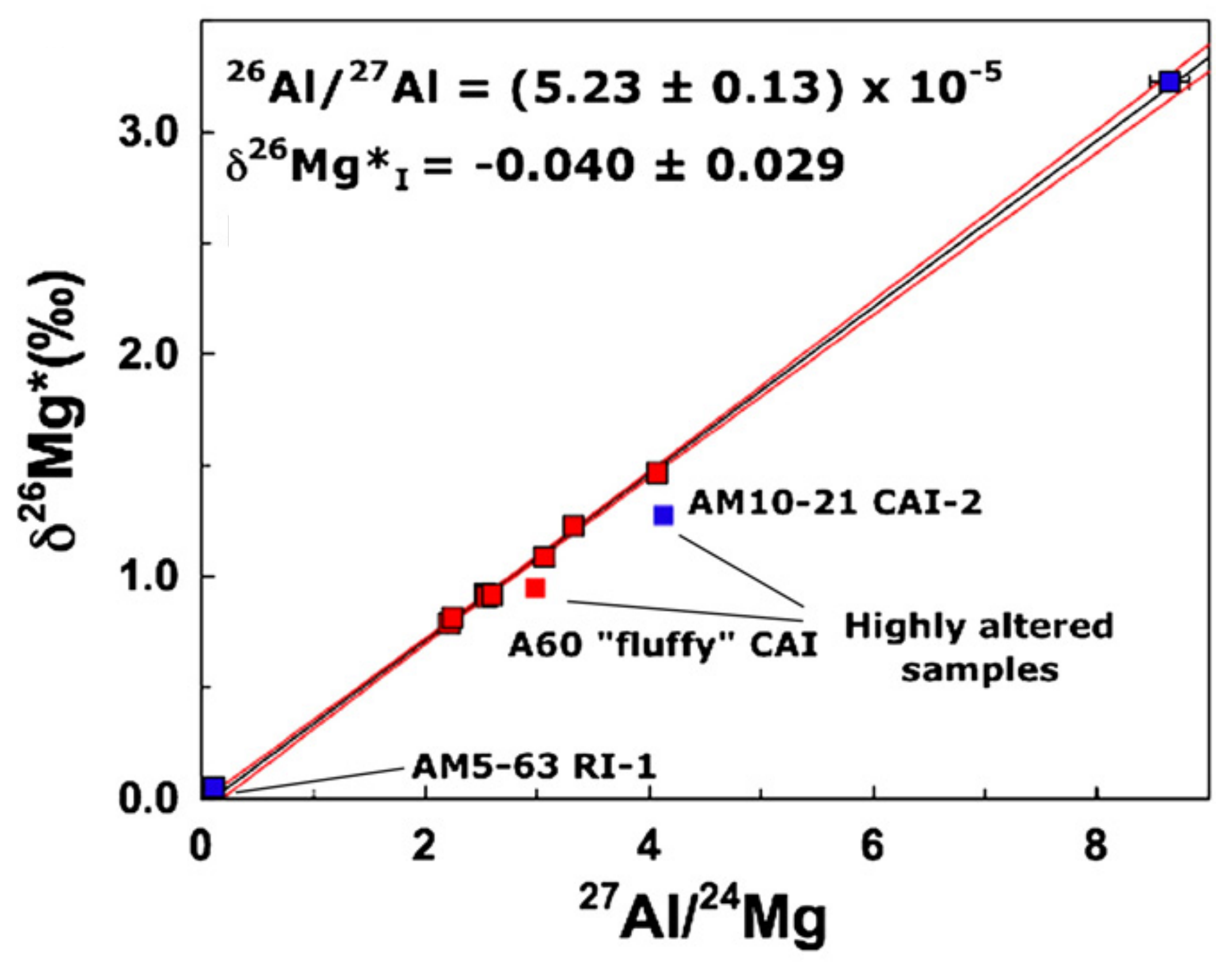}
\caption{Radiogenic \iso{26}Mg excess ($\delta$\iso{26}Mg\*, represented as deviations in parts per 1000 from a terrestrial standard) versus the \iso{27}Al/\iso{24}Mg ratio
in CAIs from the Allende CV3 carbonaceous chondrite. Red squares are from \citet{jacobsen2008} and the blue diamonds from \citet{bizzarro04}. The lines represent the isochrones derived from fitting the data points. The slope of the lines represent the initial \iso{26}al/\iso{27}Al ratio at the time of the formation of the CAIs, and the intercept at \iso{27}Al/\iso{24}Mg, the initial \iso{26}Mg/\iso{24}Mg ratio.  \citep[From][by permission]{jacobsen2008}).
\label{fig:isochrone}}
\end{figure}

One of the most interesting events in the Universe in relation to measuring the \iso{26}Al isotope is the birth of our own Solar System.  It was predicted in the 1950s that \al should have been part of the newborn Solar System as a heating source \citep{urey55}, and in 1970s, \citet{lee77} discovered that a relatively large abundance of \iso{26}Al was in fact present when the Sun formed.
Since such primordial \iso{26}Al is now decayed, its original abundance 4.6 Gyr ago can only be inferred indirectly, from excesses in its daughter isotope \iso{26}Mg, and particularly from the correlation of such excess in the form of \iso{26}Mg/\iso{24}Mg versus Al/Mg ratios of the analysed samples.
The \iso{26}Al/\iso{27}Al ratio present at the time of the formation of a given sample can be derived through the slope of the linear $isochrone$ that can be constructed by connecting data points obtained from inclusions of different chemical composition (i.e., different Al/Mg) with their varying \iso{26}Mg excess (see Figure~\ref{fig:isochrone}).

Calcium-aluminium-rich inclusions (CAIs) are believed to be the oldest solids to have formed in the early Solar System that have been recovered so far from meteorites.  The  \iso{26}Al/\iso{27}Al ratio derived from such CAIs is $(5.23 \pm 0.13) \times 10^{-5}$ \citep{jacobsen2008}. This is more than one order of magnitude higher than the ratio expected to be present in the interstellar medium at the time of the formation of the Sun \citep[see, e.g.,][and Figure~\ref{fig:26Al27Alratios}]{huss09,cote19a}.
An explanation for this requires investigation of the temporal and spatial heterogeneity of the interstellar medium (see Section~\ref{sec:solarSysEnv}) and/or invoking extra local stellar sources of \iso{26}Al beyond those that contribute this isotope to the galactic interstellar medium. Such extra sources should have contributed close in time and space to the formation of the Sun.

\subsubsection{Stardust in meteorites}  
\label{sec:observations26Al:stardust}
\begin{figure}[t]
\includegraphics[width=\columnwidth]{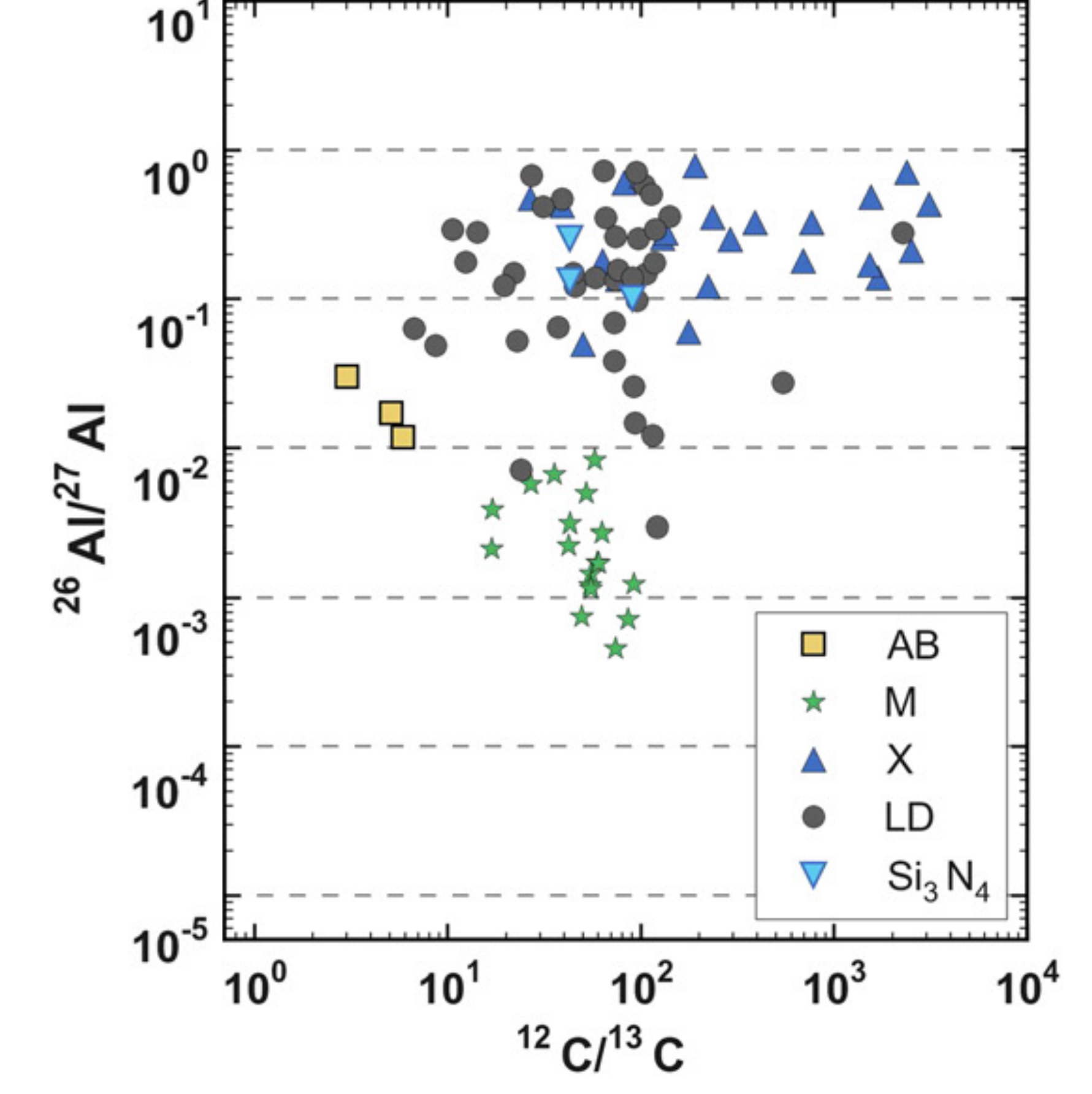}
\caption{Ranges of \iso{12}C/\iso{13} and \iso{26}Al/\iso{27}Al ratios in presolar low-density graphite (LD) and Si$_{3}$N$_{4}$ grains from supernovae, and SiC grains of different populations: M (mainstream) grains from AGB stars, X grains from supernovae, and AB grains of unclear origin. \citep[From][by permission]{groopman15}.
\label{fig:Groopman}}
\end{figure}

Stardust grains found in meteorites formed around stars and supernovae carry the undiluted signature of the nucleosynthesis occurring in or near these cosmic sites via their isotopic abundances, as measured in the laboratory \citep{lugaro05,zinner14}. Many types of stardust grains have been recovered so far, both carbon and oxygen rich. Both these classes include types of minerals that are rich in Al, in particular silicon carbide (SiC) and graphite in the case of C-rich grains, and corundum Al$_2$O$_3$ in the case of O-rich grains.
These grains are also poor in Mg, which allow us to identify
the measured abundance of \iso{26}Mg as the initial abundance of \iso{26}Al at the time of their formation in stellar outflows. 
In this simple, traditional approach the derivation of the initial \iso{26}Al/\iso{27}Al for stardust grains is not based on an isochrone, as for Solar System materials (Sec.~\ref{sec:ESS}), but on abundance of \iso{26}Mg. This is approximately correct because Mg is not a main component
of neither SiC, graphite, corundum (Al$_3$O$_2$), nor hibonite (CaAl$_{12}$O$_{19}$) grains.
Mg is, however, a main component of spinel (MgAl$_2$O$_4$).
Therefore, for this type of stardust grains the initial abundance of \iso{26}Al needs to be disentangled from the initial abundance of \iso{26}Mg.
Stochiometric spinel contains two atoms of Al per each atom of Mg, which corresponds roughly to 25 times higher than in the average Solar System material. In single stardust spinel grains, however, this proportion may vary and such variation needs to be taken into account when attempting to derive the initial \iso{26}Al/\iso{27}Al ratio. A more recent study by \citet{groopman15}, applyed the isochrone method also to stardust grains and demonstrated that the \iso{26}Al/\iso{27}Al abundance ratios for stardust derived from such analysis result in more accurate measurements, and generally higher ratios than previously estimated (see~Figure~\ref{fig:Groopman}).

Overall, the C-rich grains believed to originate from core-collapse supernovae show very high abundances of \iso{26}Al, with inferred \iso{26}Al/\iso{27}Al ratios in the range 0.1 to 1 (see~Figure~\ref{fig:Groopman}, grey and light and blue symbols), higher than theoretical predictions.
They apparently require some extra production mechanism for \iso{26}Al at work in core-collapse supernovae beyond those described in Section~\ref{sec:al26_from_CCSN}; possibly, this is related to ingestion of H into the He-burning shell and subsequent explosive nucleosynthesis \citep{pignatari13c}.
The grains that originated in AGB stars show somewhat lower \iso{26}Al abundances (see~Figure~\ref{fig:Groopman}), green symbols), with \iso{26}Al/\iso{27}Al in the range $10^{-3}$ to $10^{-2}$. These numbers can also be used for comparison and constraints to the nucleosynthesis models and the nuclear reactions involved in the production of \iso{26}Al in AGB stars of low mass, which become C-rich and are the origin of most of the SiC grains \citep{vanraai08}. The AGB stardust most rich in \iso{26}Al is represented by O-rich grains belonging to the specific Group II. These grains show strong depletion in \iso{18}O/\iso{16}O and have relatively high \iso{26}Al/\iso{27}Al ratios (up to 0.1). These features are both a product of efficient H burning, attributed to some kind of extra mixing in low-mass AGB stars \citep{palmerini11}, and recently also connected to hot-bottom burning in massive AGB stars \citep[see Section~\ref{agb1} and][]{lugaro17}.

\subsubsection{Gamma rays from interstellar $^{26}$Al}  
\label{sec:observations26Al-gammas}

\begin{figure}  
\centering
\includegraphics[width=1.0\columnwidth]{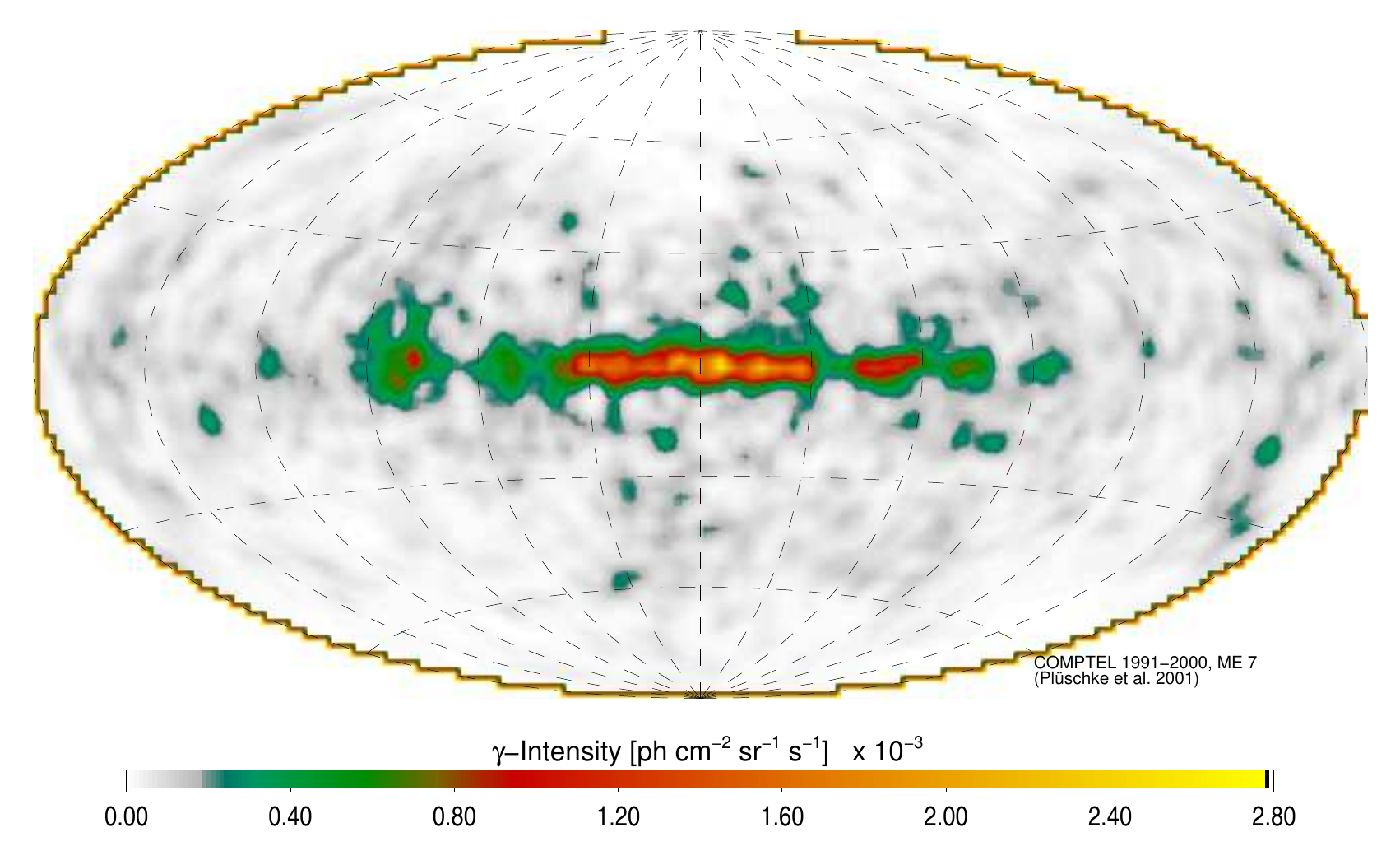}
\caption{The $^{26}$Al sky as imaged with data from the COMPTEL telescope on NASA's Compton Gamma-Ray Observatory. This image \citep{Pluschke:2001c} was obtained from measurements taken 1991-2000, and using a maximum-entropy regularization together with likelihood to iteratively fit a best image to the measured photons.}
\label{fig:Almap}
\end{figure}   

The direct observation of \Al~ decay in interstellar space through its characteristic gamma rays with energy 1808.65~keV was advertised as an exciting possibility of $\gamma$-ray astronomy in its early years \citep{Lingenfelter:1978}.
The HEAO-C satellite then obtained sufficiently-sensitive measurements from galactic $\gamma$-rays in 1978/1979, which were published in 1982 as a first and convincing detection of \Al~ decay in interstellar space and therefore a proof of currently-ongoing nucleosynthesis in our Galaxy \citep{Mahoney:1982}.

The HEAO-C finding was confirmed by balloon-borne observations in the 1980s \citep[e.g.][]{Ballmoos:1986}; but the sparse signals and poor sensitivity and spatial resolution could only reveal 1.8 MeV emission from the general direction towards the galactic center.
The observed emission location prompted  speculations throughout that decade about the origin of the emission, attributed to various sources: novae \citep{Clayton:1987}, AGB stars \citep{Bazan:1993}, WR stars \citep{Prantzos:1986} supernovae \citep{Woosley:1988} and even a supermassive star in the Galactic center \citep{Hillebrandt:1987}.
Uncertainties in the yields - and even the frequencies - of all those sources made it difficult to conclude.
It was thus suggested that  the spatial profile of the gamma-ray emission as function of galactic longitude could reveal the underlying \Al\ sources, being more peaked towards the galactic centre in some cases (where an old stellar population, such as novae or low mass AGB stars would dominate) than in others (where a young population, such as massive stars and their core-collapse supernova would dominate).
\citet{Prantzos:1991} suggested that the smoking gun for the case of massive stars would be the observation of enhanced $\gamma$-ray emission from the directions tangent to the spiral arms of the Galaxy, where most massive stars form and die.

The  exciting prospect to measure radioactive decays from recently-synthesised cosmic nuclei gave a new impetus to observational $\gamma$-ray astronomy.
The NASA Compton Gamma-Ray Observatory was agreed upon as a flagship mission. Even though its high-resolution spectrometer (GRSE) instrument had been abandoned in the late planning phase due to technical and financial problems, the Compton Gamma-Ray Observatory mission from 1991 until 2000 collected valuable data that provided the first all-sky survey in $\gamma$-rays, including lines from cosmic radioactivity.
The Imaging Compton Telescope (COMPTEL, one of four instruments aboard this observatory) sky survey provided a sky image in the \Al\ $\gamma$-ray line (see Fig.~\ref{fig:Almap}), which showed structured \Al\ emission, extended along the plane of the Galaxy \citep{Pluschke:2001c,Prantzos:1996a,Diehl:1995b}, in broad agreement with earlier expectations of \Al\ being produced throughout the Galaxy and mostly from massive stars and their supernovae.
We caution here that this argument is based on massive stars showing up in clusters, while AGB stars and novae evolve on longer time scale and thus can move away from their birth sites before ejecting $^{26}$Al; but this may not hold for the higher-mass end of AGB stars (evolving within $\sim$3~10$^7$~yr) and high-luminosity novae, so that their $^{26}$Al contribution would be degenerate with that of massive stars beyond 8~\Msol and their core-collapse supernovae.

\begin{figure}  
\centering
\includegraphics[width=0.8\columnwidth]{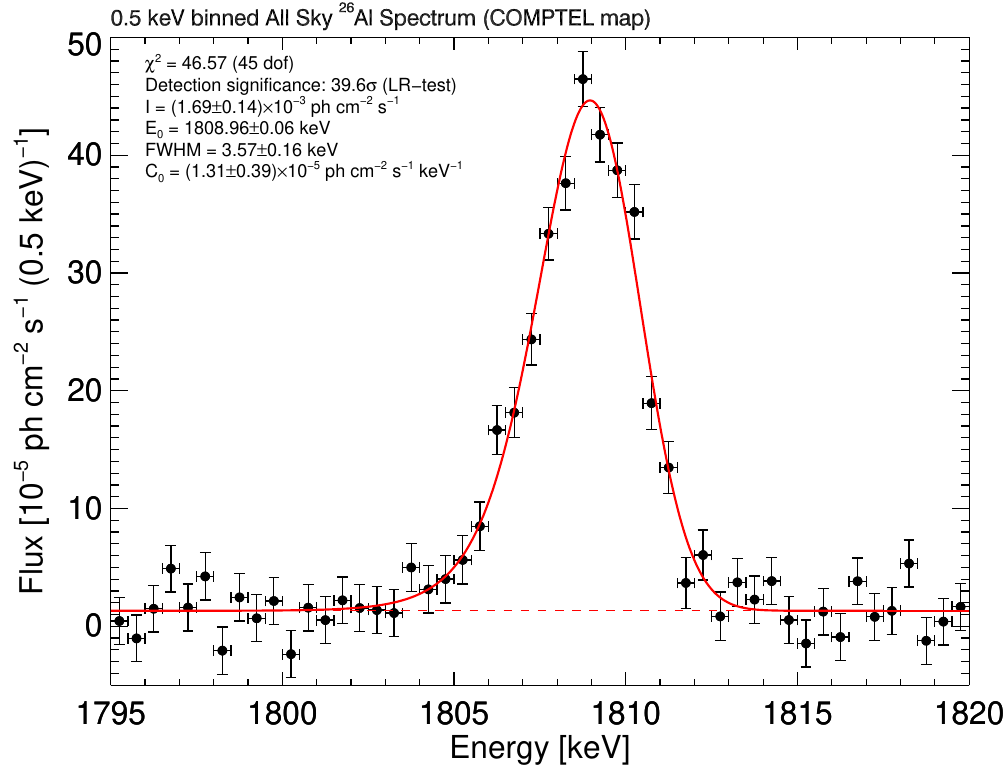}
\caption{The \Al\ line as seen with INTEGRAL high-resolution spectrometer SPI and 13 years of measurements integrated \citep{Siegert:2017}.
}
\label{fig:almap-spec}
\end{figure}   

While the COMPTEL detectors lacked the spectral resolution required for line identification and spectroscopic studies (with \about~200~keV instrumental resolution, compared to \about~3~keV for Ge detectors, at the energy of the \Al~ line), a 1995 balloon experiment also carrying high-resolution Ge detectors provided an indication that the \Al~ line was significantly broadened to 6.4~keV \citep{Naya:1996}. This implied kinematic Doppler broadening of astrophysical origin of 540~km~s$^{-1}$. Considering the $1.04 \times 10^6$~y mean life of $^{26}$Al, such a large line width would naively translate into kpc-sized cavities around \Al\ sources, or alternatively major fractions of  \Al\ should be condensed on grains so that they travel ballistically and are not decelerated by cavity shells \citep{Chen:1997,Sturner:1999}.

The ESA INTEGRAL space observatory \index{INTEGRAL} with its Ge-detector based spectrometer SPI, launched in 2002, provided a wealth of high-quality spectroscopic data accumulated over its more than 15-year long mission. The sensitivity of current $\gamma$-ray surveys penetrates to intensities as low as 10$^{-5}$~ph~cm$^{-1}$s$^{-1}$ or below, with typical exposure times of at least a month, and correspondingly longer for deeper exposures.
This allowed higher precision Galaxy-wide study of \Al\ \citep{Diehl:2006d}, aided by the spectral information from the $^{26}$Al line, which now is resolved with a width of $\sim$~3~keV (Figure~\ref{fig:almap-spec}).
These deep observations also provide a solid foundation for the detailed test of our understanding of the activity of massive stars within specific and well-constrained massive star groups \citep[see][for a review of astrophysical issues and lessons]{Krausea21a}, with specific results for the Orion \citep{Voss:2010a} and Scorpius-Centaurus \citep{Krause:2018} stellar groups.

\begin{figure}  
\centering
\includegraphics[width=0.6\columnwidth]{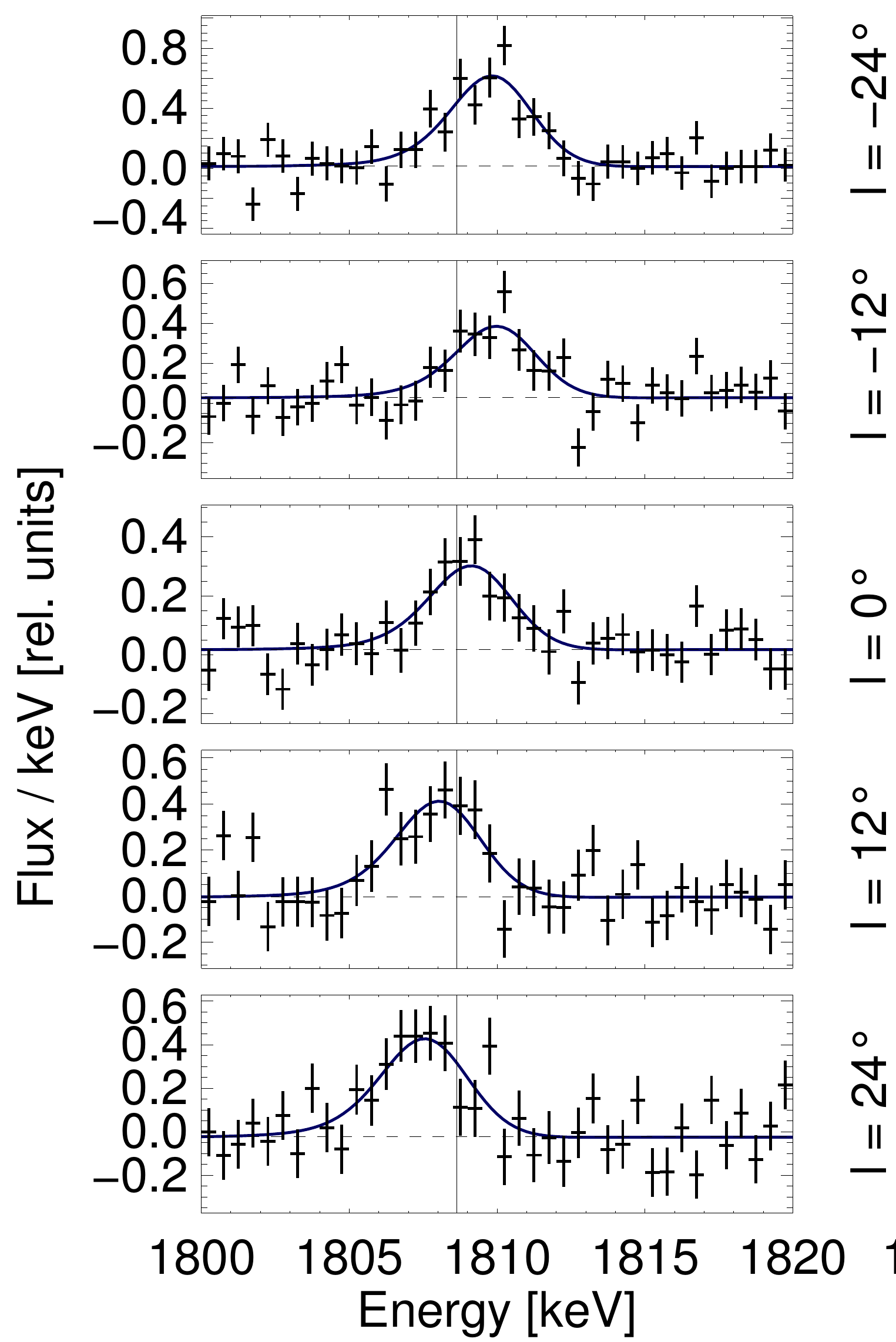}
\caption{The \Al\ line as seen towards different directions (in Galactic longitude) with INTEGRAL's high-resolution spectrometer SPI. This demonstrates kinematic line shifts from the Doppler effect, due to large-scale Galactic rotation \citep{Kretschmer:2013}. }
\label{fig:al_longitudes}
\end{figure}   

It is not straightforward to interpret measurements of $\gamma$-ray line emission. As Figure~\ref{fig:Almap} shows, the inherent blurring of the measurement by the response function of the $\gamma$-ray telescopes, and their inherently-high background, require sophisticated deconvolution and fitting methods to obtain the astrophysical result. Bayesian methods need to be applied, and simple methods of subtracting background and inverting data with the instrument response matrix are not feasible. Comparisons to predictions can be made in different ways. One may consider the \Al\ amount in the Galaxy and compare measurements with theoretical predictions; this has been done up to the early 2000s \citep[see][and discussions therein]{Diehl:2006d}.
For example, the total mass of \al\ within the Galaxy can be estimated from observations, and compared to theoretical estimates.
Observational estimates are based on the measured flux of $\gamma$-ray photons, but the \al\ mass depends on the assumed distance to the \al\ that produced those photons upon decay \citep[see][for a discussion of such bias]{Pleintinger:2019}. We believe now that observations constrain this to fall between 1.7 and 3.5~\Msol, with a best current estimate of 2~\Msol\ \citep{Pleintinger:2020}.
Similarly, theoretical estimates based on predicted, theoretical \al\ yield for a given type of nucleosynthetic source also depend on extrapolation to the galactic mass, i.e., on how many sources of such type have contributed to the current \al\ mass in the galactic interstellar medium.
For example, Wolf-Rayet winds have been estimated to contribute (0.9$\pm 0.5$)~\Msol\ of \al\ per Myr \citep{Meynet:1997}, the number later updated to 0.6--1.4~\Msol\ per Myr \citep{Palacios:2005}. 
For massive stars altogether (winds plus supernova), \citet{Limongi:2006a} provided a thorough discussion of uncertainties, and give a best estimate of 2.0~\Msol\ of \al  for the Galaxy as a whole\footnote{In such galactic-mass estimates, an initial-mass function is assumed, and integrated for total yields $dM$, over a time $dt\approx$1~Myr. This is called \emph{steady state}, i.e. $dM/dt$ within $\tau_{26Al}\approx$1~Myr.}.

\begin{figure}  
\centering
\includegraphics[width=0.9\columnwidth]{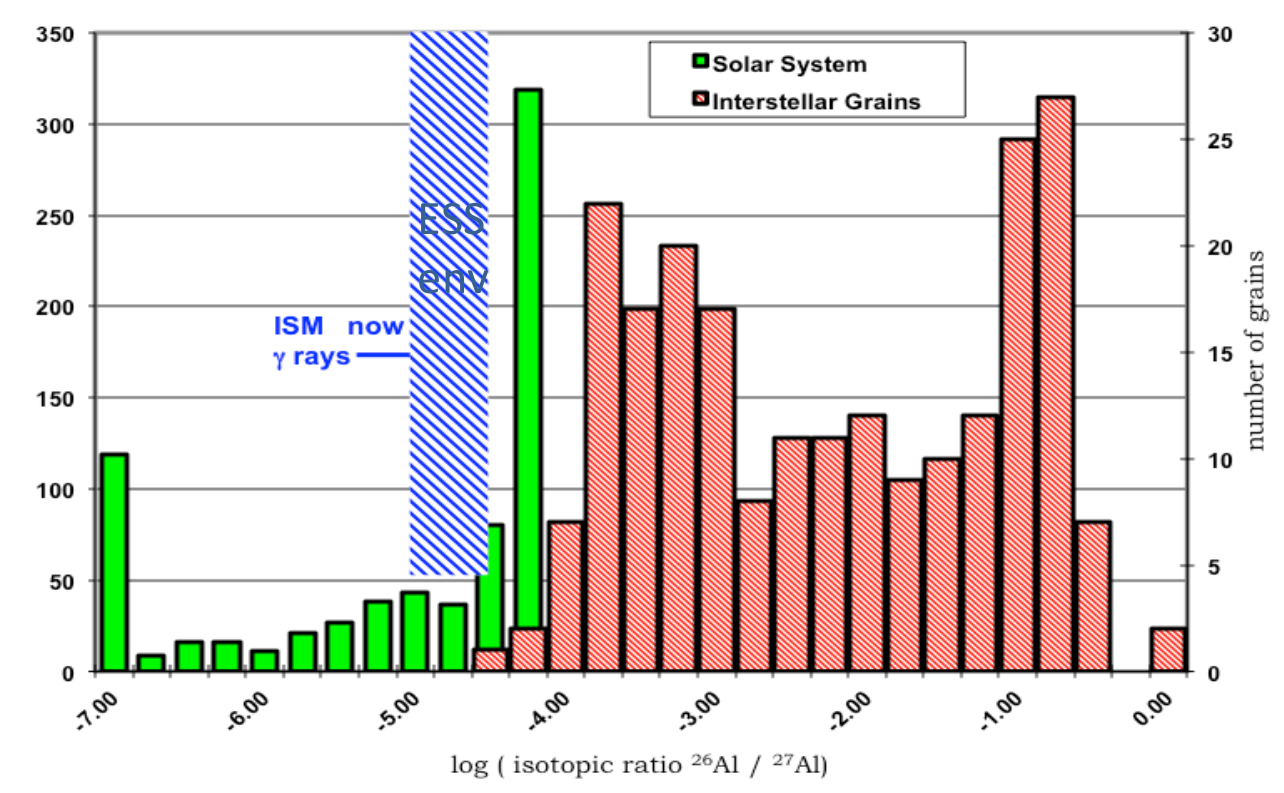}
\caption{The $^{26}$Al/$^{27}$Al ratio measured from $\gamma$~rays in the current Galaxy, extrapolated to the early Solar System (hatched area), as compared to measurements from the first solids that formed in the Solar System (Section~\ref{sec:ESS}) and in stardust grains (Section~\ref{sec:observations26Al:stardust}). (From a presentation by Roland Diehl at the 2013 Gordon conference). }
\label{fig:26Al27Alratios}
\end{figure}   

We may use the $\gamma$-ray measurement to represent the \al content in the current Galaxy, and translate this into an isotopic ratio for $^{26}$Al/$^{27}$Al, which has been discussed above for stardust grains and for the early Solar System.
If we assume a total interstellar gas mass of 4.95$\times$10$^9$~\Msol\ \citep{Robin:2003} and an abundance of $^{27}$Al taken from solar abundances of 6.4 \citep{lodders10}, and obtain a value of 6$\times$10$^{-6}$ for $^{26}$Al/$^{27}$Al. If the interstellar medium abundance of $^{27}$Al scales linearly with time, 4.6~Gy before present, this ratio would have been roughly 3 times higher, but still roughly a factor of two lower than the canonical early Solar System value obtained from CAI inclusions in meteorites of $5 \times 10^{-5}$ \citep{jacobsen2008}, as shown in Figure~\ref{fig:26Al27Alratios}. It should be noted that this is a conservative upper limit estimate as there is no evidence that metal abundances grew with time in the past 5 Gyr, but rather there is a large spread at each age \citep[see, e.g.][]{Casagrande:2011}.


As theoretical predictions have become more sophisticated in predicting directly the expected observational signatures \citep{Fujimoto:2018,Rodgers-Lee:2019}, more accurate comparisons can be made.
However, biases in both the observations and the theoretical predictions require great care in drawing astrophysical conclusions \citep{Pleintinger:2019}. In particular, theoretical predictions often need to make assumptions about our Galaxy and its morphology. These are particularly critical for the vicinity of the solar position, as nearby sources would appear as bright emission that may dominate the signal \citep{Fujimoto:2020} .

Kinematic constraints could be obtained from the \Al~ line width and centroid \citep{Kretschmer:2013} and have added an important new aspect to the trace of nucleosynthesis by radioactivities. Kinematics of nucleosynthesis ejecta are reflected in these observables from high resolution spectroscopy through the Doppler effect (Figure~\ref{fig:al_longitudes}); velocities down to tens of km~s$^{-1}$ are accessible.
Their analysis within multi-messenger studies using the stellar census and information on atomic and molecular lines from radio data, as well as hot plasma from X-ray emission \citep{Voss:2009,Voss:2010a,Krause:2013,Krause:2014a}, have taught us that the ejection of new nuclei and their feeding into next-generation stars apparently is a much more complex process than the instantaneous recycling approximation assumed in most 1D chemical evolution models (see Section~\ref{sec:GCE}). 
Thanks to observational constraints such as the $\gamma$-ray line measurements of \Al, ejection and transport of new nuclei can now be studied in more detail. These
\Al\ observations and their analyses have led to an \emph{\Al\ astronomy} within studies of stellar feedback and massive-star nucleosynthesis \citep[e.g.,][]{Krausea21a}.

\begin{figure}  
\centering
\includegraphics[width=\columnwidth]{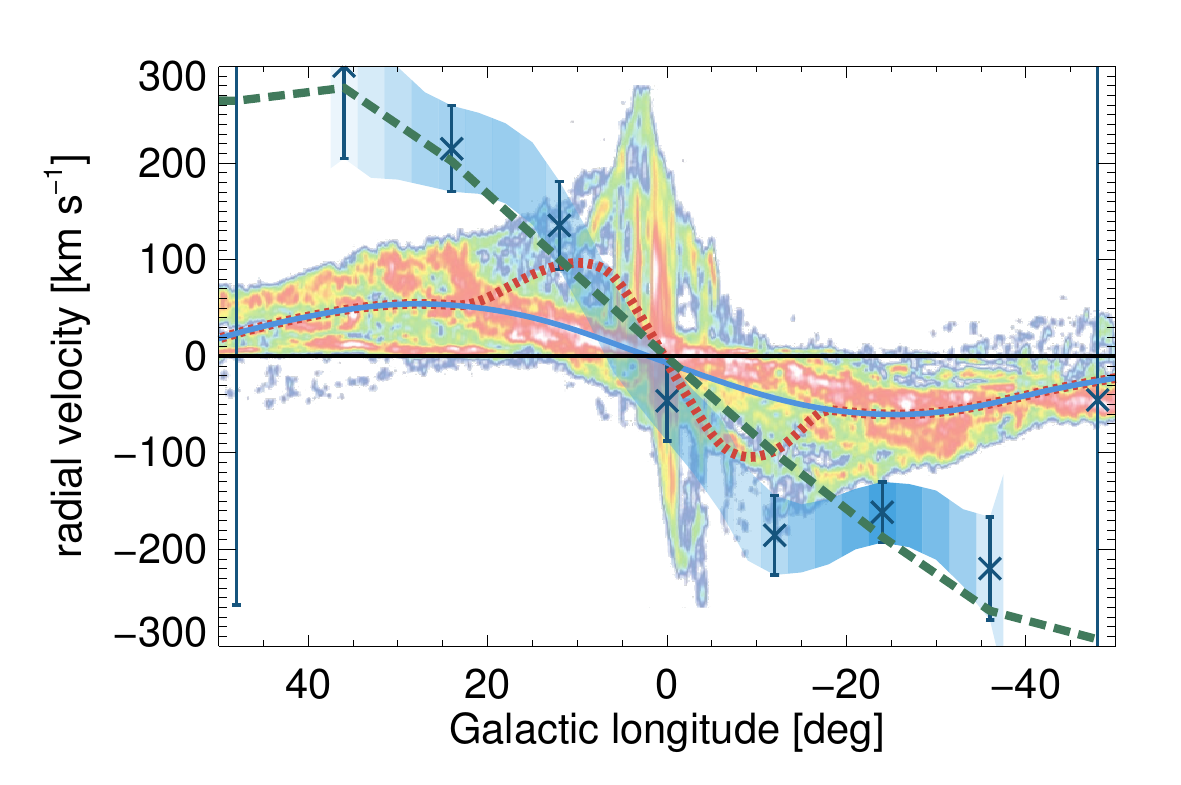}
\caption{The line-of-sight velocity shifts seen in the \Al\ line versus Galactic longitude, compared to measurements for molecular gas, and a model assuming \Al\ blown into inter-arm cavities at the leading side of spiral arms \citep{Kretschmer:2013,Krause:2015}. }
\label{fig:al_long-velocity}
\end{figure}   

In summary, the lessons learned from the \al\ $\gamma$-ray observations can be summarised as follows:
\begin{itemize}
    \item $^{26}$Al is abundantly present throughout our Galaxy, both in the inner galactic regions as well as through outer spiral arm regions.
    \item The sources of $^{26}$Al cluster in specific regions. This is not compatible with $^{26}$Al ejections from classical novae, because the few times 10$^7$ individual sources expected from the galactic nova rate and spatial distribution would make a smooth and centrally-bright appearance of the $^{26}$Al $\gamma$-ray sky. Also intermediate- and low-mass AGB stars are not favoured sources, from the same smoothness reasoning.
    \item The accumulated current mass of $^{26}$Al in the Galaxy is estimated between about 1.7 and 3 \Msol. This estimate depends on the spatial distribution of sources assumed throughout the Galaxy because the source distance determines the $\gamma$-ray brightness, with a 30\% uncertainty arising from this assumption. As discussed by \citet{Pleintinger:2019}, nearby sources are also important in such estimate.
    \item The galactic $^{26}$Al mass can be used to estimate the rates of core-collapse supernovae and star formation, when relying on $^{26}$Al yields from theoretical models. \citet{Diehl:2006d} discussed this in detail. Adopting an origin from massive stars alone, they derived a core-collapse supernova rate of 1.9$\pm$1.1 per century. This is in agreement with other supernova rate estimates, most of which rely on data either from the solar vicinity or from other galaxies assumed to be similar to the Milky Way galaxy. Therefore, despite the large (systematic) uncertainty quoted above, this measurement is significant, because it is based on $^{26}$Al $\gamma$ rays as a more penetrating tracer, measured from the entire Galaxy. The systematic errors are related to nuclear astrophysics and specifically the stellar yields, rather than to occultations and their corrections. Accounting for better understandings about nearby source regions and their contributions to the measured $\gamma$-ray flux, this estimate has been updated to a supernova rate of 1.3$\pm$0.6 events per century \citep{Diehl:2018}. This has been converted to a star formation rate in our Galaxy of 2--5~\Msol~yr$^{-1}$ \citep{Diehl:2006d,Rodgers-Lee:2019},
    when adopting an initial-mass distribution of stars and a mass limit for core-collapses to occur.
    Note that the probability of massive-star explosions versus direct collapse, and as function of initial mass, has been much discussed in the recent years \citep[][see also Section~2.2 and 4.2]{OConnor:2011}.
    \item Several individual regions within the Galaxy have been identified as sources of $^{26}$Al. These are: Cygnus \citep{Knodlseder:2002,Martin:2008,Martin:2009}, Carina \citep{Voss:2012,Knoedlseder:1996b}, Orion \citep{Voss:2010a,Diehl:2003e}, Scorpius-Centaurus \citep{Krause:2018,Diehl:2010}, and Perseus \citep{Pleintinger:2020}.
    Their observations provide more stringent constraints on massive-star models because in these regions the stellar population is known to much better precision, often even allowing for age constraints. Multi-wavelength tests of $^{26}$Al origins have been made, using $^{26}$Al $\gamma$ rays to understand nucleosynthesis yields, interstellar cavity sizes as seen in HI data for kinetic energy ejections, stellar census seen in optical for the stellar population scaling, and interstellar electron abundance estimated from pulsar dispersion measures characterising the ionization power of the massive star population \citep[see][for more detail on the population synthesis method]{Voss:2009}.
    \item The high apparent velocity seen for bulk motion of decaying $^{26}$Al \citep{Kretschmer:2013} shows that $^{26}$Al velocities remain higher than the velocities within typical interstellar gas for 10$^6$ years, and have a bias in the direction of Galactic rotation. This has been interpreted as $^{26}$Al decay occurring preferentially within large cavities (superbubbles; see Section~\ref{superbubbles:text} and Figure~\ref{fig:al_long-velocity}), which are elongated away from sources into the direction of large-scale Galactic rotation. \citet{Krause:2015} have discussed that wind-blown superbubbles around massive-star groups plausibly extend further in forward directions away from spiral arms (that host the sources), and such superbubbles can extend up to kpc (see also \citep{Rodgers-Lee:2019,Krausea21a}).
\end{itemize}

\noindent
The main open issues in $\gamma$-ray measurements of $^{26}$Al are:
\begin{itemize}
    \item Imaging resolution of measurements is currently inadequate to locate source regions such as massive-star groups to adequate precision. This also constrains the measurements of the latitude extent of $\gamma$-ray emission from $^{26}$Al.
    \item Sensitivities of current instruments are too low to measure $^{26}$Al in regions of the galactic halo (and the superbubbles extending there), or in regions where contributions from lower-mass stars and/or novae may dominate $^{26}$Al locally, as well as in external nearby galaxies such as the LMC or M31.
    \item Methods to compare measured data to model predictions are indirect. It is desirable to have models predict measured data; rather, currently we are restricted to deconvolve the measured data under some assumptions about the $^{26}$Al sources, and define comparisons of models to data in parameters that are subject to systematics from how models were constructed as well as how data have been deconvolved \citep[see][]{Pleintinger:2020}.
\end{itemize}

\subsubsection{Other electromagnetic radiation}  

Supernova remnants can be studied through X-ray spectroscopy. The hot plasma with temperatures around 10$^7$ K produces highly-ionized atoms, such that metals like Si, Mn, Fe have as few electrons as H and He 
\citep{Vink:2012}. These measurements have shown enhancements in metals that are signposts of recently-enriched gas in these objects, as expected.
It is not as straightforward as in $\gamma$ rays to interpret these data in terms of absolute abundances of new nuclei, however, because the degree of ionisation is difficult to assess in such a dynamic, hot, and tenuous plasma.
Nevertheless, metal ratios and unusual enrichments have been determined \citep{Yamaguchi:2010aa}, which provide a diagnostic of supernova nucleosynthesis.
$^{26}$Al, however, is inaccessible to X-ray spectroscopy, being a rather light nucleus.

A few years ago, advances in sub-mm spectroscopy have been reported with the first instruments for the ALMA sub-mm observatory, and corresponding advances in laboratory studies to identify lines for molecules including radioactive species.
Rotational lines of $^{26}$AlF could be measured from a point nova-like source called CK Vul \citep{Kaminski:2018}, which represents a breakthrough, as spatial resolution allows to pinpoint a source directly.
However, it is also clear that molecule production such as in this case will only occur under very special conditions. While this bias makes it difficult to derive conclusions on $^{26}$Al sources generally, learning about special sources will be a very fruitful complement of $^{26}$Al observations in $\gamma$ rays, stardust, and cosmic rays, especially for molecule-rich environments such as AGB stars and even proto-planetary discs.

\section{The cosmic trajectory of $^{60}$Fe}
\label{sec:Fe60}

\begin{figure}
\centering
	\includegraphics[width=\linewidth]{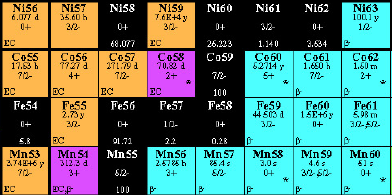}
	\caption{The table of isotopes in the neighbourhood of ${}^{60}$Fe. (see Figure~\ref{fig:isotopes_Al-region} for legend details on the information per isotope, and original reference).
	}
	\label{fig:isotopes_Fe-region}
\end{figure}

\subsection{Nuclear properties, creation and destruction reactions}
\label{sec:fe60nuclear}
\subsubsection{Nuclear properties of $^{60}$Fe}
\begin{figure}
\centering
	\includegraphics[width=\linewidth]{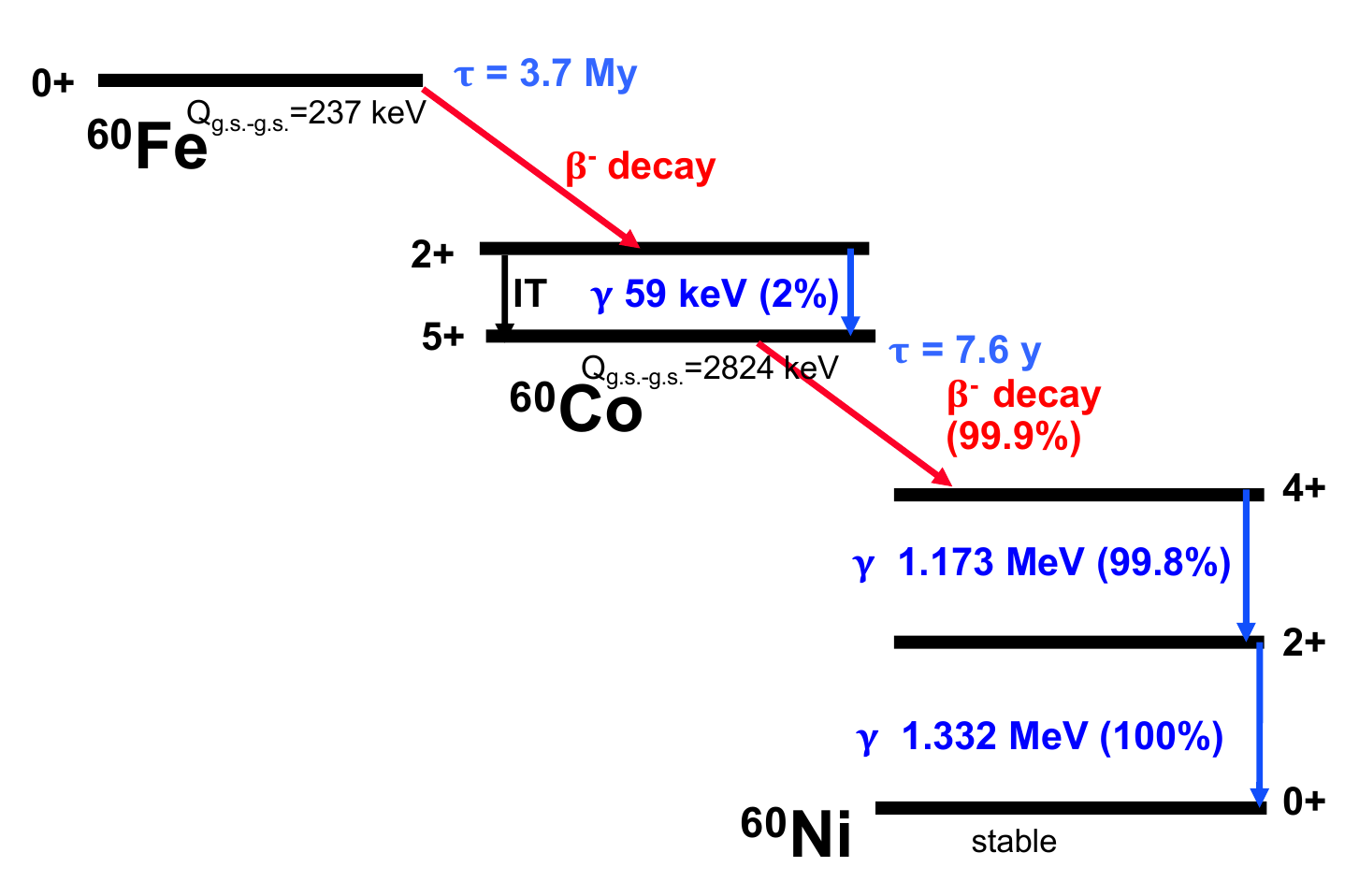}
	\caption{The decay of ${}^{60}$Fe with its level scheme. Red arrows indicate $\beta$ decay, blue lines $\gamma$ transitions, and the black arrow marks an internal transition.}
	\label{fig:fe60_decay}
\end{figure}

$^{60}$Fe is a neutron rich isotope with 8 excess neutrons.
The nuclide chart is shown in Figure~\ref{fig:isotopes_Fe-region} illustrates the nuclides that may be involved in the production or destruction of $^{60}$Fe.
$^{60}$Fe is unstable with a relatively long half-life of 2.62 Myr.
This terrestrial half-life is well determined, with two recent experiments \citep{wallner15a,ostdiek17} confirming the half-life of 2.62 Myr presented by \citet{rugel09}, and, initially surprising, 75\% longer than the previous standard value \citep{Kutschera:1984}.
The decay and level scheme is shown in Figure \ref{fig:fe60_decay}. The ground state of ${}^{60}$Fe has spin and parity of 0$^{+}$. Its long lifetime is due to the higher multipole $\beta$~decay transitions to the 2$^+$ state of ${}^{60}$Co (E$_x$=58.59 keV). This state rapidly decays to the ground state through an internal transition mostly, with a $\gamma$ transition fraction of 2\%.
${}^{60}$Co is an unstable nucleus with a half-life of 5.2714 years. It decays to ${}^{60}$Ni dominantly by emitting two characteristic gamma rays with energies of 1.173 and 1.333 MeV, respectively.

\subsubsection{Production and destruction of $^{60}$Fe}
\begin{figure}
\centering
	\includegraphics[width=0.8\linewidth]{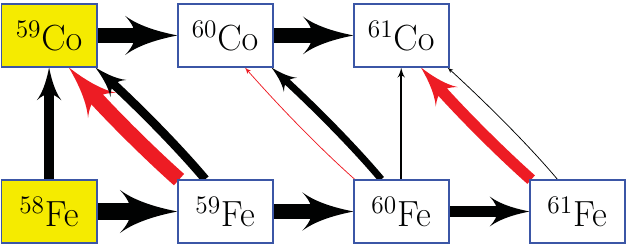}
	\caption{Integrated reaction flow chart for the $^{60}$Fe nucleosynthesis in the C/Ne shell burning calculated with the 1-zone code NUCNET. As in Figure~\ref{fig:flow_shell_burning}, the thickness of the arrows correspond to the intensities of the flows; red and black arrows show $\beta$ interactions and nuclear reactions, respectively.}
	\label{fig:flow_fe60_shell_burning}
\end{figure}

\begin{figure}
\centering
	\includegraphics[width=0.8\linewidth]{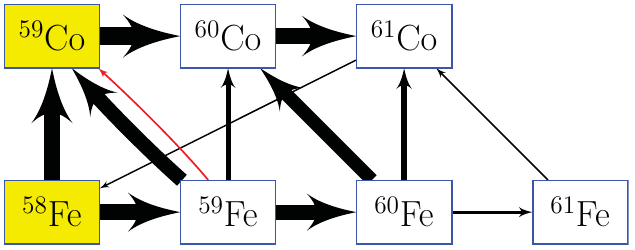}
	\caption{Same as Figure~\ref{fig:flow_fe60_shell_burning} but for Ne/C explosive burning.}
	\label{fig:flow_fe60_expl_burning}
\end{figure}

Differently from \iso{26}Al, which is made mostly via proton captures, $^{60}$Fe is produced by a series neutron captures from the stable iron isotopes (see Figure~\ref{fig:isotopes_Fe-region}).
The reaction flow chart for $^{60}$Fe nucleosynthesis in two example environments is shown in Figure\ref{fig:flow_fe60_shell_burning} and Figure\ref{fig:flow_fe60_expl_burning}.
$^{60}$Fe is produced through $^{58}$Fe($n$,$\gamma$)$^{59}$Fe($n$,$\gamma$)$^{60}$Fe, and destroyed by $^{60}$Fe($p$,$n$)$^{60}$Co and ${}^{60}$Fe($n$,$\gamma$)${}^{61}$Fe. The branching point at $^{59}$Fe is crucial for the production of $^{60}$Fe. The rates of these neutron captures have to be within a suitable range to produce enough $^{59}$Fe and $^{60}$Fe before  $^{59}$Fe  is destroyed by  $\beta$~decay or (p,n) reaction, but does not destroy $^{60}$Fe significantly through further neutron captures.
Estimates of suitable neutron densities fall in the vicinity of $10^{10-11}$cm$^{-3}$.
These neutron captures are expected to occur mainly in the He and C shells, and possibly also during explosive burning, within massive stars.
At the C/Ne shell burning temperature ($>$1 GK), the $^{59}$Fe($n$,$\gamma$)$^{60}$Fe
competes with $^{59}$Fe $\beta$-decay.
During Ne/C explosive burning, the temperature is even higher, about 2 GK. Then the $^{59}$Fe($p$,$n$)$^{59}$Co reaction becomes faster than the ${}^{59}$Fe $\beta$~decay,
and dominates the destruction of ${}^{59}$Fe; again, it competes with the $^{59}$Fe($n$,$\gamma$)$^{60}$Fe reaction.
Other than in massive stars, $^{60}$Fe may also produced during helium burning in AGB stars (Section~\ref{sec:fe60_cosmic_environment}).

${}^{60}$Fe is primarily destroyed by the ${}^{60}$Fe($n$,$\gamma$)${}^{61}$Fe reaction in hydrostatic shell-burning environments, and also by the ${}^{60}$Fe($p$,$n$)${}^{60}$Co reaction during explosive burning.
Besides these destructive reactions, ${}^{60}$Fe can also undergo $\beta$-decay within these production sites, due to an enhanced rate at T$>$2~GK, as compared to terrestrial conditions.


\subsubsection{Reaction rate uncertainties related to $^{60}$Fe}
The temperature dependency of $\beta$-decay rate of \iso{59}Fe has been studied using both theoretical modelling and experimental data.
At T$_9$=1.2, a typical carbon shell burning temperature, the stellar decay rate of \iso{59}Fe is two orders of magnitude faster than terrestrial rate.
The rate estimate based on large-scale shell-model calculation with KB3 interaction  \citep{LM-2000} (LMP) has been widely used in stellar-evolution codes.
A recent study based on a new shell model calculation \citep{lika16} finds decay rates are about 3 times higher than LMP, at C-shell burning conditions.
This calculation also finds that the transition \iso{59}Fe ($5/2^-$, 472 keV)$\rightarrow$ \iso{59}Co ($7/2^-$ g.s.)
plays an important role in the \iso{59}Fe $\beta$-decay.
This calls for a better determination of this transition by experiment.
A charge exchange experiment could determine the weak interaction strength, helping the estimate for stellar $\beta$-decay \citep{Cole2012}.
The \iso{59}Co(n,p)\iso{59}Fe reaction was studied to determine the weak-interaction Gamow-Teller strengths B(GT) for the allowed transitions from the \iso{59}Co ground state to \iso{59}Fe. Limited by the poor energy resolution of the neutron beam, the uncertainties of B(GT) came out quite large, $\sim$40\%) \citep{co59np2}.  The \iso{59}Fe(571 keV) state had been assigned a spin/parity of $3/2^-$ from a (n,p) experiment, while  a $\gamma$ multiplicity measurement \citep{zhu-fe59}  prefers a 5/2$^-$ assignment of this state; then this state could also decay via an allowed transition
to the \iso{59}Co ground state, and contribute to the stellar $\beta$~decay of \iso{59}Fe.
A high-resolution measurement of the \iso{59}Fe(t,\iso{3}He)\iso{59}Co reaction, performed at NSCL \citep{fe59beta_gao_2021}, found the stellar decay rate to be 3.5(1.1) times faster than the LMP rate, confirming previous shell model calculations \citep{lika16}, and the contribution of 571keV could be neglected (although its $J^{\pi}$ was still not confirmed).
Stellar-evolution calculations show that the \iso{60}Fe production of an 18~\Msun\  star is decreased by 40\% when using the new rate \citep{fe59beta_gao_2021}.

Also the effective decay rate of $^{60}$Fe in stellar environments is affected by the thermal population of excited states.
Due to the lower level density in a even-even nucleus, only the first excited state of $^{60}$Fe (2$^+$, 824keV) is expected to be notably populated at a temperature of T$\leq$2~GK.
From currently-accepted shell model calculations \citep{LM-2000}, a half-life of 3.14~yr is suggested at 1.3~GK.
However, a different  earlier approach, based on a single particle approximation \citep{FFN}, had suggested a much shorter half-life of 0.14~yr.
Such a difference has significant impact  on the $^{60}$Fe yield from massive stars.


The temperatures in explosive burning are higher than that in carbon
shell burning, and $\beta$-decay rates such as of \iso{59}Fe and \iso{60}Fe would be even faster.
However,the \iso{59}Fe(p,n)\iso{59}Co reaction supersedes the \iso{59}Fe $\beta$-decay at $T_9>2$, and the
uncertainty in $\beta$-decay only plays a minor role for the \iso{60}Fe yields (Figure~\ref{fig:flow_fe60_expl_burning}).

In the He burning shell, instead, the temperature never
exceeds $T>0.4$GK, and the \iso{59}Fe and \iso{60}Fe $\beta$-decay rates remain  at the terrestrial values.




The \iso{59}Fe(n,$\gamma$)\iso{60}{Fe} reaction is  crucial to the production of \iso{60}Fe. A direct measurement of this reaction rate is challenged by the short half-life of \iso{59}Fe.
Experimental results
obtained from a Coulomb dissociation experiment of \iso{60}Fe~\citep{fe60ng} found
a rate is roughly 20\% higher than the value estimated from statistical theory as reported in the REACLIB database \citep{reaclib}.
However, the conversion herein of the
reverse reaction rates is model dependent, with an uncertainty of about 40\%.
Clearly, a better experiment for measuring \iso{59}Fe(n,$\gamma$)\iso{60}{Fe} is needed.

The surrogate ratio method has been proposed as a practicable way to
study the neutron capture rate of unstable nuclei  \citep{escher-rmp}.
In such an experiment,
one can measure the \iso{56,58}Fe(2n,$\gamma$)\iso{58,60}Fe rate to obtain the ratio between \iso{57}Fe(n,$\gamma$)\iso{58}Fe, which  is well known, because \iso{57}Fe is a stable isotope, and \iso{59}Fe(n,$\gamma$)\iso{60}Fe.
This approach has been proven as feasible
in determination of the key s-process reaction \iso{95}Zr(n,$\gamma$)\iso{96}Zr \citep{yan-apj},
and suggests a precision of 30\% or better.
Another surrogate approach
is through the \iso{59}Fe(d,p)\iso{60}Fe reaction with inverse kinematics.
In a recent benchmark
experiment \citep{dp-prl2019}, such a technique has achieved about 25\%
precision for the \iso{95}Mo(n,$\gamma$)\iso{96}Mo rate.

The \iso{60}Fe destruction reactions include \iso{60}Fe(n,$\gamma$)\iso{61}Fe,
\iso{60}Fe($p$,$n$)\iso{60}Co. The \iso{60}Fe(n,$\gamma$) \iso{61}Fe rate has
been measured directly with an uncertainty of $\sim$30\%  \citep{fe60ng}.
The \iso{60}Fe($p$,$n$)\iso{60}Co reaction becomes dominant for destruction in explosive burning.
This rate is only estimated from statistical models, it needs to be tested within nucleosynthesis models, and possibly followed up by experimental efforts.

We note that not only the reactions relating to \iso{59,60}Fe but also the neutron source reactions, discussed in Section~\ref{sec:al26_nuclear_uncertainties},
\iso{22}Ne($\alpha$,n)\iso{25}Mg, \iso{12}C(\iso{12}C,n)\iso{23}Mg, are
crucial to the overall \iso{60}Fe yields.


\subsection{Stellar Nucleosynthesis Environments}
\label{sec:fe60_cosmic_environment}





\subsubsection{Low- and Intermediate-Mass Stars}\label{fe60agb}

In AGB stars the high neutron flux required for $^{60}$Fe production can occur within the thermal pulse convective zone, due to the activation of the $^{22}$Ne($\alpha$,n)$^{25}$Mg neutron source.
This reaction requires temperatures in excess of 300 MK and occurs in stars of masses $\geq$ 2-3\msun \citep{Karakas:2014}.
For efficient $^{60}$Fe production the neutron density must exceed $\approx$ 10$^{11}$ n/cm$^3$ \citep{lugaro12a}. The $^{60}$Fe must then be mixed from the intershell to the surface via third dredge-up, and hence the yield relies on both the amount and efficiency of third dredge up and also the mass loss rate which dictate the number of third dredge up events that may occur. As mentioned in Section~\ref{agb1} the mass-loss rate of AGB stars is uncertain, in particular for more metal-poor and massive AGB stars \citep{Hofner:2018}.
In addition, the efficiency (and even the occurrence at all at higher initial masses) of third dredge up is dependent on uncertain physics, in particular the treatment of convective boundaries \citep{frost96}. Due to this, there is no consensus on whether AGB stars of masses $\geq$ 5 \msun\ undergo efficient third dredge up (e.g.\citealt{karakas10a,ritter2018}), inefficient third dredge up \citep{cristallo2015}, or even no third dredge up at all \citep{ventura2011,siess2010}. This constitutes the main uncertainty for $^{60}$Fe production within AGB stars.
Reaction rate uncertainties associated with the neutron producing reaction $^{22}$Ne($\alpha$,n)$^{25}$Mg \citep{longland2012}, and the neutron capture reactions $^{58}$Fe(n,$\gamma$)$^{59}$Fe, and $^{59}$Fe(n,$\gamma$)$^{60}$Fe discussed in  Section~\ref{sec:fe60nuclear} can also impact AGB star $^{60}$Fe yields.

\begin{figure}
	\includegraphics[width=\columnwidth]{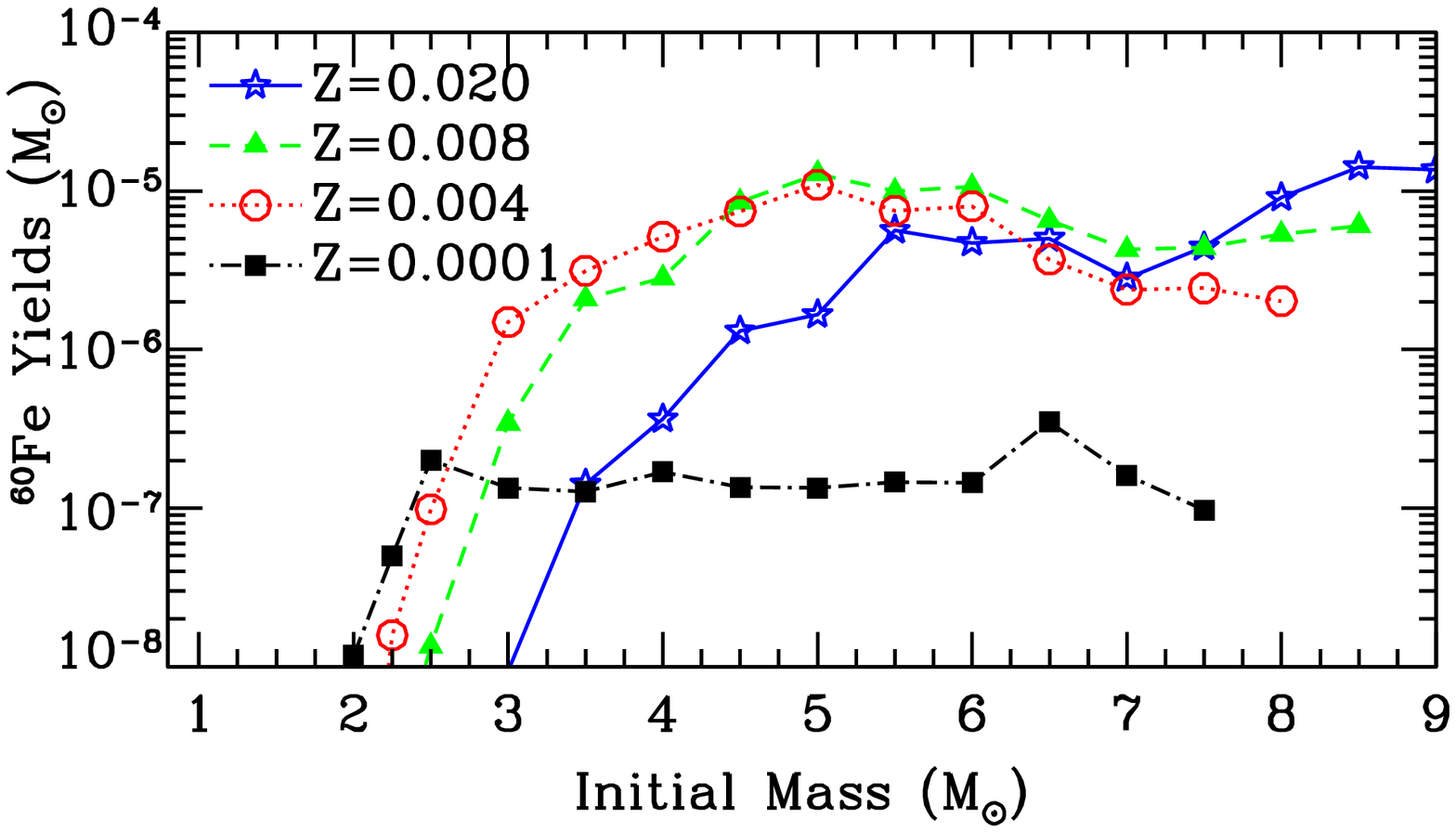}
    \caption{Stellar yields of $^{60}$Fe for the range of metallicities (Z$=$0.02$-$0.0001) as a function of initial mass. Results taken from \cite{karakas10a} and \cite{doherty2014a,doherty2014b}}
    \label{fig:yieldagb2}
\end{figure}

Figure ~\ref{fig:yieldagb2} shows the $^{60}$Fe yields for a range of metallicites (Z=0.02-0.0001) as a function of initial mass from the Monash set of models  \citep{karakas10a,doherty2014a,doherty2014b}. Although individual intermediate-mass AGB stars can make a sizeable amount of $^{60}$Fe ($\approx$ 10$^{-5}$-10$^{-6}$\msun), low and intermediate mass AGB stars make only a very small contribution to the overall galactic inventory of this isotope \citep{lugaro2008,Lugaro:2018}, as with the case of $^{26}$Al.
There are two main effects that make the $^{60}$Fe dependent not only on the mass, as discussed above, but also on the metallicity. The mass at which $^{60}$Fe production becomes significant decreases with decreasing metallicity. This is because there is a metallicity dependence for the activation of the $^{22}$Ne source. Stars of lower metallicities attain higher thermal pulse temperatures, and hence more efficient $^{22}$Ne($\alpha$,n)$^{25}$Mg activation, than their more metal-rich counterparts for the same initial mass. The other effect is due to the fact that
in AGB stars $^{22}$Ne can be of both primary and secondary origin. This nucleus derives from conversion of $^{14}$N in the intershell via the reaction chain  $^{14}$N($\alpha$,$\gamma$)$^{18}$F($\beta^+$,$\nu$)$^{18}$O($\alpha$,$\gamma$)$^{22}$Ne.
The secondary component derives from the initial CNO abundance in the star converted into N via H burning via the CNO cyle
within the H shell or at the base of the convective envelope during hot-bottom burning. The primary component derives from the extra $^{12}$C produced within the thermal pulse and contributing to the CNO abundance in the star. Therefore, due to the primary nature of the $^{22}$Ne neutron source, the maximum amount of $^{60}$Fe produced within the intershell is limited primarily by the amount of available $^{56,58}$Fe seeds, which depends on the initial stellar metallicity.
This can be seen comparing the yields of Z=0.0001 models with those of Z=0.008 over the intermediate-mass regime $\approx$ 4-6 \msun.
In both cases, there is a flat trend for the $^{60}$Fe yield over this mass range due to the relatively constant amount of third dredge up material mixed to the stellar surface.
Interestingly, within this mass range the total amount of third dredge up enrichment is also similar for these two metallicities, and hence the offset of $\approx$ 50-100 in the $^{60}$Fe yields is related primarily to initial metallicity, varying by a factor of $\approx$~80.

Another possible source of $^{60}$Fe from intermediate-mass stars is via the explosion of super-AGB stars as electron capture supernovae. These supernovae can produce up to $\sim$10$^{-4}$~\msun\  of $^{60}$Fe per event \citep{wanajo2013}, however at solar metallicity these objects are expected to be quite rare, making up only $\sim$5$\%$ of all core collapse supernovae (e.g., \citealt{poelarends2008,doherty2015}).

\subsubsection{Massive Stars and their core-collapse supernovae}
\label{sec:massiveStars60Fe}

\begin{figure}
    \centering
    \includegraphics[width=\linewidth]{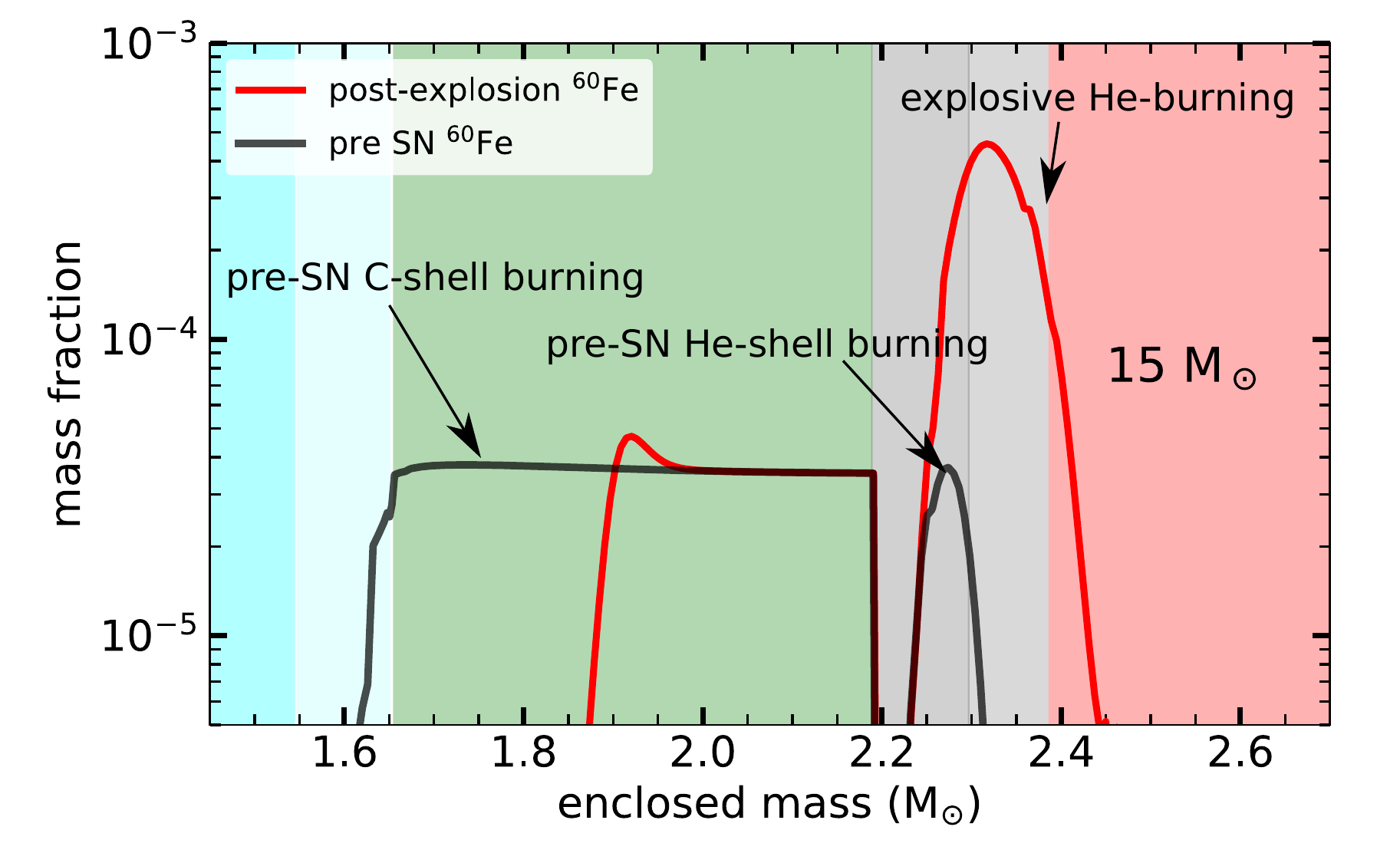}
    \caption{Mass fraction profiles of $^{60}$Fe for a $15 M_\odot$ supernova model, indicating different production regions.}
    \label{fig:fe60_production_regions}
\end{figure}

In massive stars, the neutrons required to to produce \iso{60}Fe are provided mainly by the $^{22}$Ne($\alpha$,n)$^{25}$Mg neutron source
activated during convective C and He shell burning phases as well as by the supernova explosion shock \citep{Limongi:2006,Jones2019}. Figure \ref{fig:fe60_production_regions} indicates these regions for a $15 M_\odot$ model \citep{Sieverding:2017}. The balance between these contributions is sensitive to stellar mass and the assumptions made in the stellar evolution models, in particular to convection. The bulk of pre-supernova production occurs relatively late, around 100 years before collapse \citep{Jones2019}, and relatively deep, at most at the bottom of the He-shell. This makes is unlikely for any of the material enriched in \iso{60}Fe to be ejected prior to the explosion and represent the main difference between production of \iso{26}Al and \iso{60}Fe in massive stars: while \iso{26}Al ejection occurs both in the winds and in the final supernova explosion, \iso{60}Fe can only be expelled by the supernova explosion.

The inner part of the O/C shell that is enriched in \iso{60}Fe from C-shell burning is heated to temperatures above $3$~GK by the supernova shock.
The strength of the supernova explosion determines how much of the inner \iso{60}Fe produced in C shell burning survives.
This region coincides with the
explosive contribution to \iso{26}Al discussed in Section~\ref{sec:al26_from_CCSN}.
Furthermore, the $^{22}$Ne neutron source is activated again by the explosion in the C- and lower He shell. Since \iso{22}Ne is more abundant in the He shell, the explosive production is more efficient in this region and dominates the total \iso{60}Fe yield for some models.

\begin{figure*}
\centering
\includegraphics[width=0.49\textwidth]{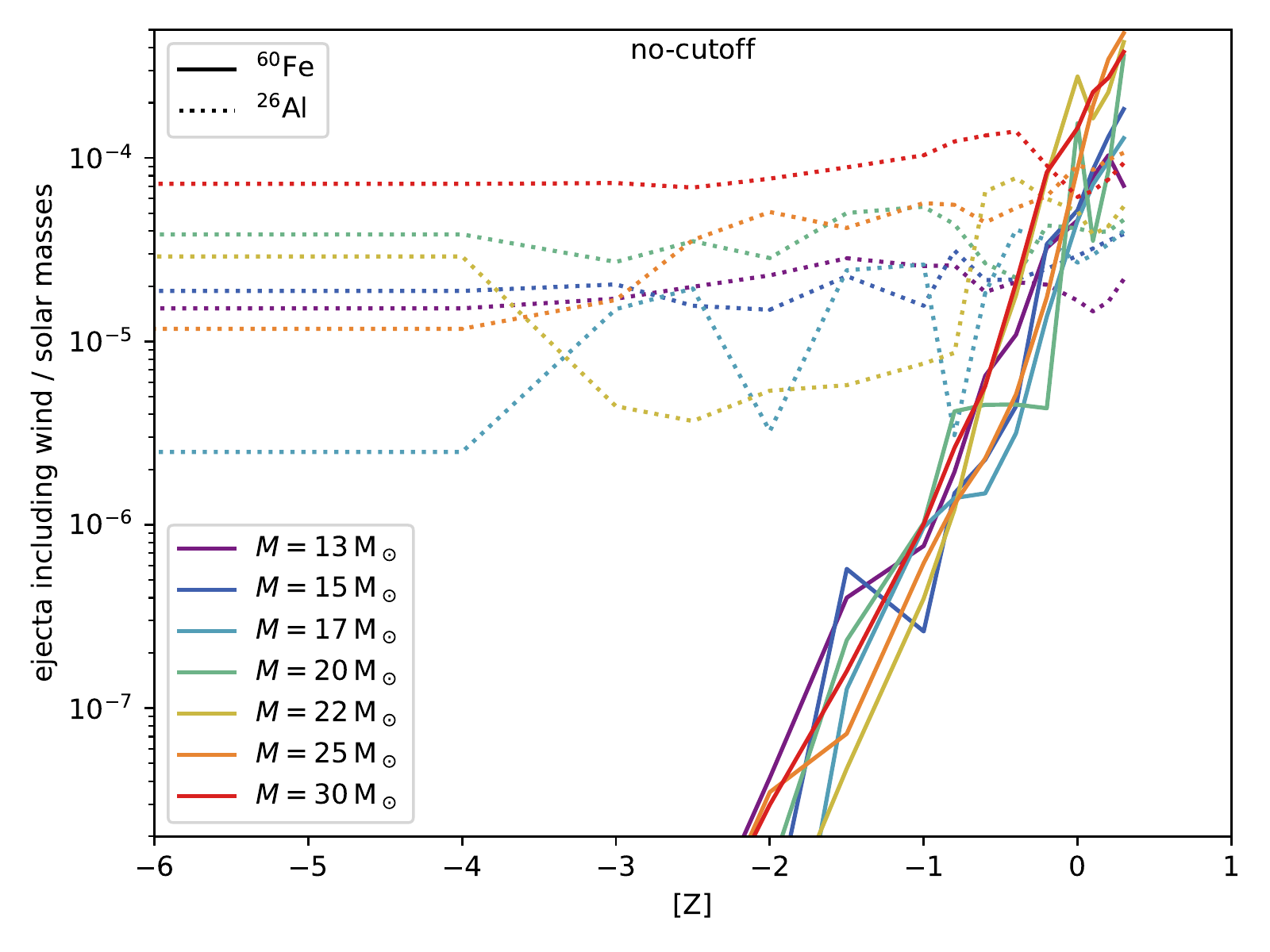}
\includegraphics[width=0.49\textwidth]{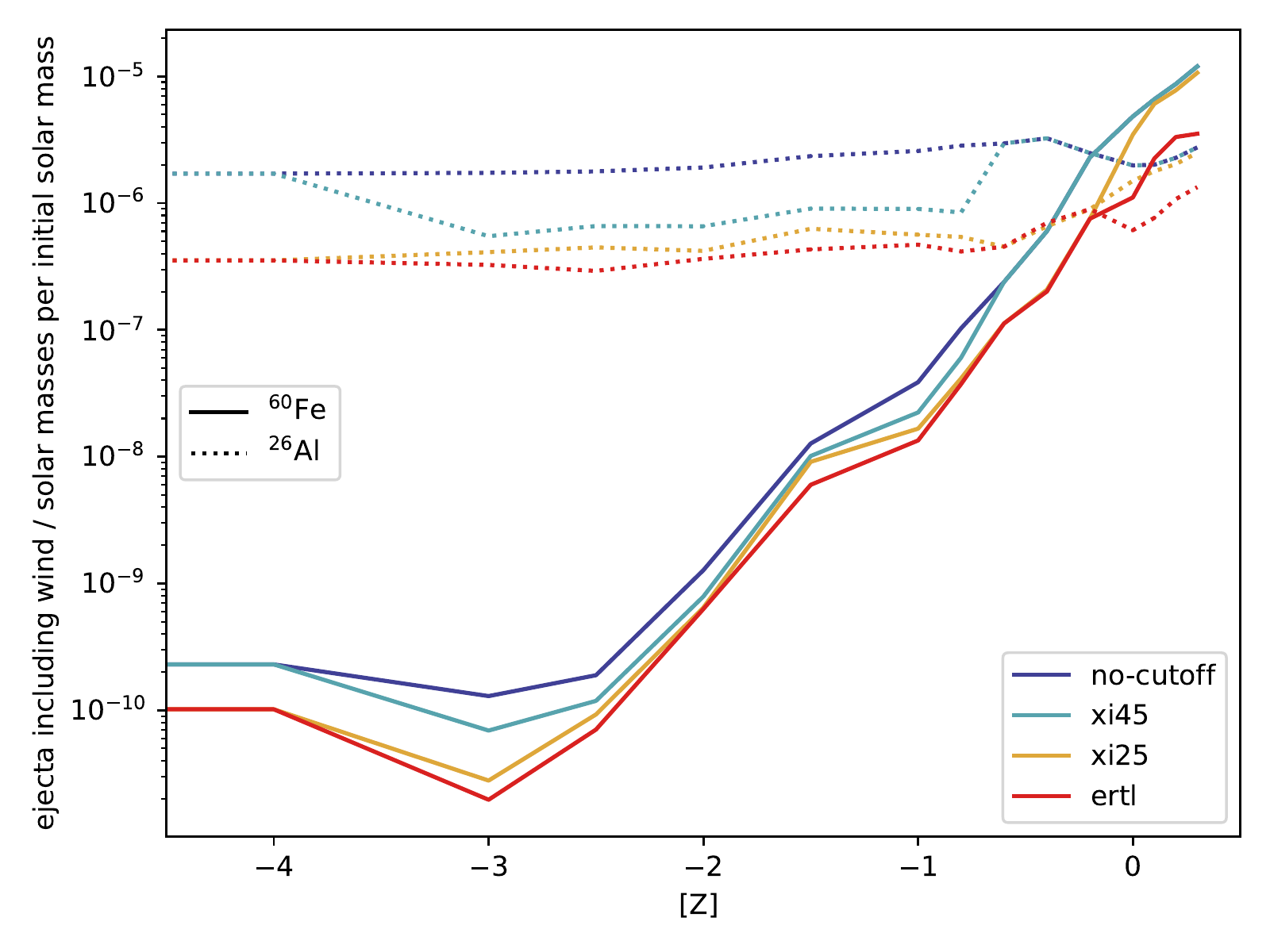}
\caption{\textsl{Left Panel:} Total yield of $^{60}\mathrm{Fe}$ and $^{26\!}\mathrm{Al}$ in units of solar masses per star
\textsl{Right Panel:} Same as the left panel but with the yields integrated over the Salpeter initial mass function (IMF) with an exponent of -1.35,  assuming a mass range for massive stars from $10\,$\ms\ to $120\,$\ms, and extrapolating the lowest and highest masses as proxies for the mass range below and above, respectively. This panel shows the results also for models that assume that stars with the compactness parameter \citep[$\xi_{2.5}$,][]{OConnor:2011} greater than 0.45 and 0.25 collapse directly to black holes and have no supernova ejecta (labels ``\texttt{xi45}'' and ``\texttt{xi25}'', respectively), and models that assume instead the explosion criterion from \citet{Ertl:2016} (label ``\texttt{ertl}'').  This order of assumed cases starts from no direct black-hole formations (``\texttt{no-cutoff}'') to increasingly larger fractions of massive stars collapsing directly into black holes, rather than exploding. The overall yields therefore decrease in that order.  Data from \citet{West:2013} and \citet{Heger:2010}.
\label{fig:alex}}
\end{figure*}

Figure~\ref{fig:alex} illustrates the production of \iso{26}Al and  \iso{60}Fe in massive stars of different masses and metallicities for a set of theoretical models of stars of seven initial masses between 13 and 30 \msun, and metallicities between Z=0 and  twice solar (Z=0.03) \citep[from the set of models used in][]{cote16}
Plotted yields include both wind and explosive contributions (with a fixed explosion energy of $1.2\,$B), except in the cases where the star is assumed to collapse directly into a black hole, in which case the explosive contribution is set to zero (as described in the figure caption).
Figure~\ref{fig:alex} demonstrates that the production of \Fe\ decreases strongly as the metallicity decreases, illustrating the secondary
nature of \Fe, which depends on the presence of neutron sources such as \iso{22}Ne, via the \iso{22}Ne($\alpha$,n)\iso{25}Mg reaction, dependent on metallicity, and of the \iso{56,58}Fe seeds. Compared to \Fe, the \Al\ yield is much less affected by metallicity. Discontinuities in the yield for some masses and metallicities are caused by stars collapsing directly to a black hole, and, in particular for \Fe, may also be due to the complex evolution of the pre-supernova shell structure that can lead to strong variations of yield from the late stellar evolution stages.

As in the case of $^{26}$Al (see Section~\ref{sec:al26_from_CCSN}), rotation within stars may change the structure of burning shells and their evolution.
Enhanced mixing from stellar rotation might strongly affect \fe\ production \citep{Brinkman:2021}.
However, most of the \iso{60}Fe is ejected during the explosion, so that the final wind yields will remain negligible compared to the supernova yields.

Also binarity and its incurred stronger mass loss probably does not influence the \Fe\ yields in a significant way. This would again be due to the fact that \Fe\ is only produced in the innermost regions of the star, too deep inside the star to be included in the mass transfer and later wind ejecta.

\subsubsection{Other explosive events}


For thermonuclear supernovae (type Ia), as we discussed in Section~\ref{sec:otheral26} nuclear burning reaches conditions of nuclear statistical equilibrium, so that the nucleosynthesis products can be characterised by nuclear binding properties, rather than tracing individual reaction parts for the fuel that may be available at the time.
For symmetric matter with similar abundance of neutrons and protons, the likely outcome is $^{56}$Ni, and in general Fe-group nuclei.
During the explosion, a more neutron-rich composition may evolve, and more neutron-rich Fe-group isotope abundances may also be typical. This has been discussed for sub-types of supernovae type Ia that appear sub-luminous because of lacking $^{56}$Ni, as the more neutron-rich isotope, and stable $^{58}$Ni is produced
(see Figure~\ref{fig:isotopes_Fe-region}). Therefore, detecting atomic lines of Ni in thermonuclear supernovae demonstrates such a neutron-rich nucleosynthesis to occur.
For Fe, the situation is similar: the symmetric isotope $^{52}$Fe also is on the proton-rich side of the valley of stability (see Figure~\ref{fig:isotopes_Fe-region}), and the most neutron-rich stable Fe isotope is $^{58}$Fe.
Thus, $\beta$-unstable $^{59}$Fe separates \fe from the stable nuclei, and more efficient neutron-capture reactions are required to overcome this.
In thermonuclear supernovae, this requires a special type of explosion scenario, which has been proposed and discussed some time ago \citep{Woosley:1997}.
In such environments, \fe yields as large as $7~\times~10^{-3}$~\msun\ could be produced, which would be roughly 100 times more than a typical core-collapse supernova.
In this case \fe\ $\gamma$-ray emission would not be diffuse, but rather point-like, for an event within our Galaxy. Future $\gamma$-ray observations hopefully  will be able to resolve emission morphologies well enough to discriminate between these two possibilities (see Section~\ref{sec:observations60Fe}).




During the $r$-process in neutron star mergers (Section~\ref{sec:otheral26}), very rapid successive neutron captures quickly process material towards very massive nuclei \citep{Thielemann:2018}.
As fission instability is encountered, the processing is halted, and fission products with typical proton numbers around 40 will supply new lighter, fuel for the neutron captures.
\fe is an isotope within the typical $r$-process reaction paths, and is produced at an equilibrium abundance as part of this efficient nucleosynthesis.
However, its abundance will most likely be low compared to the $r$-process waiting-point nuclei at the magic neutron numbers and their decay products.
Although \fe has not been included in nucleosynthesis calculations for kilonovae yet, and it is unclear how much \fe could be produced, we can make a rough estimate:
If 50\% of the nuclear products are due to waiting-point nuclei, fission nuclei, and their decay products, the remaining 50\% would be populating the abundance of nuclei along the other $r$-process reaction paths, including \Fe. If roughly 50 elements fall into the $r$-process reaction regime within the table of isotopes, from fission products to fissioning nuclei, and if the neutron-capture chain within one element includes about 20 isotopes, this implies that 10$^3$ nuclei are involved.
Therefore, as an order-of magnitude estimate, $0.5~\times~10^{-3}$ of the nucleosynthesis ashes could be in the form of \Fe.
This translates into kilonova \fe yields  of between 10$^{-6}$ and 10$^{-9}$~\msun, using currently-predicted total ejecta masses of kilonovae in the range of 10$^{-2}$ to 10$^{-5}$~\msun.
This estimate is very rough and uncertain, as freeze-out of neutron-star merger nucleosynthesis is a very complex and probably event-to-event variable environment.
Nevertheless, it demonstrates that neutron-star merger events will likely not be significant sources of \fe in our Galaxy.

Finally, we note that interstellar spallation reactions, which are significant for some \Al\ production in cosmic rays, are unimportant for producing a neutron-rich isotope such as \Fe.

\subsection{Transport into and through the gas phases}
\label{sec:fe60GCE}



As discussed above, in contrast to $^{26}$Al, which is also ejected by massive star winds, the main galactic source of $^{60}$Fe are explosions of massive stars through core-collapse supernovae.
This means that in a realistic embedded star cluster, we expect \al to diffuse earlier than \fe\ into dense gas of the birth cloud,
as star forming regions with massive stars usually become exposed before the massive stars explode
\citep[e.g.][]{Krause:2020}. \citet{Vasileiadis:2013} present a sophisticated 3D simulation of \al\ and \fe\ diffusion in an embedded star cluster. But, as discussed in Section~\ref{superbubbles:text}, since the periodic boundaries of these calculations do not allow for an escape of the dense gas, their relatively high \al\ and \fe\ abundances in dense gas should be regarded as an upper limit.


Mixing of \fe into clouds other than the formation environment of the ejecting stars is considered by
\citet{Fujimoto:2018}. They show that the \fe distribution is more diffuse and spread-out compared to \alu, as expected from the longer decay time of \fe compared to \alu.

The different ejection times of \fe and \al may have very important consequences for their transport through the interstellar medium: the \feu-ejecting core collapse supernovae accelerate the dense shells of superbubbles which makes them Rayleigh-Taylor unstable
\citep[e.g.,][]{Krause:2014}. This means that one expects particularly efficient mixing for ejecta from supernovae, like \feu, with dense gas in supershells. Simulations seem to confirm this. \citet{breitschwerdt16}
show in their simulations of superbubbles a very significant enhancement of \fe in the supershell, also for inhomogeneous background media. This is far less pronounced for \al \cite[compare Figure~\ref{fig:sbevol}]{Krause:2018}. While both teams used the same numerical code (RAMSES), these findings remain tentative until direct quantitative comparisons are made.
As shown by \citet{Fujimoto:2020b}, the implementation of stellar feedback on the structure of interstellar medium is uncertain and debatable, with significant implications for \fe transport and ingestion into star-forming regions.

It is interesting in this context that the map most-similar to the Galactic distribution of the 1173~keV radioactive decay line of \fe is that of the 4.9 micron emission, rather than that of the diffuse \al gamma-ray emission.
The 4.9 micron emission traces small dust grains and starlight from mostly low-mass M-, K-, and G-type stars. This suggest that \fe might indeed mix efficiently with dense gas in supershells relatively locally, and might take part in spiral arm and disc outflows to a lesser extent than \alu.
Advances in both
simulations and $\gamma$-ray imaging may enable us in
the future to disentangle the paths of wind and
supernova ejecta from massive stars, with implications also on the relative \al\ and \fe\ abundances in the early Solar System (Section~\ref{sec:fe60ESS}).






\subsection{Measurements of cosmic \iso{60}Fe} \label{sec:observations60Fe}

The radioactive isotope $^{60}$Fe is the only one that has so far been definitely detected in three different types of environment: in the interstellar medium (ISM) through its characteristic gamma-ray lines at 1.173 and 1.332 MeV \citep[and references therein]{Wang:2020};  in Earth's crust, fossilised bacteria, and the Moon, through dust grains, transported a few Myrs ago from recent local supernovae \citep{Wallner:2016,Ludwig:2016,Fimiani:2016}; and in the composition of Galactic cosmic rays captured and analysed by satellite instruments in interplanetary space \citep{Binns2016}.
Let us compare this to the situation for  \iso{26}Al. It is interesting to note that in many cases where it is possible to find materials enriched in \iso{26}Al from stellar nucleosynthesis, such as in stardust and CAIs, it is impossible or difficult to observe also \iso{60}Fe; this is due to the mineralogy of the sample. In fact, it is not possible to measure the abundance of \iso{60}Fe in stardust grains, because they do not include any mineral phases that are enriched in Fe and depleted in Ni, therefore, it is not possible to attribute a \iso{60}Ni excess to the radiogenic decay of \iso{60}Fe.
Vice versa, in current terrestrial and lunar inventories it is possible to observe \iso{60}Fe of clearly cosmic origins, while it is difficult to determine the original abundance of \iso{26}Al produced by cosmic nucleosynthesis, because \iso{26}Al is also abundantly produced in the Earth atmosphere by cosmic rays (Section~\ref{sec:observations26Al-particles}). The observations through $\gamma$-ray detection stand out, in that they are not affected by either such issues, and thus can serve as messengers for both isotopes and their cosmic abundances at the same time.
For the early Solar System, even if \iso{60}Fe is difficult to measure in CAIs, we also have information about the relative abundances of these two isotopes. This is because \iso{60}Fe can be measured in Fe meteorites and meteoritic inclusions such as chondrules and its initial abundance at the time of CAI formation can be derived by using the age of the meteorites relative to the age of CAIs.

\subsubsection{Interstellar dust collected on Earth}
\label{sec:60Feoceancrust}

$^{60}$Fe is a perfect candidate for search of interstellar nucleosynthesis products in terrestrial archives for the following reasons:
Earth's initial abundance of the $^{60}$Fe radionuclide has decayed to extinction over the 4.6 Gyr since formation of the solar system.
It{'}s natural production on Earth is negligible, in contrast to $^{26}$Al, therefore, its presence $-$ even at the expected low levels $-$ would be a sensitive indicator of extraterrestrial particle influx over the past $\sim${3-4} half-lives, i.e. 8-10 Myr.

How can supernova-produced $^{60}$Fe travel to Earth?
Observations suggest that the major fraction of Fe in the interstellar medium will be condensed into dust particles shortly after a supernova explosion.
Incorporated into dust grains, $^{60}$Fe can then enter the solar system and can be deposited in terrestrial archives.
For comparison, $^{60}$Fe influx could also be
(i) as highly energetic cosmic ray particles (which is orders of magnitude lower);
(ii) and $^{60}$Fe is produced in small quantities within the solar system through spallation reactions
induced by cosmic rays and is continuously deposited through interplanetary material that rains down on Earth.
However, estimations suggest a flux of only ~0.06 $^{60}$Fe atoms~cm$^{-2}$~yr$^{-1}$ evenly spread over the surface of the Earth \citep{Wallner:2016}).
 \citet{Ellis:1996} and  \citet{Korschinek:1996} suggested to search for such supernova-produced
$^{60}$Fe and other radionuclides that may become deposited on Earth  before they decay (\emph{live} detection).
Candidate terrestrial reservoirs that can incorporate extraterrestrial particles
over long time periods are deep-sea sediments, ferromanganese crusts, and nodules.
These grow slowly over millions of years, with growth rates between some cm per 1000 years (deep sea sediments) and a few mm per million years (deep-sea crusts and nodules).
The accessible time resolution for deep-sea sediment cores is accordingly ~1000 times higher than for the crusts or nodules.

The expected influx of supernova-produced radionuclides can be estimated from
(model-dependent) supernova nucleosynthesis yields and adopted typical distances.
Close-by distances are less than about 150~pc, which is considered the maximum distance where direct supernova ejecta can penetrate the solar system.
Typical influx values for $^{60}$Fe are then found between 10$^4$ and 10$^8$ atoms per cm$^{2}$ per supernova,
this would correspond to extremely low concentrations of $<$10$^{-17}$ g~g$^{-1}$ in a terrestrial archive. Analogously to the case of $^{26}$Al, the only technique sensitive enough for detecting such low traces of radioisotopes is single atom-counting using accelerator mass spectrometry.

Indeed, accelerator mass spectrometry was successfully applied in pioneering work at Munich, where live $^{60}$Fe was discovered for the first time in a ferromanganese crust from the ocean floor \citep{Knie:1999}.
Since then, this method has been further developed at TU Munich and later at the Australian National University, searching for interstellar $^{60}$Fe in terrestrial archives as well as in lunar samples \citep{Knie:2004, Wallner:2016, Fimiani:2016, Ludwig:2016, Koll:2019,  Wallner:2020, Wallner:2021}.
Measurement backgrounds of $^{60}$Fe/Fe as low as $3\times10^{-17}$, equivalent to one identified background event over one day of measurement, can be handled \citep{Wallner:2015b}.

\begin{figure}
\centering
	\includegraphics[width=\linewidth]{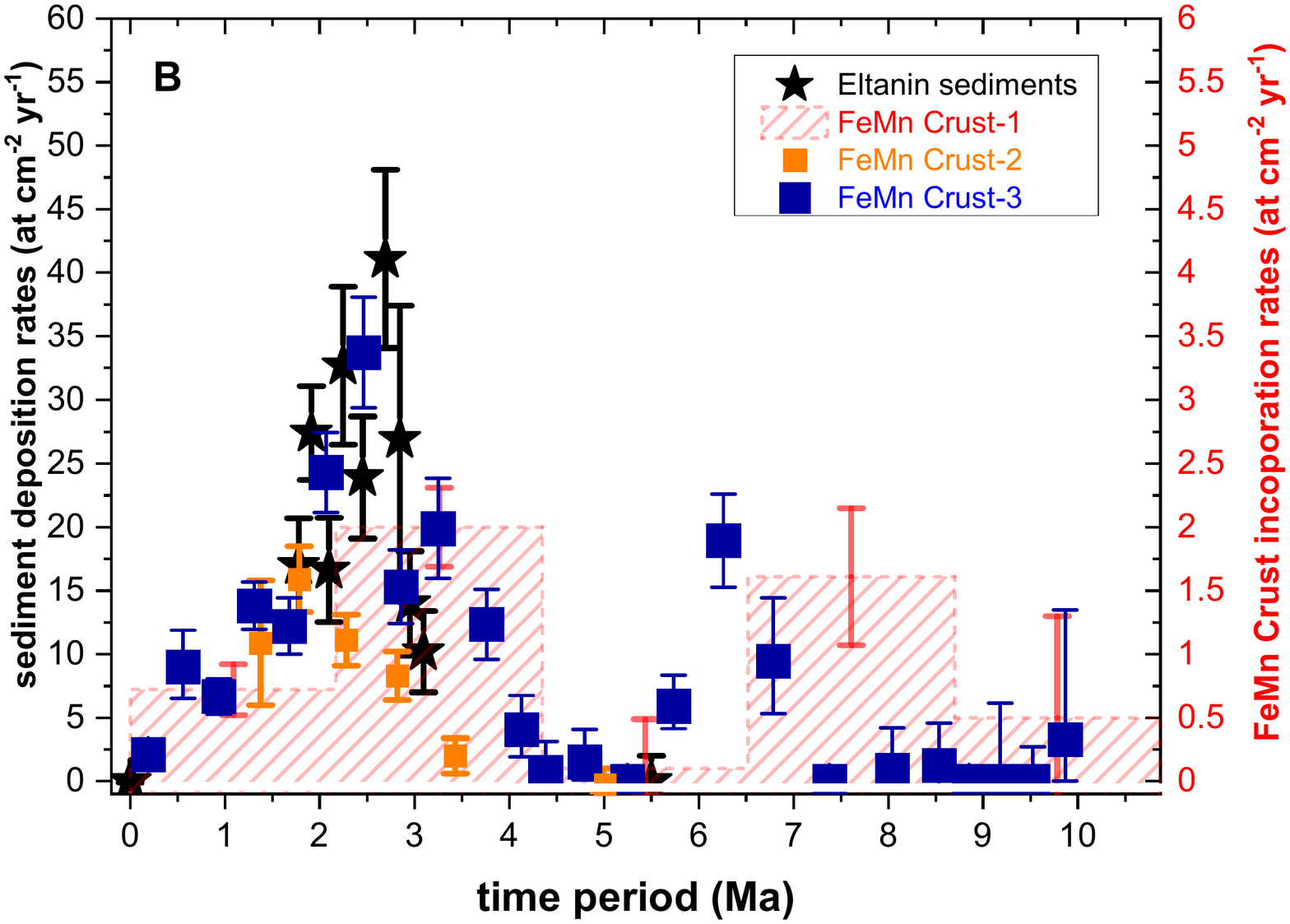}
	\caption{\fe detections in a variate set of sediments, versus sediment age \citep[updated from][]{Wallner:2016}. An enhanced exposure of Earth to cosmic \fe influx is seen around 3~Myr ago, and a second influx period is indicated earlier. }
	\label{fig:60Fe-sediments}
\end{figure}

Time-resolved depth profiles were generated for a number of archives by studying individual layers that represent specific time periods in the past.
Up to now, the $^{60}$Fe contents of three different deep-sea archives (6 sediment cores, 7 FeMn-crusts and two FeMn-nodules),
recovered from the Indian, Pacific and Atlantic Oceans respectively, were determined \citep{Knie:2004, Wallner:2016, Ludwig:2016,  Wallner:2020, Wallner:2021}. Moreover, $^{60}$Fe was found in Antarctic snow \citep{Koll:2019}, and in lunar soil \citep{Fimiani:2016}.
Figure~\ref{fig:60Fe-sediments} shows an extensive set of results from deep-sea sediments, crusts and nodules, as obtained from measurements 
\citep{Wallner:2021}.
These results demonstrate that the $^{60}$Fe signal is a global signal of extraterrestrial origin, extended in time, and from multiple events.
The two broad signals over a time of ~1.5 Myr, which are based on highly time-resolved sediment and crust data, point to a long-term influx of $^{60}$Fe, possibly caused by several close-by supernova explosions.
Alternatively, the solar system may have traversed clouds of $^{60}$Fe-enriched dust. Note that the measured flux cannot be explained by $^{60}$Fe originating from cosmic-ray spallation of Ni in (micro)meteorites \citep{Wallner:2016}.

These results demonstrate  a significantly-enhanced $^{60}$Fe deposition on the Earth: two clear $^{60}$Fe signals are observed in the sediment and crust samples.
The more-recent enhanced $^{60}$Fe influx occurred for the time-period from present to $\sim${4} Myr,
and a second enhancement is seen between 5.5 and $\sim${8} Myr  (see Figure~\ref{fig:60Fe-sediments}).
In summary, two distinctly-separated $^{60}$Fe signals are observed, with maxima between 2 and 3 Myr, and around 6 to 7 Myr.
No $^{60}$Fe above the background has been observed about 4 and 5.5 Myr and for samples older than $\sim${8}  Myr.
Between 1.7 and 3.1 Myr, the deposition rate into sediments is between $\sim${11-35} $^{60}$Fe atoms~cm$^{-2}$y$^{-1}$ (300-kyr averages in sediments).

$^{60}$Fe has also been reported in lunar material, though without time information \citep{Fimiani:2016} as well as a low present-day influx: measurements of $^{60}$Fe in antarctic snow represents influx over the last few decades \citep{Koll:2019}. Also, a set of highly time-resolved deep-sea sediments were analysed for the past $\sim${40} kyr \citep{Wallner:2020}. Both archives suggest a low but continued $^{60}$Fe deposition rate between 1 and 3.5 $^{60}$Fe atoms~cm$^{-2}$y$^{-1}$ during the past 40~kyr until present. This number, still being significantly above background, is about a factor of 10 lower than in the peak between 2 and 3 Myr.

Taking Earth's cross section into account, an $^{60}$Fe flux of $\sim${100} atoms~cm$^{-2}\,\mathrm{yr}^{-1}$ into the inner solar system is measured for the younger peak; or integrated over the $1.5,$Myr time period, this corresponds to an $^{60}$Fe fluence of $\sim{1-2}\times{10^8}$ atoms~cm$^{-2}$ at Earth orbit; the fluence for the older event is $\sim10^8$ atoms~cm$^{-2}$.

Clearly, these enhanced $^{60}$Fe data are incompatible with a constant $^{60}$Fe production or deposition and a terrestrial origin can be ruled out. $^{60}$Fe was found in all oceans and the Antarctic and the moon at similar levels, which suggests a rather uniform global distribution. An interplanetary source of, e.g., micro-meteoritic or meteoritic origin can be excluded, since the measured flux of cosmic dust is about two orders of magnitude lower than would be required (see above).

Alternatively the solar system could also have moved through clouds of $^{60}$Fe-enriched dust. The 3-4 times higher $^{60}$Fe influx for the younger time-period (see Figure~\ref{fig:60Fe-sediments})
may be explained by a less distant supernova compared to the older event;
also a different nucleosynthesis yield for $^{60}$Fe or more than one massive star explosion (compatible with the broader peak);
furthermore, a different and more efficient transport of dust to Earth is another possibility.

The measured $^{60}$Fe deposition can be linked to the interstellar $^{60}$Fe fluence\footnote{We use \emph{fluence} here because the influx may vary significantly in time, while we are interested in averages over times that are relevant for cosmic sources nearby.} and nucleosynthesis yields taking into account the survival-fraction, f$_{60}$, of $^{60}$Fe.
f$_{60}$ includes factors such as the probabilities that the $^{60}$Fe in the interstellar medium is incorporated into dust grains, but also its survival and efficiency moving into and across the solar system.
Can we deduce the distances of supernova explosions from the measured influx of $^{60}$Fe?

In a very simplified approach, we may assume that the ejected $^{60}$Fe particles are equally distributed over a surface area of the respective distance.
We then compare the measured $^{60}$Fe fluence into the inner solar system with nucleosynthesis yields.
The  distances of such events can be constrained by two arguments: Less than 20~pc is unlikely; otherwise, an intense cosmic-ray flux could have triggered a global climatic and biological disruption of which we have no indication.
We can also set an upper limit of about 150~pc, i.e. within the boundaries of the Local Bubble:
for larger distances, the supernova dust-particles would have slowed down to velocities too low to overcome the ram pressure and magnetic field of the solar system.
The minimum size and mass distribution of dust particles surviving a supernova shock may match that observed by satellite data within the solar system (see above, Section~2.4.2);
thus there would not be large filtering and deflection of supernova dust-particles by the solar system at all.
This suggests that the $^{60}$Fe flux at Earth's orbit may be not altered much from the interstellar medium.
Hence, one may assume f$_{60}$ to represent the survival fraction of $^{60}$Fe 
as ejected from the supernova.

Assuming an average $^{60}$Fe yield of $2\times10^{-5}$~\Msol (see Section 3.2), two supernovae would produce a mean concentration of $^{60}$Fe in the interstellar medium of $\sim1.5\times10^{-11}$ atoms~cm$^{-3}$,
if we assume an idealised scenario of its homogeneous distribution over a sphere of 75~pc radius.

The $^{60}$Fe influx pattern (see Figure~\ref{fig:60Fe-sediments})  suggests a supernova rate of ~0.3 core-collapse supernovae~Myr$^{-1}$ within 100~pc distance (equivalent to ~1 supernova per $^{60}$Fe lifetime within 100 pc).
This number agrees with the average rate of nearby supernovae calculated from the average core-collapse supernova rate of 1-2 per century in the galaxy (see Sec.~2.4.5), and corresponds to an average concentration $\sim4\times10^{-12}$ $^{60}$Fe atoms~cm$^{-3}$.
Models suggest a production of ($0.75\pm0.40$)~\Msol $^{60}$Fe Myr$^{-1}$, in agreement with observations of $^{60}$Fe-decay in the interstellar medium (\citep{Diehl:2013i,Wang:2020}).
This is equivalent to a total $^{60}$Fe~mass of 1--4 \Msol in the Galaxy.
But as discussed above (Sec.~3.2), supernova models vary by a large factor in their $^{60}$Fe-predictions, and, additionally, steady-state conditions may not be reached for $^{60}$Fe.
Assuming for f$_{60}$ a value of 6\% \citep{Altobelli:2005, Wallner:2015}, the measured $^{60}$Fe-fluence into the terrestrial archives of $2\times10^{8}$~atoms~cm$^{-2}$ over the last 10 Myr corresponds to an interstellar fluence of $3.3\times10^9$ atoms~cm$^{-2}$.
Assuming a simplified ballistic dynamics with a solar system velocity of 15~km~s$^{-1}$ relative to the Local Standard of Rest, about 150~pc ($4.6\times10^{20}$~cm) would be traversed by the solar system within 10 Myr.
This would lead to an averaged interstellar $^{60}$Fe concentration of $\sim7\times10^{-12}$ atoms~cm$^{-3}$.
Under these considerations, the local interstellar $^{60}$Fe concentration would not differ substantially from its average throughout the Galaxy.

Clear evidence for the deposition of extraterrestrial $^{60}$Fe on Earth has been found.
As discussed above (Sec. 2.4.1), if we assume the same origins, we can estimate from the $^{60}$Fe data also an expected $^{26}$Al influx.
However, owing to a substantially higher terrestrial production of $^{26}$Al compared to this expected influx,
the extraterrestrial $^{26}$Al signal, when distributed over the broad peaks of a width of $\sim2$ Myr,
will be hidden within the large terrestrial signal (see details above).
Accordingly, the non-detection of an interstellar $^{26}$Al signal above the terrestrial signal is in agreement with the $^{60}$Fe data \citep{Feige:2018} (for the $^{26}$Al/$^{60}$Fe ratio see Sec.~4.4 below).

\subsubsection{Cosmic rays near Earth}\label{sec:cosmicray60Fe}   

The radioactive isotope $^{60}$Fe can only be a primary (i.e, stellar nucleosynthetic) component of cosmic rays in interstellar space and near Earth, since the number of heavier nuclei is insufficient to produce it by fragmentation as a secondary product in significant amounts.
It
is the only primary cosmic-ray radioactive isotope with
atomic number Z~$\leq$~30  decaying slowly enough to potentially survive
the time interval between nucleosynthesis and detection
near Earth, with the possible exception of  $^{59}$Ni. However, while \fe\ has been experimentally confirmed, only an upper limit is available for  $^{59}$Ni  \citep{Wiedenbeck1999}.

\begin{figure}
\includegraphics[width=\columnwidth]{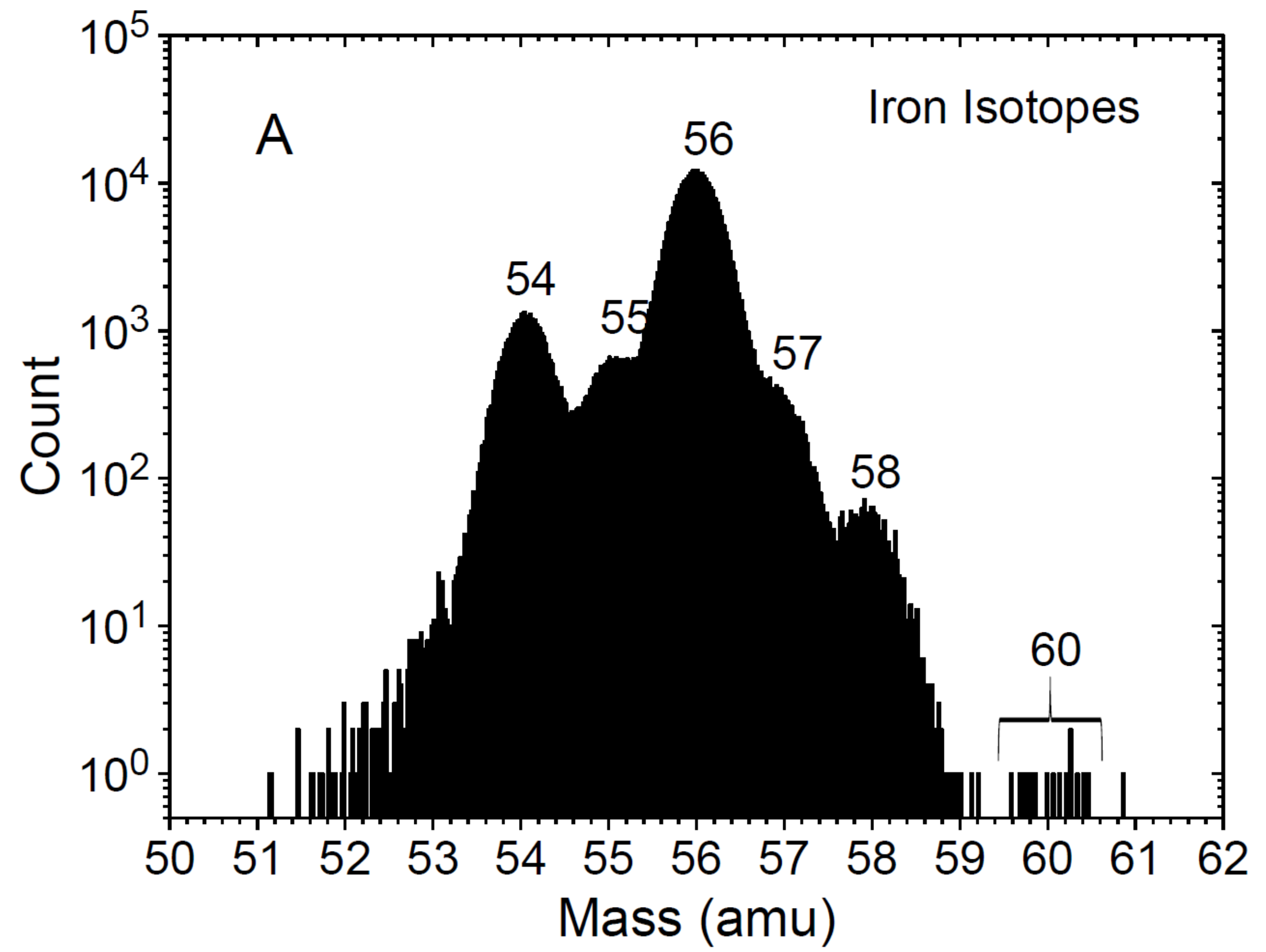}
\caption{ Mass histogram of iron nuclei detected during the first 17 years of ACE/CRIS. Clear peaks are seen for mass numbers 54, 55, 56,
and 58 amu, with a shoulder at mass of 57 amu. Centered at 60 amu are 15 events identified as
rare radioactive $^{60}$Fe nuclei. \citep[From][by permission]{Binns2016}}.
\label{fig:60FeCRIS}
\end{figure}

After 17 years of data collection by the Cosmic Ray Isotope Spectrometer (CRIS) aboard NASA's Advanced Composition
Explorer (ACE), $^{60}$Fe was detected \citep{Binns2016}.
This is thanks to the excellent mass and charge
resolution of the CRIS instrument and its capability for background rejection.
This detection came unexpected:
Explosive nucleosynthesis calculations in supernovae \citep{woosley02,Woosley:2007,Limongi2018} suggest
a small production ratio with respect to $^{56}$Fe.
With CRIS on ACE, 15 events were detected that were attributed to $^{60}$Fe nuclei, with $2.95~\times~10^5$ $^{56}$Fe nuclei (see Figure~\ref{fig:60FeCRIS}); considering background, it is estimated that
$\sim$1 of the $^{60}$Fe nuclei may result from interstellar fragmentation of heavier nuclei,
probably $^{62}$Ni or $^{64}$Ni, and also $\sim$1 might be attributed to background possibly from interactions within the CRIS instrument.
Thus, the detection of $^{60}$Fe is the first observation of a primary cosmic-ray clock. The measured ratio is  $^{60}$Fe/$^{56}$Fe = [13 $\pm$1(systematic) $\pm$ 3.9(statistical)]/2.95 10$^5$ = (4.4 $\pm$ 1.7)~$\times$~10$^{-5}$.  Correcting for interactions in the instrument and differing
energy ranges finally results in $^{60}$Fe/$^{56}$Fe = (4.6 $\pm$ 1.7)~$\times$~10$^{-5}$ at the top of the detector.

The ratio at the acceleration site can be obtained from the ratio at the top of the detector
through a model of cosmic-ray propagation in the Galaxy. In their analysis, \citet{Binns2016} adopted a simple leaky box model (see Section~\ref{sec:observations26Al-particles}) adjusted to CRIS observations of $\beta$-decay secondaries $^{10}$Be, $^{26}$Al, $^{36}$Cl, $^{54}$Mn \citep{Yanasak2001}.
All four radioactivities are fit by a simple leaky box model with the leakage parameter (called \emph{escape time}) of 15.0 $\pm$ 1.6 Myr.
Within such framework \citet{Binns2016} found that the ratio  ${\rm ^{60}Fe}/{\rm ^{56}Fe}$  at the acceleration source is (7.5 $\pm$ 2.9) $\times$ 10$^{-5}$. However, the average interstellar particle density found in that model is $n_H$ = 0.34 $\pm$ 0.04 cm$^{-3}$, i.e. $\sim$3 times smaller than that measured locally, suggesting that cosmic rays travel outside the plane of the disk, diffusing into regions of lower density. In those conditions, disk/halo diffusion models appear more physical, in particular for the case of radioactive nuclei, as suggested in \citet{Morlino2020}. Despite considerable differences between the two models - e.g. the neglecting by \citet{Binns2016} of advection and ionization losses, which are found to be relevant in \cite{Morlino2020} -  the latter model produces a quite similar number ratio at the source of (6.9$\pm$2.6) $\times$ 10$^{-5}$.

This ratio reflects the production ratio of the two isotopes at the nucleosynthesis source - i.e,  massive and exploding stars in OB associations - modulated by effects occuring before acceleration of the nuclei:  a) the radioactive decay of $^{60}$Fe in the time interval between its production and acceleration and b) the mixing (if any) of the supernova ejecta with some unknown amount of circumstellar marerial, with $^{56}$Fe and no $^{60}$Fe. Both effects reduce the interstellar ratio compared to that provided by calculations of stellar nucleosynthesis. It is reassuring therefore that such theoretical predictions provide ratios larger than that found for the cosmic-ray source, varying from several times 10$^{-4}$ \citep{Woosley:2007} to a few times 10$^{-3}$ \citep{Limongi2018}.

If effect (a) above operated alone, one could then straightforwardly estimate the time between nucleosynthesis and acceleration of $^{60}$Fe to several Myr. However, effect (b) is equally important and cannot be ignored.  It depends on the assumed scenario  for cosmic-ray acceleration, and represents one of the major unknowns in the study of Galactic cosmic-ray physics today.
Indeed, in the simplest scenario of individual supernovae as cosmic-ray accelerators,  supernova forward shock waves accelerate mostly material of the surrounding interstellar medium (with roughly solar $^{56}$Fe) while the weaker reverse shock accelerates the supernova ejecta, with both $^{56}$Fe and $^{60}$Fe. In a more realistic version of this individual accelerators, the forward shock accelerates first the wind that left the star  before the explosion (mostly piled-up in the wall of the cavity/bubble cleared up by the wind) and then, perhaps, some amount of the interstellar gas; both with solar $^{56}$Fe and no $^{60}$Fe, so the reverse shock is needed again to accelerate the radioactive nucleus within the supernova core.
In the scenario of cosmic-ray acceleration in superbubbles generated by more than one massive stars, the forward shocks of supernovae accelerate a mixture of material from previous supernova explosions and winds plus any pristine material left over in the bubble after the burst of star formation.
Overall, massive stars explode within their winds (produced either during their red supergiant phase for the less massive, or the Wolf-Rayet for the most massive) and many - but not all - of them within the superbubbles that their winds and explosions have shaped before. It is still unclear whether cosmic rays are accelerated mostly in the former or the latter site, each one having some drawbacks; see \cite{Prantzos2012a,Prantzos2012b} for a criticism of the latter, and \citet{Tatischeff2018} for a different view. The detection of $^{60}$Fe in cosmic rays does not, by itself,  help us to clarify this important issue.
In the framework of a diffusive propagation model, \citet{Binns2016} evaluated the distance from which the cosmic rays originate that arrive on Earth as: L $\sim$ ($D\, \gamma$ $\tau$)$^{1/2}$, where $\gamma$ is the Lorentz factor and $\tau$ the effective lifetime of $^{56}$Fe and $^{60}$Fe in cosmic rays (including escape, ionization losses and destruction through spallation for both nuclei, and radioactive decay for $^{60}$Fe). They found that L$<$1 kpc and noted that within this enclosed volume more than twenty OB associations and stellar sub-groups exist, including several hundred stars. Most of the $^{60}$Fe detected in cosmic rays must have originate in a large part from those stars. The closest of them, up to distances of a few tens of pc, must have given rise to the $^{60}$Fe detected in deep-sea manganese crust layers, as described in Sec.~\ref{sec:60Feoceancrust}.



\subsubsection{Early solar system material}
\label{sec:fe60ESS}

The abundance of \iso{60}Fe inferred for the early solar system is controversial because it represents an analytical challenge and because a high \iso{60}Fe/\iso{56}Fe would represent a smoking gun for a potential contribution of nearby supernovae to the early solar system.
The most recent estimates of the initial \iso{60}Fe/\iso{56}Fe ratio range from roughly $10^{-8}$ from measurements  \citep{tang12,tang15} of bulk meteorites and bulk chondrules using inductively coupled plasma mass spectrometry (ICP-MS), which does not require a local core-collapse supernova source, to $10^{-7}$-$10^{-6}$ from in-situ measurements  \citep{mishra14,telus18} of high Fe/Ni phases using secondary-ion mass spectrometry (SIMS), which would require a local core-collapse supernova source.
Note that the values reported above do not represent a possible range of variation, but two different types of measurements, with clearly-different results. Because of the possibility that the SIMS analyses are compromised by stable-isotope anomalies in Ni and/or unrecognised mass fractionation effects, a value in the lower range of $(1.01 \pm 0.27)~\times~10^{-8}$ is currently recommended.

To try to resolve the issue, \citet{trappitsch18} reported new measurements of the Ni composition of a particular chondrule (DAP1) using resonance ionization mass spectrometry (RIMS).
With this method it is possible to avoid mass interference from isotopes of the same mass but a different element because the lasers can be tuned such as only the Ni isotopes are ionised and therefore selected from the material to be analysed.
In this way it was possible to measure with high precision all the four Ni isotopic ratios (since Ni has five isotopes at masses 58, 60, 61, 62, and 64, and ratios relative to the most abundant one at mass 58 can be determined), instead of only the three isotopes with masses 60, 61, and 62, as is possible by SIMS.
This allowed \citet{trappitsch18} to use a precise \iso{62}Ni/\iso{58}Ni ratio to perform the corrections for mass-depended fractionation effects always required in these type of measurements.
In fact, SIMS measurements require normalizing to the relatively less abundant \iso{61}Ni, which inflates uncertainties and introduces correlations.
With RIMS, instead, this problem is significantly mitigated, because \iso{58}Ni can be measured and used for normalization.
Thus an initial \iso{60}Fe/\iso{56}Fe=$(6.4 \pm 11.9)~\times~10^{-8}$ was derived.
While the uncertainty (given as 2$\sigma$ here) is relatively large, the RIMS data reveal no statistically significant excesses in \iso{60}Ni beyond uncertainties, and therefore  no statistically significant \iso{60}Ni excess to be attributed to \iso{60}Fe decay.
According to this latest analysis, supernova injection of \iso{60}Fe into the early Solar System is therefore not required.
More studies of early solar system meteoritic samples are needed to resolve the long-standing debate on the abundance of \iso{60}Fe in the early Solar System.


.

\subsubsection{Gamma rays from interstellar radioactive decays}  
After the $^{26}$Al success of $\gamma$-ray astronomy (Section~\ref{sec:observations26Al-gammas}) and the theoretical predictions of co-production of $^{60}$Fe in the same massive-star source populations \citep[e.g.,][]{Timmes:1995}, it was a surprise that live \Fe\ was detected first in an ocean crust sample of the Pacific \citep{Knie:2004}, rather than via $\gamma$-ray line astronomy.
This result and subsequent detections of \Fe\ enriched material in various terrestrial as well as lunar samples discussed above, supported evidence for very nearby massive-star activity.

\begin{figure}  
\centering
\includegraphics[width=\columnwidth]{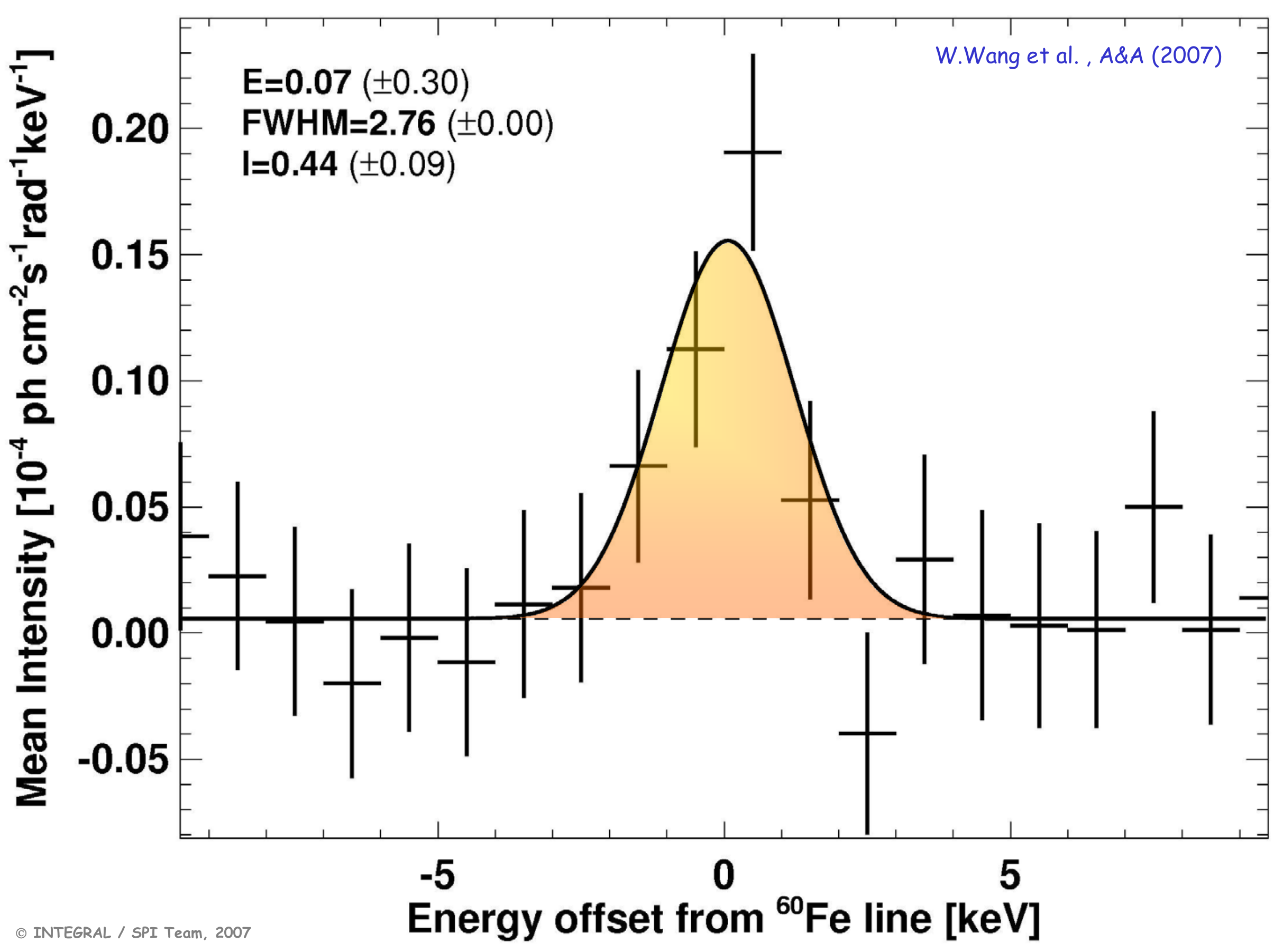}
\caption{The \Fe\ lines as seen with INTEGRAL high-resolution spectrometer SPI and 3 years of measurements integrated \citep{Wang:2007}. Here, both lines are superimposed using their laboratory energy values. (Legend: E gives the deviation of line centroids from laboratory energies in keV, with its uncertainty, FWHM is the line width at half maximum intensity in keV, and I the measured intensity in units of 10$^{-5}$ph~cm$^{-2}$s$^{-1}$rad$^{-1}$.)}
\label{fig:60Fespec}
\end{figure}

Characteristic $\gamma$ rays from interstellar \Fe\   were first reported from a marginal (2.6$\sigma$) $\gamma$-ray signal \citep{Smith:2004} with the NaI spectrometer aboard the Ramaty High Energy Spectroscopic Imager (RHESSI), a space mission aimed at solar science.
The first solid detection of Galactic diffuse \Fe\  emission was obtained from INTEGRAL/SPI measurements \citep{Wang:2007}, detecting \Fe\  $\gamma$-rays with a significance of $4.9\sigma$ after combining the signal from both lines at 1173 and 1332 keV (Figure~\ref{fig:60Fespec}).

With 15 years of INTEGRAL data, a re-analysis of diffuse \Fe\  emission was undertaken recently \citep{Wang:2020}.
The \Fe\  signal itself did not become much stronger with the additional exposure, because radioactivity activation from cosmic-ray bombardment of the spacecraft had built up radioactive $^{60}$Co, which is an emitter of the identical signal to cosmic
\Fe.
However, improved knowledge of systematic effects, as well as a much more precise model for the instrumental background built from the 15-year data set, resulted in better constraints for galactic \Fe\  line emission.
\citet{Wang:2020} obtained a detection significance for the  \Fe\  lines of $\sim 5\sigma$ with a combined line flux of $(0.31\pm0.06)~\times~10^{-3}$~ph~cm$^{-2}$~s$^{-1}$.


We can use the average flux in these lines to measure the current number of \fe decays in the Galaxy.
If we assume that the spatial distribution of sources of \fe is identical to that of \Al, as we discuss below, we may use the spatial analysis of \al emission to allow us to convert the $\gamma$-ray flux into a Galactic mass.
Using a $\gamma$-ray line intensity ratio of 0.15 (for details and constraints of this value see below), we obtain a current steady-state mass of \fe in the Galaxy of 2.85~\Msol.

Again, as discussed above for \Al\ (see Figure~\ref{fig:26Al27Alratios}), we can consider the diffuse $\gamma$-ray emission seen in the two lines at 1173 and 1332~keV as a measurement of the \emph{current average} abundance of \fe in the interstellar medium of the Galaxy.
Then, it is interesting to estimate how the isotopic ratio of the unstable to stable isotope, $^{60}$Fe/$^{56}$Fe, compares between the current Galaxy and the environment when the Sun formed, the latter measured in meteorite inclusions (Section~\ref{sec:fe60ESS}), and extrapolated from the $\gamma$-ray data.
We use here the measured intensity ratio of the $\gamma$-ray lines of radioactive decays of \fe and \al (averaged between the two lines for \Fe), as discussed below in Section~\ref{sec:60Fe26Alratio}; here we use a value of 0.15.
We again use a Galactic interstellar-gas mass of 4.96~10$^9$\Msol\ \citep{Robin:2003}, and thus arrive at an isotope ratio of $3.4\times~10^{-7}$ of $^{60}$Fe/$^{56}$Fe by number. Extrapolating a linear metallicity growth for $^{56}$Fe (the stable isotope; \fe\ is assumed to not grow and remain in steady state between production and decays), we infer an estimated isotope ratio in the interstellar medium at the time of solar system formation of $\sim5.7\times10^{-7}$ of $^{60}$Fe/$^{56}$Fe by number.
The uncertainty of this value is estimated to be about 50\%, mainly from systematics of the source distances and \fe flux.
More uncertainty may be included in the extrapolation of metallicity \citep{Schoenrich:2012}.

\subsubsection{Other electromagnetic radiation}  
The decay scheme of \fe (Figure~\ref{fig:fe60_decay}) shows another characteristic line at 59~keV from de-excitation of $^{60}$Co. This line should be accessible to hard-X-ray telescopes such as NuSTAR.
However, the branching ratio of only 2\% makes this a very faint emission. Moreover, its diffuse nature is challenging for focusing instruments such as NuSTAR, as its field-of-view size measures merely 6 arcmin at this energy, thus capturing a small fraction of \fe emission.

Molecule formation near nucleosynthesis sites with inclusion of \fe is likely \citep{Tachibana:2013}, however, no measurements of \Fe-carrying molecules in sub-mm lines have been reported so far (unlike for \Al, see Section~\ref{sec:observations26Al-gammas}).

\section{The $^{60}$Fe/$^{26}$Al ratio}
\label{sec:60Fe26Alratio} 

We showed above that significant amounts of both \Al\ and \Fe\ are expected to be produced from stars of intermediate and high mass, including core-collapse supernovae, and that this expectation is in agreement with a diverse ensemble of observations.
The observed Galactic distribution of \al suggests a massive-star origin, while the one of \fe has not yet been measured. Another piece of information can be obtained from
the observed $^{60}$Fe/$^{26}$Al ratio. This holds direct information on their sources, if the same type of source co-produces these radioisotopes, though in different quantities;
source distances and individual locations cancel out.
Then, the ratio of the $\gamma$-ray flux in the $^{60}$Fe/$^{26}$Al lines
represents directly a stellar yield ratio by number
\footnote{Following the basic radioactive exponential decay equation $dN/dt=N exp(-t/\tau$, where $dN/dt$ is equivalent to the number of $\gamma$ photons emitted per unit time, both flux values would need to be multiplied by the ratio of the mean lives $\tau$, first in order to obtain the abundance ratio from the flux ratio, and second to transform the ratio into its steady-state value expected in the interstellar medium.}.
Therefore, stellar yield ratios can be compared directly to $\gamma$-ray observations, just keeping in mind that yields given in solar masses need to be translated into abundances, which for the $^{60}$Fe/$^{26}$Al ratio can simply be done by multiplying by a factor 26/60. Of course, it needs to be kept in mind that the flux ratio derives from abundances averaged over the observed region of the sky, rather than single stellar sources.


\subsection{Stellar Nucleosynthesis Environments}
\label{sec:60Fe26Alratio-sources}
We now discuss what could be learned from an observed $^{60}$Fe/$^{26}$Al ratio, first in relation to AGB stars, and then for massive stars with their supernovae
and how current and future measurements can interpreted in this framework.

\subsubsection{Low- and Intermediate-Mass stars}

While production of $^{26}$Al in AGB stars, in particular for the intermediate-mass stars, is a fairly robust prediction with only some dependence on model detail (such as the exact stellar mass above which hot-bottom burning is activated), this is not the case for $^{60}$Fe. As mentioned in Section~\ref{fe60agb}, although $^{60}$Fe is expected to be produced quite efficiently within the He-rich intershell regions of AGB stars above $\sim$3-4 \msun, the amount which reaches the surface (and hence the total yield) depends on the amount of mixing due to third dredge up. The efficiency of this process is highly model dependent, and yields can vary from practically no production to above 10$^{-5}$ \msun\ total yield  \citep[see Section~\ref{fe60agb} and][]{cristallo2015,karakas10a}.
The predicted $^{60}$Fe/$^{26}$Al ratio is therefore very different, depending on the input physics adopted.

\begin{figure}
	\includegraphics[width=\columnwidth]{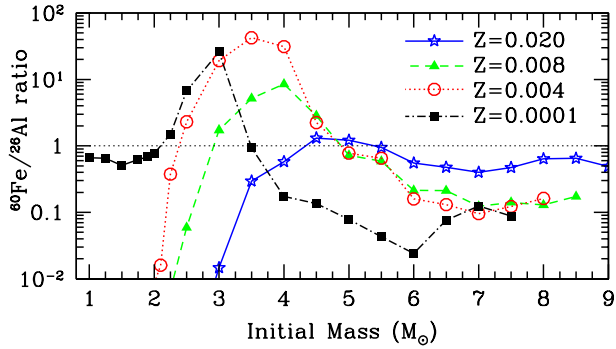}
    \caption{Yield Number ratios of $^{60}$Fe/$^{26}$Al for the range of metallicities (Z$=$0.02$-$0.0001) as a function of initial mass \citep[from][]{karakas10a,doherty2014a,doherty2014b}.}
    \label{fig:yieldagb3}
\end{figure}

For the models that predict $^{60}$Fe production shown in Figure ~\ref{fig:yieldagb3}, stars of masses lower than about 4.5 \msun  the $^{60}$Fe/$^{26}$Al ratio can be much greater or much less than unity, by orders of magnitude.
However, the absolute production of both isotopes (i.e. their yields) in this mass range is very small. For the more massive AGB stars, where the yields are more significant, the $^{60}$Fe/$^{26}$Al ratio ranges from $\sim$ 0.001 to 2, with the value decreasing with increasing mass for the intermediate metallicity cases.
This reflects the general trend of $^{26}$Al increasing  with initial mass, while $^{60}$Fe production remains relatively constant as a function of initial mass.
Unlike their massive star counterparts, which are able to enrich the interstellar medium with $^{60}$Fe and $^{26}$Al quite rapidly, these massive mass stars are longer-lived and take $\sim$ 30-200 Myr to reach the AGB phase, where the radioactive isotopes are expelled as stellar winds become strong. Stardust grains represent a strong constraint for $^{26}$Al production in AGB stars, however, $^{60}$Fe cannot be measured within these grains (Section~\ref{sec:fe60obs})
therefore we currently have no constraints for  the $^{60}$Fe/$^{26}$Al ratios from AGB stars.

\subsubsection{Massive stars and their core-collapse supernovae}

As explained above, the ratio from the $^{60}$Fe/$^{26}$Al flux can be directly compared to the yields of the massive stars that produce them in the Galaxy. Most massive star models over-predict such ratio by about a factor 3 to 10,  \citep[see, e.g.][]{Woosley:2007b,Sukhbold:2016aa,austin17}.
However, there are uncertainties in the nuclear reaction rates, and in the stellar physics of stars of high mass \citep{Woosley:2007b}, and some models have been able to match the observed flux ratio \citep{Limongi:2006}.
Moreover, as discussed in Section~\ref{sec:massiveStars60Fe}, a strong initial-mass and
metallicity-dependence of the absolute \Fe\ yields is typically present, as compared to their behaviour for \Al\ (Figure~\ref{fig:alex}). Due to this, the flux ratio of
\Fe/\Al\ could be lowered if galactic star formation relevant to the observed flux was dominated by low-metallicity.
Stars of low metallicity were relatively much more common in the early Universe than today, so this is unlikely, except if some regions of high metallicity may dominate \Fe\ production, whereas other regions may dominate \Al\ production.
A reduction of the \fe contributions from massive stars due to direct black-hole collapses for most-massive stars may also affect the \Fe/\Al\ ratio.
Considering this effect, \citet{Sukhbold:2016aa} accepted that their model ratio came out three times higher than observed.
Overall, current single-star models are still uncertain and missing many contributions to either isotopes may be present, for example, the enhancement in \iso{26}Al produced by models with proton ingestions  \citep[needed to match the stardust grains from supernovae, as discussed in Section~\ref{sec:observations26Al:stardust}][]{pignatari2015}, would also predicted  lower \iso{60}Fe abundance due to less neutrons from the \iso{22}Ne source, which would preferably capture protons than $\alpha$ particles. More investigations are needed, also comparing these 1D model results to current 3D supernova modelling.



\subsection{Population synthesis modelling}
\label{sec:popsyn}

\begin{figure}
\centering
\includegraphics[width=\columnwidth]{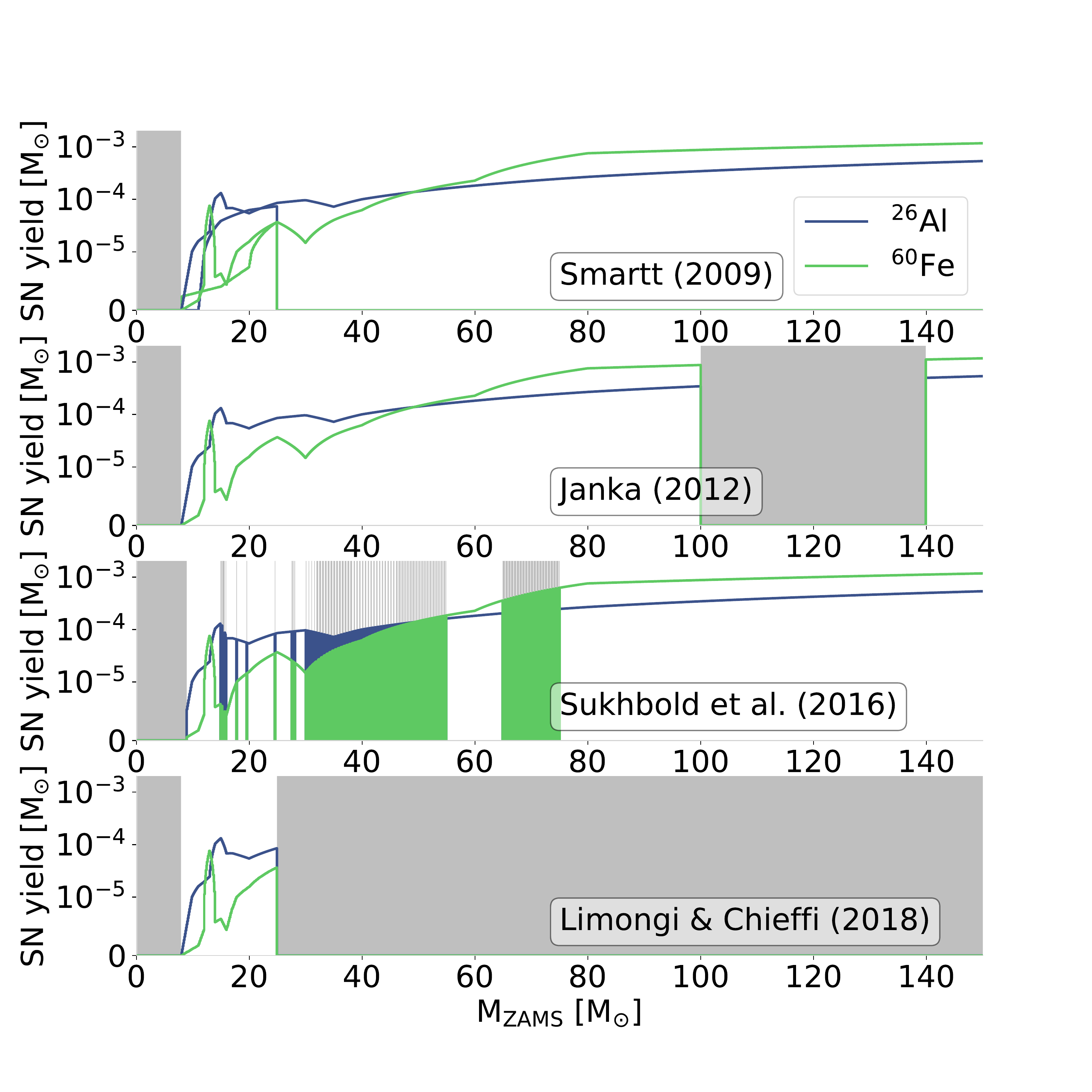}
\caption{Massive-star yields of \al (blue) and \fe (green). Model yields of \citet{Limongi:2006a} are shown for different explodability models. Grey shading indicates initial stellar masses for which no core collapse supernovae occur in a given explodability study; green areas appear as islands of explodability follow each other closely in mass.
\citep[From][]{Pleintinger:2020}}
\label{fig:mp_exp1}
\end{figure}

Another level of complexity and accuracy in the comparison between models and observations can be added by when considering massive stars as populations.
In particular, since \fe is ejected only by the explosions while \al by both explosion and winds (Sections \ref{sec:al26_from_CCSN} and \ref{sec:massiveStars60Fe}),
the \feu/\al ratio is strongly affected by the ability of single stars to actually undergo explosions, also called \emph{explodability}.
Three-dimensional hydrodynamic supernova simulations indicate that not all massive stars with $M > 8$\,\msun\ explode in a supernova but some immediately collapse into black holes without significant mass ejection  \citep{OConnor:2011}.
The explodability of massive stars shows a complex profile with rather irregular islands and gaps, especially for initial stellar masses $M \geq 20$\,\msun \citep[e.g.,][]{Ugliano12,Ertl:2016,Sukhbold:2016aa,Ebinger19}.
Explodability  and its effect in yields is illustrated in Figure~\ref{fig:alex}.
The impact of four different explodability models on \al and \fe yields are shown in Figure \ref{fig:mp_exp1}; explodability variants are applied here to stellar evolution models by \citet{Limongi:2006a}, as the stellar population across the initial-mass range is sampled. This illustrates that the \feu/\al ratio also carries an imprint of supernova explosion dynamics.

\begin{figure}
\centering
\includegraphics[width=\columnwidth]{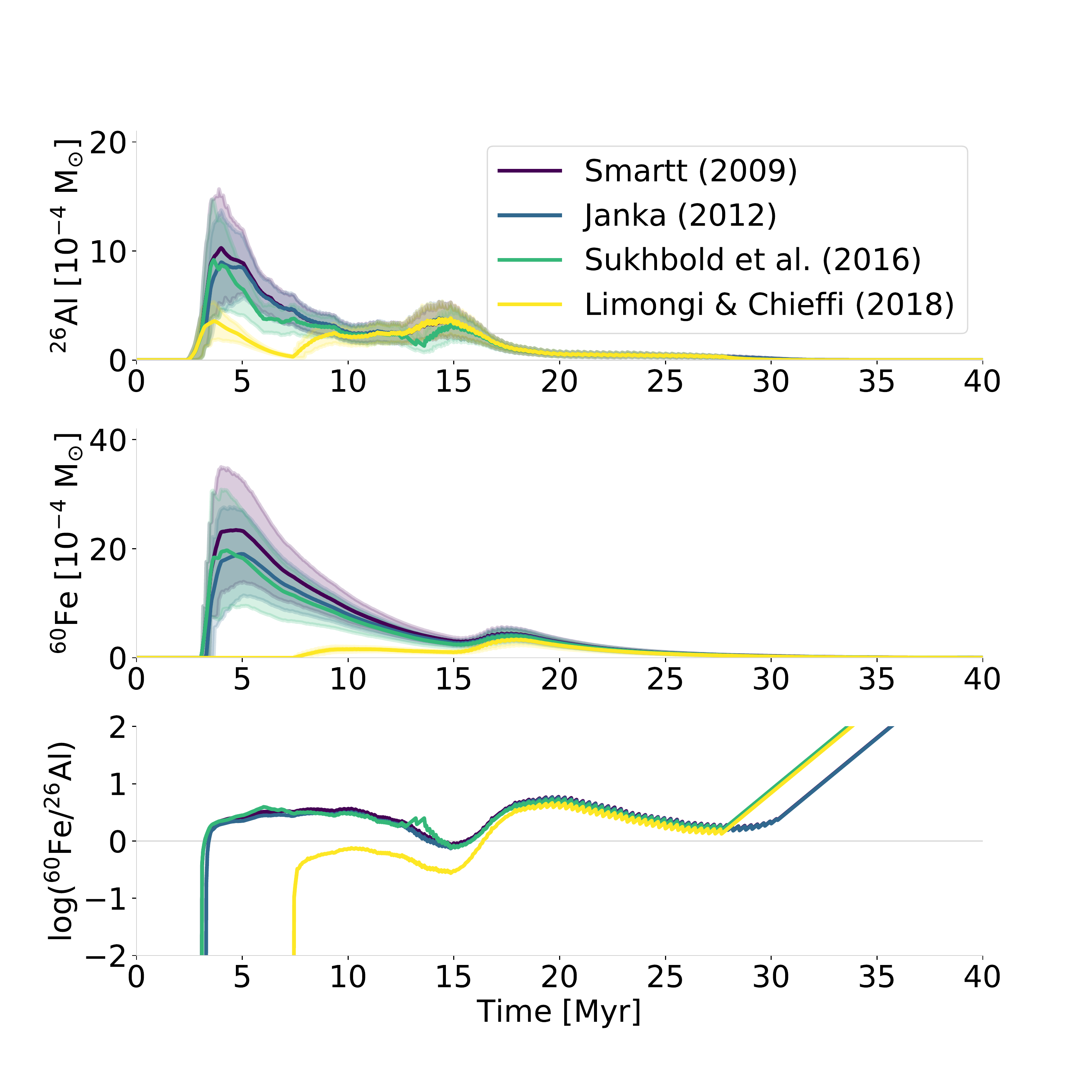}
\caption{Population synthesis of nucleosynthesis ejecta \al (upper), \fe (middle), and their average mass ratio (lower), in a source from a $10^4$\,\msun\ star group. Calculations are based on stellar yield models by \citet{Limongi:2006a} and assuming a standard IMF \citep{Kroupa:2001im}. Different colours denote different explodability model assumptions \citep{Smartt:2009cc,janka12,Sukhbold:2016aa,Limongi2018}.
\citep[From][]{Pleintinger:2020}
\label{fig:mp_exp2}}
\end{figure}

As mentioned above, over the entire Galaxy the \feu/\al ratio reaches an equilibrium state and the effects of single-star evolution parameters, e.g., rotation and explodability, on the  nucleosynthesis yields average out.
To observe the direct effects of such parameters, individual star-forming regions and massive star groups constitute the smallest scale currently accessible by $\gamma$-ray observations, as discussed in Section~\ref{sec:observations26Al-gammas}.
If we consider entire populations of stars, and account for the variations with initial mass by weighting with the abundance of stars of each initial mass, through a population-synthesis approach \citep[see][and references therein for such an approach, often used for spectral analysis of galaxies, but here including nucleosynthesis yields for the first time]{Voss:2009}, we can trace a time profile of ejected nucleosynthesis products (Figs~\ref{fig:mp_exp1} and \ref{fig:mp_exp2}, with contributions from stars of each initial-mass value arising successively, as stellar evolution tracks tell us. This assumes that all stars of such a given population group are coeval, i.e., born at the same time.
Figure~\ref{fig:mp_exp2} shows population synthesis calculations of \al and \fe ejection from a $10^4$\,\msun\ group of stars, with the different explodability assumptions shown in Figure~\ref{fig:mp_exp1}.

The explodability models by \citet{Smartt:2009cc}, \citet{janka12}, and \citet{Sukhbold:2016aa} are consistent between each other, within the statistical uncertainties that are expected for different numbers of stars, here implemented from Monte Carlo sampling.
As expected, the nucleosynthesis yields are generally smaller, if explosions occur for fewer stars.
Therefore, the most significant difference is obtained when comparison to the explodability-case of \citet{Limongi2018}, which does not allow supernova explosions and thus product ejections for stellar masses above $25$\,\msun.
In this case, the first \al abundance peak from a stellar group after several Myr is reduced to its stellar wind component, and the first \fe appearance is delayed by $\sim 6$\,Myr.
If explosions for $M > 25$\,\msun\ are omitted, the \feu/\al ratio is dominated by \al for over $16$\,Myr, otherwise, this phase only lasts for about $3$\,Myr.
This result illustrates the potential of the \feu/\al ratio as an observational target for investigations of massive-star explodability and the formation of stellar-mass black holes.

\begin{figure}
\centering
\includegraphics[width=\columnwidth]{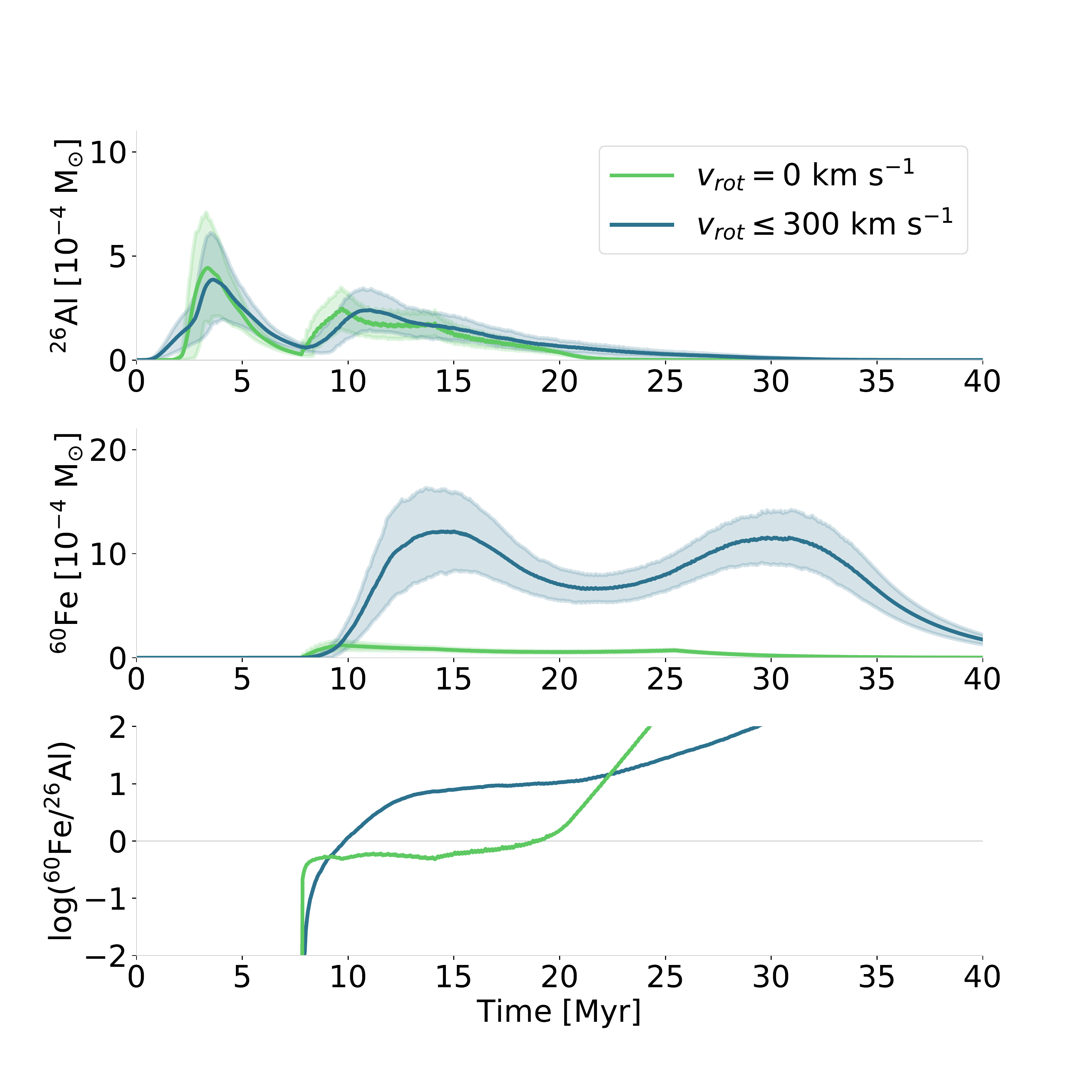}
\caption{Same as Figure \ref{fig:mp_exp2} based on stellar yield models by \citet{Limongi2018} for non-rotating stellar models (green) and models including stellar rotation $0 \leq v_{\text{rot}} \leq 300$\,\kms\ (blue). \citep[From][]{Pleintinger:2020} \label{fig:mp_vrot}}
\end{figure}


Stellar rotation has been discussed above (Sections \ref{sec:al26_from_CCSN} and \ref{sec:massiveStars60Fe}), as it enhances convective regions and the transport of raw material from nuclear reactions throughout a star, and therefore also affects the nucleosynthesis feedback in massive star groups.
Its temporal scope is generally shifted to an earlier onset of wind phases from rotating stars. Additionally, stellar evolution is overall prolonged which delays and extends later evolutionary phases and the respective nucleosynthesis feedback implications.
Different modeling approaches for rotating massive stars show a consistent enhancement of light elements from C to Al \citep{choplin2020} and increased s-process contributions \citep{prantzos2019, banerjee2019}. However, in the mass range of AGB stars, which is particularly relevant for \al, this effect of rotation appears to be negligible \citep{denHartogh:2019}. Since the outer H and He layers of massive stars are mostly convective, irrespective of rotation, the overall \al yield of a stellar group is not strongly affected by this parameter compared to initial mixing, for example \citep{bouret2021}.
However, the implications of stellar rotation for \fe in the deeper layers are more striking. Due to an increase of neutron sources as result of an enlarged C-burning shell, \fe ejection in stellar groups can be enhanced by a factor of $10$ because of the contribution from fast-spinning stars. This emphasizes the \feu/\al ratio as important tracer of the connection between stellar rotation and nucleosynthesis feedback.
To illustrate the relevance of this ongoing research, a population synthesis example including the effects of stellar rotation is shown in Figure \ref{fig:mp_vrot}. The model refers to a $10^4$\,\msun\ massive-star group and is based on evolutionary tracks by \citet{Limongi2018}. Particularly within the large uncertainties in the \fe reaction networks, these are generally consistent with other recent models \citep[e.g.][]{prantzos2019, banerjee2019, choplin2020} Rotational velocities $0 \leq v_{\text{rot}} \leq 300$\,\kms\ are included following the observed distributions from O- and B-type stars \citep{Glebocki:2005vr}, and assigned according to the respective individual spectral classifications in the stellar-evolution library that underlies the population synthesis.
The  effects from stellar rotation are clearly visible in the time evolution of the \feu/\al ratio within the group:
While the dominance of the ratio by \al lasts for $\sim 18$\,Myr with only non-rotating stars, this phase is shortened to only $\sim 10$\,Myr if rotation is taken into account.
We note, however, that stellar rotation and its impacts can be implemented in models in different ways, none of which can be shown to be \emph{realistic}.

When zooming into more specific regions of the Galaxy, and considering more or less incomplete mixing, the $^{60}$Fe/$^{26}$Al is predicted to be more variable than its steady-state value.
As already pointed out in Section~\ref{superbubbles:text}, models for the varieties and evolution of the $^{60}$Fe/$^{26}$Al ratio have been calculated within the general galactic interstellar medium \citep{Fujimoto:2018,Fujimoto:2020b} and in giant molecular clouds \citep{Vasileiadis:2013,Kuffmeier:2016}.
The main result is that a range of $^{60}$Fe/$^{26}$Al ratios may occur, and the concept of a galactic \emph{average} may be misleading.
This is relevant to specific locations in the Galaxy, such as, for example, the time and place of the formation of the Sun, as discussed at the end of the next Section~\ref{sec:fe60obs}.


\begin{figure}
	\includegraphics[width=\columnwidth]{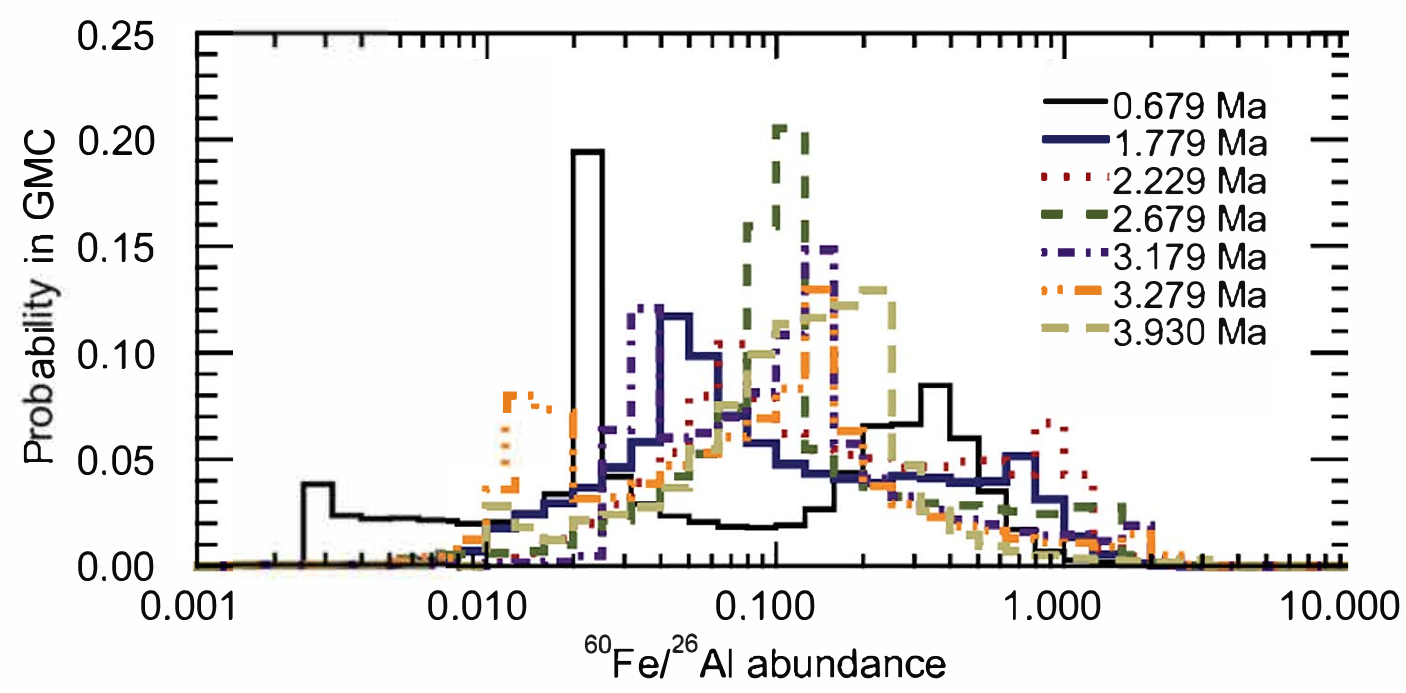}
    \caption{Simulations of $^{60}$Fe/$^{26}$Al within a giant molecular cloud \citep{Kuffmeier:2016}. Shown is  a histogram of ratio values inferred for stars forming at different times and locations in a giant molecular cloud. \citep[From][ by permission]{Kuffmeier:2016}.
    }
    \label{fig:60Fe26Alratio_GMCsim}
\end{figure}

\subsection{Interpreting observational constraints}
\label{sec:fe60obs}

\subsubsection{The star-forming interstellar medium}



$^{60}$Fe from the interstellar medium had been clearly identified in terrestrial archives (see Section~\ref{sec:observations60Fe}). In contrast, measurements of $^{26}$Al in the same deep-sea sediments did not reveal any significant interstellar medium influx, owing to a dominant cosmogenic terrestrial production background (see  Section~\ref{sec:observations26Al-particles}). However, combing both data sets nevertheless allows to deduce a lower limit for the $^{60}$Fe/$^{26}$Al isotope ratio in this terrestrial archive.

As shown in Section~\ref{sec:observations26Al-particles} and Figure~\ref{fig:Al_terrestrial},  \citet{Feige:2018} had analysed in detail the time period between 1.7 and 3.2 Myr for $^{26}$Al influx from the interstellar medium; i.e. the time period that is also characterized by the interstellar medium influx of $^{60}$Fe during its younger and broad influx (see Figure~\ref{fig:60Fe-sediments}) \cite{Wallner:2016,Wallner:2021}. They found an average lower limit for the ratio $^{60}$Fe/$^{26}$Al = 0.18$^{+0.15}_{-0.08}$, i.e. suggesting a range in the lower limit between 0.1 and 0.33. In summary, a $^{60}$Fe/$^{26}$Al ratio at 0.1 or lower would have been visible in their experimental data \citep{Feige:2018}. Owing to the shorter half-life value of $^{26}$Al of 0.7 Myr compared to the 2.6 Myr for $^{60}$Fe, there is presently no chance to find $^{26}$Al together with $^{60}$Fe in the older peak between 5.5 and 7 Myr (see Figure~\ref{fig:60Fe-sediments}).

Under the assumption that the ratio found in the terrestrial archive has not been altered during transport and deposition, it would represent the local interstellar medium conditions. If one further neglects any non-isotropy in the supernova ejecta, the experimental lower limit could then be related to the $^{60}$Fe/$^{26}$Al ratios in recent local supernova events. We note, such data do not reflect steady-state conditions of the interstellar medium and they cannot be simply compared with the galactic average $\gamma$-ray flux ratio of $\sim$ $0.2- 0.4$ \citep{Wang:2020}. The sediment data do provide nevertheless a constraint on the supernova-associated $^{60}$Fe/$^{26}$Al isotope ratio in the solar environment in the recent past. Despite these deficiencies in a direct comparison, the observations in the $\gamma$-flux and the independent measurements in the deep sea sediments agree with each other.

Besides $^{26}$Al and $^{60}$Fe, other longer-lived radionuclides are also produced in supernovae and massive stars and will be present in the interstellar medium. Similarly to the cases of $^{26}$Al and $^{60}$Fe, they can be found in terrestrial archives and their relative ratios to $^{26}$Al and $^{60}$Fe can add important information on understanding the journey of theses nuclides from production to their incorporation into terrestrial archives. In particular, accelerator mass spectrometry had been applied for search of $^{53}$Mn and $^{244}$Pu. 
$^{53}$Mn in the interstellar medium would be of pure supernova origin. Terrestrial production is low, i.e. not significant here, but this nuclide is produced within the solar system by cosmic-ray induced spallation reactions on the abundant Ni and Fe in planetary objects (cosmogenic production).
By combining the $^{53}$Mn data obtained from four deep-sea ferromanganese crusts,  \citet{Korschinek:2020} present an overabundance of $^{53}$Mn/Mn $\sim 4 \times 10^{-14}$ over that expected from cosmogenic production.
This  corresponds to a $^{53}$Mn/$^{60}$Fe ratio of about 14, in agreement with some nucleosynthesis models. We note however
that the incorporation efficiency into crusts and the transport through the interstellar medium may be substantially different for the two isotopes.
$^{244}$Pu (half-life 80.6 Myr) is of particular interest, because it is produced in the r process, for which
neutron star mergers and a subset of supernovae are prime candidates.
With its much longer half-life compared to $^{60}$Fe, $^{244}$Pu could originate also from older r-process events, not limited to those that formed the local bubble and the supernovae that were responsible for the $^{60}$Fe. 
Similarly to $^{60}$Fe, using dust particles as vehicles, $^{244}$Pu could enter the solar system, either being ejecta from one supernova or a sample of interstellar abundances.
For the ratio of these isotopes, some supernova models suggest values of $^{244}$Pu/$^{60}$Fe between 10$^{-3}$ and well below 10$^{-5}$.
But from other models, it is not clear that supernovae produce actinides at all \citep[see discussion in][]{Goriely16,Cowan:2021}.
%
%
With a major improvement in accelerator mass spectrometry detection efficiency in the past few years,  a clear signal of $^{244}$Pu, well above anthropogenic production was found recently (\citep{Wallner:2021}).
Despite being based on samples with a low time resolution (two samples, each integrating over the full time-periods of enhanced $^{60}$Fe influx, (see Figure~\ref{fig:60Fe-sediments}), the influx of $^{244}$Pu appears correlated with the $^{60}$Fe influx pattern.
Additional measurements with a higher time resolution are required to confirm this signal being of interstellar medium origin, but also for a better understanding of a possible correlation of $^{244}$Pu and  $^{60}$Fe influx.


\begin{figure}  
\centering
\includegraphics[width=0.8\columnwidth]{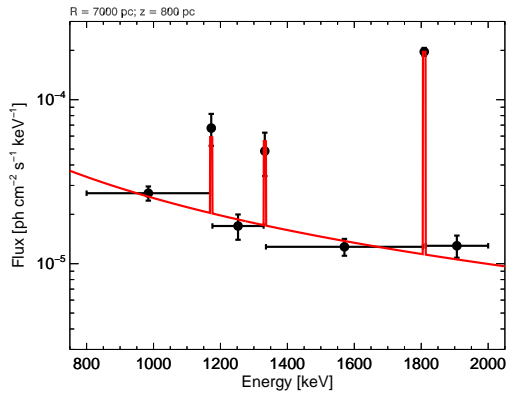}
\caption{The \Al\ and \Fe\ lines as seen with INTEGRAL's high-resolution spectrometer SPI and 15 years of measurements integrated \citep{Wang:2020}. }
\label{fig:60Fe26Al-spec}
\end{figure}   

As anticipated above, from diffuse $\gamma$-ray emission we have obtained constraints on the \fe/\al ratio from the large-scale Galaxy, which averages over many individual sources, and probably even a diversity of regional source ensembles with possibly-different environments such as metallicity.
A first upper limit had been derived from the pioneering RHESSI result on \Fe\  \citep{Smith:2004} that has been discussed above. The result was a value of $\sim 0.4$ for the flux ratio of $^{60}$Fe/$^{26}$Al $\gamma$-ray emissions.
The subsequent INTEGRAL/SPI result on \Fe\ \citep{Wang:2007} reported a flux ratio of the range of $0.09- 0.21$.
Subsequent analysis of more INTEGRAL/SPI data with a different analysis method similarly suggested a ratio in the range $\sim 0.08-0.22$ \citep{Bouchet:2011,Bouchet:2015}.
In all cases, the \Fe\ signal itself is much weaker than the \Al\ signal, and insufficient for imaging studies.

In the re-analysis of \Fe\ signals from 15 years of data \citep{Wang:2020}, model fitting of a range of spatial distribution models for \Fe\ and \Al\ emissions from the same data from one and the same instrument had provided a more reliable basis for flux ratio determinations.
These studies across a broad range of spatial distribution models clearly showed that the \Fe\ $\gamma$-ray emission is of diffuse nature, rather than possibly allocated to rare but bright individual point sources, and somewhat similar, yet not identical in spatial distribution to \Al.
Exploiting the uncertainty information as well as the fit qualities of models,  \citet{Wang:2020} obtained a flux ratio constraint of (18.4 $\pm$~4.2)\%. How can we understand this constraint?

\begin{figure}  
\centering
\includegraphics[width=\columnwidth]{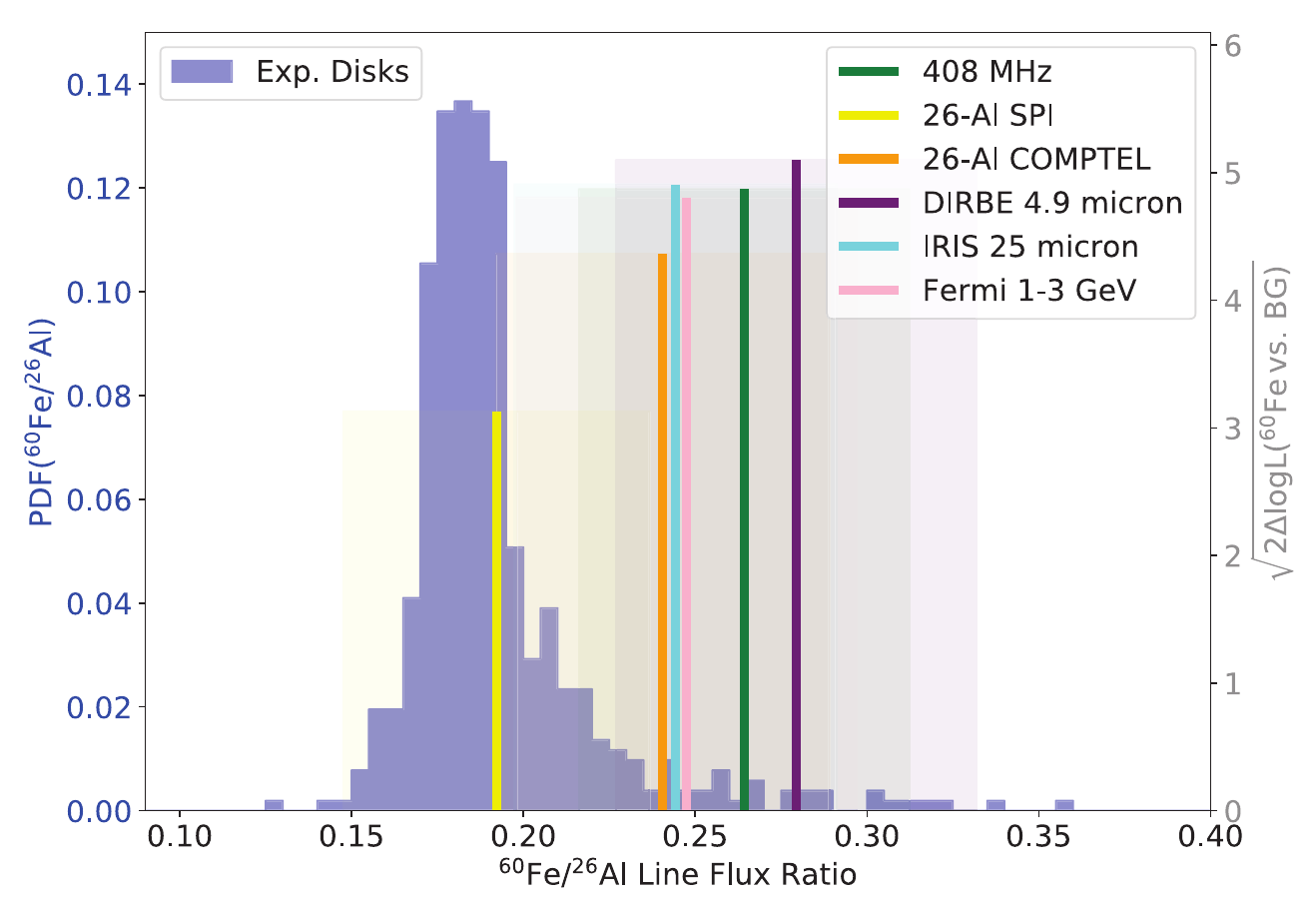}
\caption{The flux ratio  constraints from INTEGRAL's high-resolution spectrometer SPI and 15 years of measurements \citep{Wang:2020}, re-assessed accounting for astrophysical biases and systematics uncertainties as discussed in the text.
Here, the measurement uncertainties are illustrated through the distributions of flux ratio values obtained for a broad range of different exponential-disk models for the Galaxy 
(blue histogram).
Alternatively, the flux ratios derived from a set of tracer maps are shown as vertical lines (significances can be read off the righthand axis, their uncertainties are indicated as shaded bands).}
\label{fig:60Fe26AlFluxRatio}
\end{figure}   

As discussed in \citet{Wang:2020}, systematic uncertainties lead to difficulties in deriving a well-defined observational constraint on the $^{60}$Fe/$^{26}$Al ratio from the $\gamma$-ray measurements.
Figure~\ref{fig:60Fe26Al-spec} shows the spectrum with \al and \fe  results obtained in this study: for a particular (most-plausible) spatial distribution model, size parameters were varied within uncertainties to determine a histogram of the $^{60}$Fe/$^{26}$Al ratio values; additionally, the ratio was derived for a variety of plausible tracers of the massive-star group sources, also indicating uncertainties (in shaded bands).
Tracer maps are a way to reduce uncertainties from the morphology, which cannot be well measured in $\gamma$ rays.
The 25 micron IR map shows the largest improvement above a background-only description for both lines consistently, so that a flux ratio estimate from this map serves as a measure of the systematic uncertainty.
Thus, we obtain a good idea of systematic uncertainties from the width of the histogram, as well as from differences among different source tracers.
Altogether, we thus have constraints for the flux ratio to a range 0.2--0.4, as concluded also by \citet{Wang:2020}.
This is in agreement with the lower limits of 0.10 and 0.33, for the more and the less conservative cases, respectively, derived from deep-sea sediments data \citep{Feige:2018}. While this ratio is a popular constraint considered in studies of core-collapse supernova models, there is currently no agreement on its predicted value between different sets of models. For example, several models predict higher ratios \citep{Sukhbold:2016aa,austin17}, while others predict ratios closer to, or lower than the observations \citep{Limongi2018}. Such differences highlight that our understanding of the nucleosynthesis of these two isotopes in core-collapse supernovae is not settled yet.

\subsubsection{The origin of the solar system}

Using all the available information, we can constrain the $^{60}$Fe/$^{26}$Al isotope ratio in the early Solar System, and compare it to the abundance ratio derived from the flux ratio. This can provide a stronger constraint on the origin of these two nuclei in the early Solar System, since the evolution in the Galaxy of the abundances of \iso{27}Al and \iso{56}Fe is not needed to derive such ratio. A flux ratio 0.2--0.4 corresponds to an abundance ratio of
0.7 -- 1.5. In the early Solar System, instead we have the $^{60}$Fe/$^{26}$Al = $^{60}$Fe/$^{56}$Fe $\times$ $^{27}$Al/$^{26}$Al $\times$ $^{56}$Fe/$^{27}$Al $\simeq (10^{-8}) \times (2 \times 10^{4}) \times 11 = 0.0022$. This is between 2 and 3 orders of magnitude lower than the ratio derived from the interstellar flux ratio, when considering the $^{60}$Fe/$^{56}$Fe reported by ICP-MS and confirmed via RIMS. Even if we considered the higher reported SIMS values for this ratio, of $10^{-7}$-$10^{-6}$, the early Solar System $^{60}$Fe/$^{26}$Al ratios would still be of roughly an order of magnitude lower than that from interstellar $\gamma$-ray spectroscopy.
This clearly indicates that the environment of the Sun's formation did not collect the average $^{60}$Fe/$^{26}$Al in the Galaxy, and that specific sources or strong deviations from \emph{the average} sampled by the flux ratio are required.
These specific sources may range from the winds of massive stars, which carry $^{26}$Al and not $^{60}$Fe, and which also re-shape the surroundings of the sources, to ejecta from supernovae that experience processes such as proton ingestion in the He shell \citep{pignatari2015}, which may favour production of $^{26}$Al over $^{60}$Fe.
The $^{60}$Fe/$^{26}$Al therefore provides us with one of the strongest constraints to understand the place of the birth of the Sun within the Galaxy.


\begin{figure}
\includegraphics[width=\columnwidth]{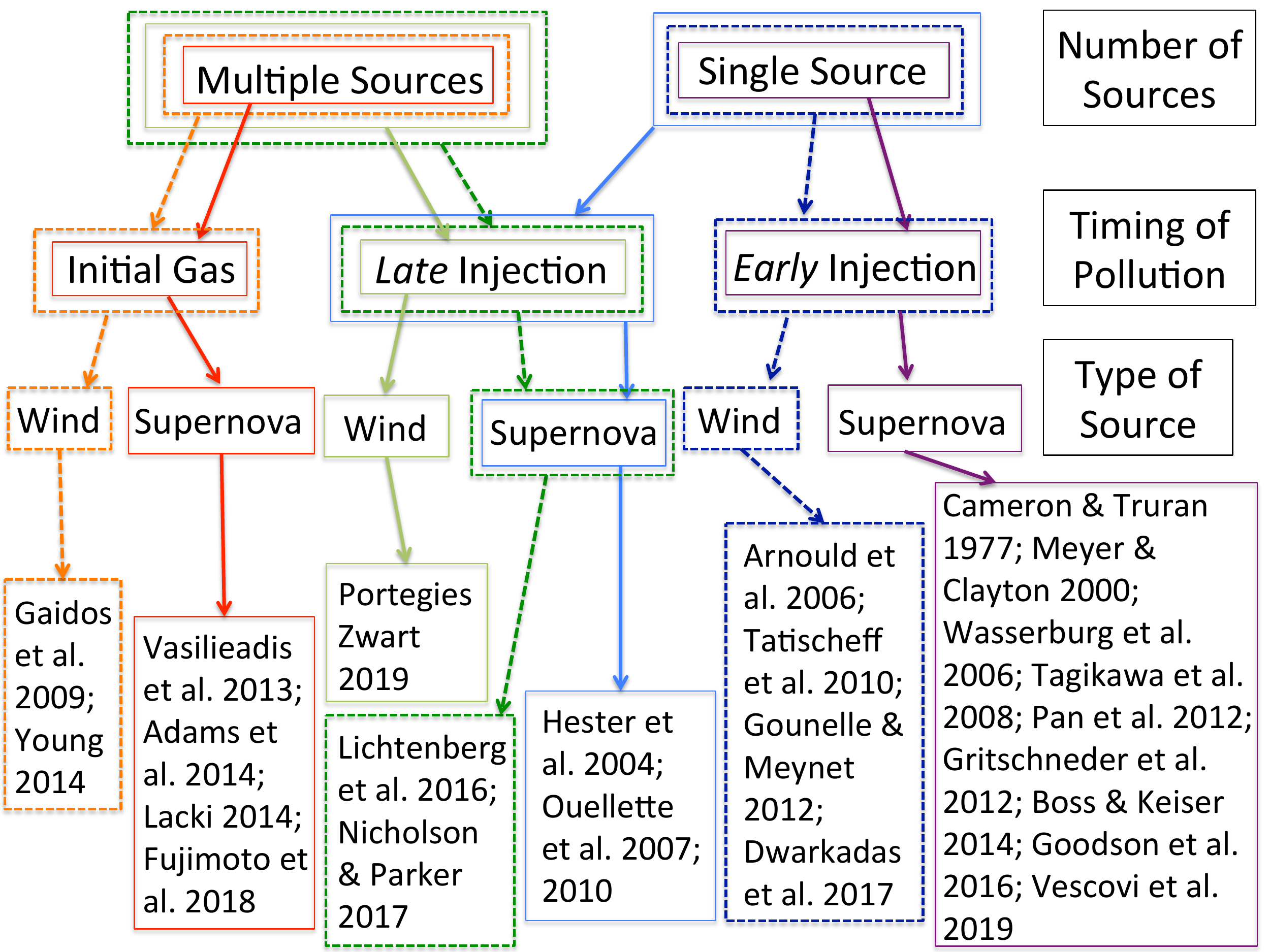}
\caption{Schematic summary of the current scenarios proposed to explain the abundance of \iso{26}Al in the early Solar System and selected references. See text for details. \label{fig:solarsystem}}
\end{figure}

Figure~\ref{fig:solarsystem} presents a schematic diagram of the varieties of scenarios that have been invoked for the presence of \iso{26}Al in the early Solar System\footnote{This figure includes references to the relevant literature, and represents a brief summary, aimed at giving an idea of the complexity of the question of the origin of \iso{26}Al found in the early Solar System and of its possible solutions. A more detailed discussion can be found in Section~5 of a recent review paper by \citet{lugaro18a}.}.
Al Cameron presented the first idea to address the discovery of \iso{26}Al  \citep{cameron77}, and proposed a single supernova being responsible.
Because the likelihood of such an event coincident with Solar-System formation was seen to be extremely small, it was suggested that this same supernova event should also have triggered the collapse of the pre-solar cloud. 
In that case, the injection of \iso{26}Al with the supernova ejecta into the material mix that formed the Solar System would have been an \emph{inevitable} consequence of a supernova triggering solar system formation.
While the idea of a single source is still investigated, more complex scenarios of self-enriched star-forming environments are also currently discussed. In this case, multiple nucleosynthesis sources from different stellar birth events at different times within the galactic medium or within one star-forming giant molecular cloud could have enriched the ambient medium with \iso{26}Al, and the Solar System would then have been born from such enriched cloud material. 
Models of this kind have been discussed and simulated in detail by \citet{Vasileiadis:2013,Kuffmeier:2016,adams14,lacki14,Fujimoto:2018}, and in a late-disk enrichment variant by \citet{lichtenberg16a,nicholson17}. 
Such models
are supported by observations of \iso{26}Al enrichment in star-forming regions via $\gamma$-ray detection, for example, the case of Scorpius Centaurus \citep[see Section~\ref{superbubbles:text} and][] {Krause:2018}.
It should be noted, however, that it is difficult to compare these observations of a total amount of \iso{26}Al in a certain region to the specific \iso{26}Al/\iso{27}Al ratio in the early Solar System: for translation of the two observables to the same scale, the total mass of the observed star-forming region is needed, which is difficult to derive.

A second level of complexity in the different scenarios is opened up as \emph{pollution} of particular timing is separately invoked for the immediate environment of the formation of the solar system. 
In Figure~\ref{fig:solarsystem}, ``Initial gas''/``$Early$ injection'' refer to pollution into the pre-solar cloud before its collapse, 
while ``$Late$ injection'' refers to pollution into an already-formed solar protoplanetary disk. 
``Initial gas'' here refers to pollution into the pre-solar cloud before its collapse, with no causal relation, while ``$Early$ injection'' refers to injection combined with triggering the collapse of the protosolar cloud.
Finally, one may distinguish specific sources of the polluting material. 
Lower-mass stars on the AGB phase have been considered as polluters \citep{wasserburg06,trigo09,lugaro12a,wasserburg17,vescovi18}. 
However, these are more long-lived and hence less likely to be found in star-forming regions.
More likely polluters are Wolf-Rayet stars with their winds and core-collapse supernovae, both originating from short-lived massive stars that are typical present in star-forming regions. 
Core-collapse enrichments have been presented by \citet{cameron77,meyer00,wasserburg06,takigawa08,pan12,Gritschneder12,boss14,goodson16,vescovi18} (``early'') and by \citet{hester04,ouellette07,ouellette10} (``late'').
As discussed above, the explanation from core-collapse supernova sources is challenged by the relatively high abundance of \iso{26}Al in combination with the much lower abundance of \iso{60}Fe. As demonstrated by $\gamma$-ray measurements (see Section~\ref{sec:observations26Al-gammas}), these sources produce both these isotopes at significant abundances. 
Thus, several authors have considered massive-star winds as the possible source of \iso{26}Al in the early Solar System, because \iso{60}Fe is not ejected in such winds.
Wolf-Rayet wind enrichments have been discussed by  \citet{arnould06,tatischeff10,gounelle12,dwarkadas17} (``early'') and by \citep{gaidos09,young14,portegies19} (``late'').

It should also be mentioned that \iso{26}Al in the early Solar System could also have been produced by spallation reactions from accelerated solar particles, mainly protons and \iso{3}He \citep[see e.g.][]{gounelle06,gaches20}.
However, several challenges are present in this explanation, including insufficient available energy to produce the whole \iso{26}Al inventory in the disks \citep{duprat07}, as well as possible inconsistencies with the production of \iso{10}Be, another short-lived radioactive isotope present in the early solar system, which is efficiently produced via spallation \citep[e.g.,][]{jacquet19}.

There is currently no consensus in the community on which scenario(s) is(are) to be favoured.
While the idea of a single source represented the generally accepted solution for several decades, its main problem is the need to be fine tuned. The particularity in terms of exact distance and timing results in low probabilities associated with such single source scenarios \citep{gounellemeibom08,williamsgaidos07,gounelle15}. 
The Solar System may thus be a special case of a planetary system; although this interpretation lacks consensus and may be not readily accepted.
Therefore, chemical-evolution models based on multiple sources naturally occurring within a given star-forming environment have become popular in the past decade or so. 
In these models, the ``early injection'' scenario is not investigated as it is of even lower probability as the single source case, again asking for special circumstances in space and time.



In term of type of sources, stellar winds appear more favourable relatively to supernovae because they can produce self-consistently the abundances of the most short-lived radioactive nuclei found in the early Solar System -- not only \iso{26}Al, but also the shorter-lived \iso{41}Ca and \iso{36}Cl, with half lives 0.1 and 0.3 Myr, respectively \citep{arnould06,Brinkman:2021}, without producing \iso{60}Fe.
The presence of the other, more long-lived, radioactive nuclei, such as \iso{53}Mn, \iso{60}Fe, and  heavier isotopes from \iso{182}Hf to \iso{129}I in the early Solar System, can be attributed to the general chemical evolution of the Galaxy \citep{lugaro14science,cote19a,cote21science,trueman21}, 
without any injection requirements. 

From recent simulations of chemical evolution with improved resolutions in space and time, a better estimate could be obtained from the interplay of massive-star outputs and the thermodynamic states of interstellar gas, in particular its star-forming potential. 
In these, interesting feedback processes put in question the above scenarios of gas transport, i.e., ejections of nucleosynthesis products and how they might end up in newly-forming stars. 
In general, nucleosynthesis ashes are hot and very dynamic interstellar gas, which easily expand into large cavities, as discussed in Sections 2.3.2. and 2.4.5. 
Then, we saw from the current $^{60}$Fe measurements in ocean crusts (Sect. 3.4.1) and its interpretation that \emph{advection} or \emph{sweeping-up} of ashes may occur in a bubble-dominated interstellar medium, which may produce locally enhanced abundances of short-lived radioisotopes -- the $^{60}$Fe measurements are consistent with the Sun's encounter with the wall of the Local Bubble (see Section~3.4.1.). 
Variations of radioisotope abundances by one order of magnitude have been found \citep{Rodgers-Lee:2019}, confirming earlier 
theoretical estimates \citep{Vasileiadis:2013}. This points into the direction of interstellar medium transport from hottest (superbubble) to coldest (star-forming molecular gas) phases being a key factor to possibly control the local abundance ratios in a similar fashion as closeby ``injections'' might.

Overall, a complete scenario for the environment of the birth of the Sun that satisfies all requirements and is in agreement with observations and modelling of star forming region is still missing. 



Whichever was the birth environment of the Solar System, it resulted in its initial \iso{26}Al abundance being enhanced with respect to typical average expectations. This has crucial and far-reaching implications for planets. 
The \iso{26}Al acted as an energy source in the early Solar System, heating the interiors of the planetesimals that formed within the first few million of years. In planetesimals that formed beyond the snow line, this resulted in more water to escape than it would have occurred if the planetesimals were poor in \iso{26}Al  \citep{lichtenberg16b}. Because such planetesimals are expected to have contributed to the building of the terrestrial planets, the initial amount of \iso{26}Al in a protoplanetary disk has an impact on the water content of habitable planets \citep{ciesla15}.
The impact of \iso{26}Al available or not at stellar birth is being discussed today also in relation to exoplanet research and the possibility of life supported by the existence of water \citep{lichtenberg19}.

\section{Conclusions}
\label{sec:conclusions}


We have described the journey of freshly-synthesised $^{26}$Al and $^{60}$Fe  nuclei from their nuclear-reaction production sites within cosmic objects through the interstellar medium to reach us observers within our Solar System and on Earth.
The journey begins with a description of the $^{26}$Al nuclear structure.
Its special spin configuration of 5+ has two important consequences, both arising from less-likely transitions to other nuclear states with their less-special configurations:
(i) $^{26}$Al is long-lived, because a high-multipolarity transition is required for the transition to $^{26}$Mg, and (ii)
$^{26}$Al has an excited state, an isomer, with distinctively different nuclear-reaction properties, which must be considered as a separate isotope within nuclear reaction networks.
In combination, nuclear reactions towards $^{26}$Al production and destruction are more diverse, and more dependent on the reaction environment, as thermal excitations are of more significant influence.
Even if nuclear paths are known for the production of $^{26}$Al, finding the right conditions for this production in cosmic sites is therefore also affected by uncertainties related to stellar and stellar-explosion astrophysics.
For example, we know too little about mass loss, mixing processes, the impact of rotation, and binary interaction; all of these are understood to have impact on stellar structure and evolution, hence on the nuclear-reaction environments and on transport among these before ejection.
The simple effects of removal of outer stellar envelopes, by enhanced mass loss such as Wolf-Rayet phase, or by gravitational pull from a companion such as in close binaries, has been explored as being significant.
But that may just be the tip of the iceberg.

$^{60}$Fe adds an important alternative observational tool.
Its nuclear properties and reactions towards production and destruction are far better constrained by theory and experiment, being less diverse with neutron captures and $\beta$~decays only, as compared to $^{26}$Al.
However, substantial uncertainty remains, as the temperature dependence of the  $\beta$~decays of $^{59,60}$Fe are difficult to measure, and depend on details about Gamov-Teller transitions that are still a challenge for nuclear theory.

A key factor for massive-star yields of both $^{26}$Al and $^{60}$Fe is the recently-discovered issue of supernova explodability, i.e., which stars explode as supernovae at the end of their lives, and which stars instead collapse directly to a black hole.
$^{26}$Al ejections can also occur without a supernova explosion (from the earlier wind phase), while $^{60}$Fe ejection hinges on the explosion to happen.
Therefore, the ratio \fe/\al that would be expected in the interstellar medium surrounding massive-star groups and their supernovae depends on this unknown astrophysics of core-collapses; we illustrated this with examples from population synthesis.

The fact that $\gamma$-ray emissions from \Al\ and from \Fe\ have been detected from extended galactic diffuse emission on an apparently large scale is a confirmation of the predominant origin in massive-star groups and their supernovae.
The observed amounts and spatial distributions of \Al\ appear largely in agreement with massive-star origins and their modeling.
In contrast, \Fe\ abundance as seen in diffuse $\gamma$~rays is typically below expectations from models.
This may be related to the explodability issue discussed above.

But the astronomy of \Al\ and \Fe\ is richer than that:
The Solar System is an astronomical instrument of its own, having received environmental isotope imprints at the time of its formation, and receiving flows of interstellar gas in recent history.
Substantial detail is available from such data, which complement classical telescope astronomy with telescopes that measure cosmic radiation.
The conditions of the Solar System formation are preserved in meteorites, and these show traces of the presence of both \al and \Fe.
A crucial question that we can try to answer using \al and \fe thus is the origin of the Sun.
Is it \emph{typical}? Or has it been shaped by one ore more special nearby events?
We do not know, but we need to face the rich observational material on the isotopic composition here, and compare these observations with expectations for all the possible scenarios.
\Al\ abundances here appear clearly enriched with respect to predictions of massive-star models and its large-scale appearance in our Galaxy, while for \fe
no deviations can be claimed.
Recent improvements on the accuracy and precision of measurements of stable isotope anomalies of nucleosynthetic origin in meteoritic materials are also being exploited to investigate the evolution of the solar nebula within its formation environment \citep[see, e.g.][]{Mezger:2020,kleine20,Bermingham:2020}. 
Combining these new advances on stable isotopes  with the lessons coming from the radioactive nuclei, we expect better constraints to the environment of the Sun's birth and how it shaped the formation and evolution of the Solar System.

\Al\ also is observed in stardust recovered from meteorites, which preserve abundances of nucleosynthesis sources as this stardust formed in the atmosphere or immediate neighbourhood of these.
AGB stars are copious dust producers, and so the \Al\ fractions in dust grains that carry an isotopic signature of AGB stars have helped to point to efficient \Al\ production here, as $^{26}$Al/$^{27}$Al ratios are measured to be higher than predicted by models.
Similarly, grains attributed to supernova origins have shown $^{26}$Al/$^{27}$Al ratios higher than predicted by models.
Thus, stardust indicates that models may be deficient; the astrophysics of dust formation around AGB stars and supernovae, however, is very uncertain, so that the link to mainstream models and to bulk \Al\ and \Fe\ as seen in diffuse $\gamma$~rays remains uncertain. Overall, such material samples from more than 4.6~Gyr ago provide us an impressive detailed view on common stellar sources of nucleosynthesis.

At much more recent times, of the order of Myr, the Solar System appears to have been exposed to an influx of \Fe\ from nucleosynthesis.
Lunar material showed live \Fe, and sediments on Earth have revealed time-resolved data on \Fe\ influx.
Clearly, a prominent such \Fe\ influx period occurred around 3~Myr before present, and is more extended in time than expected from a single supernova ejecta cloud passing over the Solar System.
Traces of influx can be found also out to $\sim$8~Myr.
Detecting \Al\ in such sediments is challenging due to a high background from cosmic-ray production in Earth's atmosphere, that is relatively minor for the case of \Fe.
There is some tension with measurements, if the ratio \fe/\al as measured in diffuse $\gamma$~rays from the Galaxy is adopted to represent also current nearby nucleosynthesis.

Finally, interstellar cosmic rays that can be detected by  instruments on space satellites have been shown to also include \Fe. This is a surprise and proves a stellar/supernova origin, because \Fe\ is not expected to be produced by interstellar spallation as cosmic~rays penetrate interstellar gas, unlike \Al\ production that has been known from such cosmic-ray measurements earlier.
But also for such current-day cosmic rays, the ratio \fe/\al cannot be measured with sufficient precision for further inferences.
Together, all these Solar-System data show interesting deviations from a generic large-scale picture about \Al\ and \Fe\ origins as it has been outlined from theory and from Galactic $\gamma$-ray data.
So the question keeps coming back to us: How special is the solar environment?

Upon more precise investigation, \Al\ and \Fe\ $\gamma$ emissions should differ for several reasons:
\Al\ is ejected also from the earlier stellar winds, whereas \Fe\ ejection occurs only from the later phase of core-collapse and explosion.
Circumstellar environments for interstellar propagation of \Al\ and \Fe\ thus may differ.
Furthermore, the longer lifetime of \Fe\ implies that there is more time available to propagation in the interstellar environment during the emission of the characteristic $\gamma$-ray signal.
%
Standard galactic chemical evolution models usually assume that the interstellar medium is instantaneously mixed, while radioactive nuclei with short half lives such as \Al\ and \Fe\ are  sensitive to spatial and temporal heterogeneities and transport within the interstellar medium.
%
\al and \fe radioactivities allow to study several unknowns of the different environments that may shape the paths of ejecta from nucleosynthesis towards star-forming gas within a galaxy.
\Al\ $\gamma$ rays have shown that the ejecta flow remains fast over time scales longer than simple models of the interstellar medium and the embedding of nucleosynthesis sources would suggest. Large-scale cavities and bubbles obviously play a role in re-cycling ejecta from massive-star and core-collapse supernova nucleosynthesis.

We see that the tracks of cosmic nucleosynthesis can be complex, and the study of cosmic nuclear abundance evolution is a challenging and essential field of astrophysics. We want to know our origins, and the origins of the chemical matter that makes up our environment and ourselves.
Radioactivity obviously adds significantly to this study of cosmic chemical evolution\footnote{The astronomical exploitation of radioactivities, also described in great detail in a recent textbook \emph{Astrophysics with radioactive Isotopes} \citep{Diehl:2018d}, thus provides an unique complement to astrophysical research.}:
It adds radioactivity as a cosmic clock. This has profound implications.
Material flows can be tracked from dense and occulted nuclear-processing sites into the interstellar medium, then through the interstellar medium, and finally from the Solar-System environment to the material probes that we can measure.

\begin{acknowledgements}

This review paper has emerged from the work of a scientific team who met twice in Beijing (in 2017 and 2019) thanks to the support of the International Space Science Institute (ISSI) in Beijing. We thank Reto Trappitsch, Friedrich-Karl Thielemann, Michael Paul, Toshitaka Kajino, Zhenwei Liu, Zihong Li, Bernhard M\"uller, and Davide Lazzati for discussion. 
We thank the referee for suggestions that helped to improve our presentation and discussion of the science status.
This work is supported by the ERC via CoG-2016 RADIOSTAR [11] (Grant Agreement 724560) and from work within the ``ChETEC'' COST Action (CA16117), funded by the COST program (European Cooperation in Science and Technology).
A.S. acknowledges support by the US Department of Energy [DE-FG02-87ER40328 (UM)].
A.H. acknowledges support by the Australian Research Council (ARC) Centre of Excellence (CoE) for Gravitational Wave Discovery (OzGrave) project number CE170100004, by the ARC CoE for All Sky Astrophysics in 3 Dimensions (ASTRO 3D) project number CE170100013, and US National Science Foundation under Grant No.\ PHY-1430152 (JINA Center for the Evolution of the Elements).

\end{acknowledgements}

\bibliographystyle{pasa-mnras}

\bibliography{referencesISSI}

\end{document}